\documentclass{aa}  

\usepackage{graphicx}
\usepackage{txfonts}
\usepackage{amsmath,amssymb}
\usepackage{url}
\usepackage{lipsum}
\usepackage{subcaption}
\usepackage{lscape} 
\usepackage{placeins}
\usepackage{booktabs}

\titlerunning{Tightening cosmological constraints from GRBs}
\authorrunning{Wei Hong et al.}

\begin{document}
\title{Tightening cosmological constraints within and beyond $\Lambda$ cold dark matter using gamma-ray bursts calibrated with Type Ia supernovae}
\author{Wei Hong\inst{1,2,3}
\and Luca Izzo\inst{1}\fnmsep\thanks{luca.izzo@inaf.it}
\and Massimo Della Valle\inst{4}
\and Orlando Luongo \inst{5,6,7,8,9}
\and Marco Muccino\inst{5,9,10}
\and Tong-Jie Zhang\inst{2,3}\fnmsep\thanks{tjzhang@bnu.edu.cn}
}
        
\institute{INAF - Osservatorio Astronomico di Capodimonte, Salita Moiariello 16, I-80131 Napoli, Italy
\and Institute for Frontiers in Astronomy and Astrophysics, Beijing Normal University, Beijing 102206, China
\and Department of Astronomy, Beijing Normal University, Beijing 100875, China
\and INAF - Osservatorio Astronomico di Padova, Vicolo dell'Osservatorio 5, I-35122 Padova, Italy
\and Universit\`a di Camerino, Via Madonna delle Carceri, Camerino, 62032, Italy
\and INAF - Osservatorio Astronomico di Brera, Milano, Italy
\and Istituto Nazionale di Fisica Nucleare, Sezione di Perugia, Perugia, 06123, Italy
\and SUNY Polytechnic affiliation, 13502 Utica, New York, USA      
\and Al-Farabi Kazakh National University, Al-Farabi av. 71, 050040 Almaty, Kazakhstan
\and ICRANet, Piazza della Repubblica 10, 65122 Pescara, Italy}
           
\date{Received XX XX, 20XX}
 
\abstract
{Gamma-ray bursts (GRBs) reach redshifts beyond Type Ia supernovae (SNe Ia) and can extend distance measurements into the early Universe, but their use as standardisable candles is limited by the circularity problem in the cosmology-dependent calibration of GRB luminosity correlations.}
{We present a methodology for reducing this calibration circularity by combining the Pantheon$+$ SH0ES SNe Ia calibrated through Cepheid and parallax measurements, a distance reconstruction based on artificial neural networks, and two GRB correlations (Amati and Combo) into a distance ladder from low to high redshift, while separately testing the model stability and cosmological consistency of the adopted GRB samples.}
{We used ReFANN to reconstruct the Pantheon$+$ SH0ES SN Ia Hubble diagram at low redshifts without imposing a cosmological model. The hyper-parameters were optimised via approximate Bayesian computation rejection and a risk function. Before calibration, we jointly fitted the GRB correlation and cosmological parameters in four background models and compared the GRB-only constraints with observational Hubble data (OHD). We then calibrated the Amati and Combo relations with low-redshift ($z<1$) GRBs and inferred distances for GRBs at $z \geq 1$. Finally, a joint Bayesian analysis constrains the cosmological and GRB-correlation parameters.}
{The Amati sample shows no significant correlation-parameter dependence on the tested cosmologies, with a maximum joint displacement of $0.24\sigma$, and the A118 sample provides a consistent benchmark. Its GRB-only cosmological constraints remain broad and are moderately shifted towards higher $\Omega_m$, while the OHD value of $\Omega_m$ remains within the 95\% credible interval of the Amati result. In contrast, the Combo/C174 sample is not consistent with the external matter density constraint and is therefore retained only as an exploratory cross-check. Plausible changes in the SN Ia absolute distance zero point mainly shift the Amati intercept, leaving the remaining parameters stable within $0.12\sigma$.}
{We present a method that mitigates cosmology-dependent circularity in GRB calibration and extends the GRB Hubble diagram based on the Pantheon$+$ SH0ES sample to $z \sim 9$. The results support the use of the present Amati sample as a high-redshift distance indicator, but we emphasise that its independent cosmological constraints remain weak and that the inferred absolute scale is conditional on the adopted SN Ia calibration.}

   \keywords{cosmological parameters --
                Gamma-ray burst: general --
                supernovae: general --
                dark energy
               }

   \maketitle
   \nolinenumbers

\section{Introduction}
Observational cosmology has recently entered into the era of precision measurement \citep{2013PhR...530...87W,2015arXiv150104076M}. The cosmological parameters, including the Hubble constant ($H_0$), the matter density parameter ($\Omega_m$), and the dark-energy equation-of-state parameters ($w_0$ and $w_a$) in the $w_0w_a$CDM model, can now be measured with high precision using a range of probes. The $w_0w_a$CDM model adopts a time-dependent dark-energy equation of state, $w(z)=w_0+w_a z/(1+z)$. However, significant tensions remain among different observations, including the difference between local distance-ladder measurements and values inferred from the cosmic microwave background (CMB). A central question in modern cosmology is how to evaluate the constraining ability of various data sets on cosmological parameters in a model-independent manner \citep{2023MNRAS.518.2247L,2023MNRAS.523.4938M,2020A&A...641A.174L}. Among the many cosmological probes, Type Ia supernovae (SNe Ia) play a crucial role at low redshifts owing to their well-established standard-candle properties \citep{1993ApJ...413L.105P}. SNe Ia provide a well-calibrated Hubble diagram at low redshifts, but their samples become increasingly sparse towards the high-redshift regime \citep{1922MeLuF.100....1M}. Extending the distance ladder to higher redshifts therefore requires introducing intrinsically brighter objects as complementary probes. Several `beyond-standard' probes have been developed, as reviewed by \citet{2022LRR....25....6M}. 

Among these, gamma-ray bursts (GRBs) are highly energetic transients that can be detected at redshifts up to $z\sim 9$ \citep{2004RvMP...76.1143P,2006RPPh...69.2259M,2007RSPTA.365.1377T,2009Natur.461.1254T,2009Natur.461.1258S,2009ARA&A..47..567G,2011ApJ...736....7C}. Their main cosmological role is not to compete with SNe Ia at low redshifts but to extend distance measurements into a regime where SNe Ia are scarce \citep{2004ApJ...612L.101D,2004ApJ...616..331G,2005ApJ...633..611L,2007ApJ...660...16S,2009A&A...508...63I,2009MNRAS.400..775C,2015NewAR..67....1W,2017A&A...598A.112D,2017ApJ...848...88D}. At the same time, GRBs involve complex emission physics and exhibit intrinsic luminosities that span several orders of magnitude with a highly dispersed distribution. Hence, one must rely on empirical correlations, such as the Amati relation \citep{2002A&A...390...81A,2006MNRAS.372..233A,2008MNRAS.391..577A} or the Yonetoku relation \citep{2004ApJ...609..935Y,2005MNRAS.360L..45G,2010PASJ...62.1495Y}, to find a relationship between the equivalent isotropic radiated energy and cosmology before using GRBs for cosmological parameter constraints \citep{2007ApJ...660...16S,2013IJMPD..2230028A,2021Galax...9...77L}.

The Amati relation connects the rest-frame spectral peak energy ($E_{p,i}$) with the isotropic-equivalent energy ($E_{\mathrm{iso}}$) and has been a widely used empirical correlation for GRBs \citep{2002A&A...390...81A,2006MNRAS.372..233A,2008MNRAS.391..577A}. Given a GRB's redshift ($z$), $E_{p,i}$, and fluence ($S_{\mathrm{obs}}$), one can calibrate its luminosity distance ($d_L$) using the Amati parameters, enabling the construction of a high-redshift ($z \geq 2$) Hubble diagram to constrain cosmology \citep{2007ApJ...660...16S,2008A&A...487...47M,2011MNRAS.415.3580D}. However, its standard application suffers from the circularity problem \citep{2008ApJ...685..354L,2010A&A...519A..73C,2011MNRAS.415.3580D,2011NewA...16...57G,2019MNRAS.486L..46A}: calibrating the correlation with luminosity distances computed under a fiducial cosmology leaves the subsequent cosmological constraints dependent on that assumed model. External distance information, such as the SN Ia Hubble diagram, can mitigate this calibration circularity. However, removing the cosmological model from the calibration is distinct from establishing whether a particular GRB sample is standardisable. The latter also requires checking whether the correlation parameters are stable against changes in the assumed cosmology and whether the corresponding GRB-only constraints are compatible with established cosmological probes. Such simultaneous correlation-cosmology tests have been applied to Amati samples by \citet{2020MNRAS.499..391K} and subsequently to several Amati and Combo samples by \citet{2021JCAP...09..042K}. The A118 sample has also been combined with more established cosmological probes to investigate constraints on the background expansion and dark-energy models \citep{2021JCAP...09..042K,2021MNRAS.501.1520C,2022MNRAS.513.5686C,2023PhRvD.107j3521C}.

The Combo relation provides an alternative empirical correlation by combining prompt and afterglow observables \citep{2008MNRAS.391..577A,2012MNRAS.425.1199B,2015A&A...582A.115I,2021ApJ...908..181M}. It connects the rest-frame X-ray afterglow plateau luminosity ($L_0$) to the spectral peak energy ($E_{p,i}$), the plateau duration ($\tau$), and the post-plateau decay index ($\alpha_{PL}$) through $\log L_0=\log A+\gamma \log E_{p,i}-\log [\tau/|1+\alpha_{PL}|]$ \citep{2013MNRAS.428..729M,2014A&A...565L..10R}. Once calibrated, both the Amati and Combo relations can provide high-redshift distance estimates. Their different observables also allow the two calibration approaches to be compared. Their use is, nevertheless, affected by intrinsic scatter, outliers, detector thresholds, and selection effects \citep{2009ApJ...694...76B,2015ApJ...800...31D,2015ApJ...806...44P,2018PASP..130e1001D}. A complete treatment of these effects requires modelling the GRB population and detector response. In this study, we instead treated the remaining selection effects as a systematic limitation and did not include them explicitly in the likelihood. The Combo relation has been applied to constrain the dark energy equation of state $w(z)=w_0+w_az/(1+z)$ \citep{2001IJMPD..10..213C,2003PhRvL..90i1301L}. Due to the scarcity of low-redshift GRBs, a redshift-binned approach is often adopted \citep{2021ApJ...908..181M,2021Galax...9...77L}. Although this approach yields preliminary constraints, they are affected by binning choices and by the propagation of uncertainties from low to high redshift \citep{2003PhRvL..90c1301H,2004PhRvL..92x1302W,2012JCAP...06..036S}.

To alleviate the aforementioned issues, a common approach is to use the SN Ia distance scale to non-parametrically reconstruct the relationships $d_L(z)$ and $\mu(z)$, which then calibrate low-redshift GRB distances for empirical relations like Amati and Combo \citep{2006MNRAS.366.1081S,2008ApJ...685..354L,2010A&A...519A..73C}. Such reconstructions, without assuming a background cosmology, have been explored using Gaussian processes \citep{2006gpml.book.....R,2012PhRvD..85l3530S,2012JCAP...06..036S,2013ascl.soft03027S,2015PhRvL.114j1304S,2021ApJ...915..123S,2023JCAP...02..014H,2023ApJS..266...27Z,2024MNRAS.534...56Z}, non-parametric regression \citep{2006MNRAS.366.1081S,2007MNRAS.380.1573S,2010PhRvD..82j3502H,2011PhRvD..84h3501H,2024ApJS..270...23Z}, and machine-learning techniques \citep{2021MNRAS.503.4581L,2023EPJC...83..956D,2023EPJC...83..304G,2023ApJS..268...67H,2025EPJC...85.1005C}. In this study, we employed artificial neural networks (ANNs) for three main reasons. They provide flexible reconstructions of both $d_L(z)$ and $\mu(z)$, scale favourably to large and heterogeneous supernova samples, and allow network and hyper-parameter uncertainties to be incorporated into our approximate Bayesian computation (ABC) framework. Unlike standard Gaussian process reconstructions that depend sensitively on kernel choices, feed-forward ANNs learn the mapping directly from data without assuming a specific covariance form. Unlike generic non-parametric regression methods such as splines or local polynomials, ANNs also provide a convenient framework for imposing smoothness and monotonicity and for propagating reconstruction uncertainties into the subsequent GRB calibration. Using the Pantheon$+$ SH0ES sample \citep{2022ApJ...938..110B,2022ApJ...938..113S} and ReFANN \citep{2020ApJS..246...13W}, we reconstructed $d_L(z)$ and $\mu(z)$ without assuming a cosmological model, such as $\Lambda$ cold dark matter ($\Lambda$CDM). To reduce the dependence on network architecture and hyper-parameter choices, we combined risk-function evaluation \citep{1999ITNN...10..988V,2001astro.ph.12050W} with an ABC rejection scheme. For model comparison within ABC, we used goodness-of-fit statistics \citep{2017A&C....19...16J,2018MNRAS.477.2874A,2019MNRAS.488.4440A,2021JCAP...08..027B,2023ApJS..266...27Z}, including a composite metric similar to the log marginal likelihood (LML) \citep{10.3389/fbuil.2017.00052,2018JCAP...04..051G}. The ABC framework itself is informed by established sampling and model selection principles \citep{2008arXiv0805.2256B,2008ConPh..49...71T,2008Bioin..24.2713C,2013ApJ...764..116W,2019AnRSA...6..379B}. All candidate networks were evaluated using the same risk function and ABC rejection criterion, allowing the uncertainty associated with hyper-parameter selection to be propagated into the reconstructed distance relations.

Before applying this calibration, we first tested the exact GRB samples used in this study. We jointly constrained their correlation and cosmological parameters in flat and non-flat $\Lambda$CDM and X cold dark matter (XCDM) models and used the A118 Amati sample as an independent benchmark. In the XCDM parametrisation, dark energy is described as a fluid with a constant equation-of-state parameter ($w_X$). We further compared the GRB-only cosmological constraints with the observational Hubble data (OHD). These tests distinguish the model stability of a GRB correlation from the subsequent SN Ia calibration. After this validation, we calibrated the Amati ($A_A, \gamma_A, \sigma_{\text{cal}}^A$) and Combo ($A_C, \gamma_C, \sigma_{\text{cal}}^C$) relations with the ANN-reconstructed distances of low-redshift GRBs and inferred distances for high-redshift events. Their parameters were determined through Bayesian inference \citep{2001AJ....121.2879B} with Markov chain Monte Carlo sampling \citep{1953JChPh..21.1087M,1970Bimka..57...97H, 2005physics..11182D,2007ApJ...665.1489K}, including observational uncertainties and intrinsic scatter. We then jointly sampled the cosmological and GRB-correlation parameters in flat $\Lambda$CDM ($H_0, \Omega_m, \Omega_{\Lambda}$) and $w_0w_a$CDM ($H_0, \Omega_m, w_0, w_a$) \citep{1982ApJ...263L...1P,2001IJMPD..10..213C, 2003PhRvL..90i1301L}. Because the GRB distance scale is inherited from the Pantheon$+$ SH0ES reconstruction, the inferred $H_0$ is conditional on that absolute calibration and is not an independent local measurement. We therefore also tested the sensitivity of the results to alternative absolute distance zero points. The role of the GRBs in this analysis is primarily to extend the calibrated Hubble diagram to higher redshifts.

The rest of the paper is organised as follows. In Sect.~2, we describe the Pantheon$+$ SH0ES SN Ia data and GRB samples. In Sect.~3, we present the ANN reconstruction of $d_L(z)$ and $\mu(z)$, together with ABC rejection sampling and risk-function evaluation. In Sect.~4, we calibrate the Amati and Combo relations with the ANN-reconstructed distances, construct the high-redshift GRB Hubble diagram, and report cosmological constraints under flat $\Lambda$CDM and $w_0w_a$CDM. Section~5 discusses the main findings, systematic uncertainties, limitations, and future extensions. Finally, Sect.~6 summarises the conclusions.

\section{Data sets and pre-processing}
This section provides a detailed description of the observational datasets used in this paper and the corresponding pre-processing procedures. Our analysis is based on two main data sources: the Pantheon$+$ SH0ES SN Ia sample and the \textit{Fermi} Gamma-ray Burst Monitor (GBM) and \textit{Swift}-X-Ray Telescope (XRT) GRB samples \citep{2009ApJ...702..791M,2012ApJS..199...18P,2014ApJS..211...13V,2016ApJS..223...28N,2020ApJ...893...46V,2012ApJS..199...19G,2014ApJS..211...12G,2021ApJ...913...60P}. From the Pantheon$+$ SH0ES data release \citep{2022ApJ...938..110B,2022ApJ...938..113S}, we used for each supernova the redshift ($z$), preferentially the Hubble-diagram redshift ($z_{\mathrm{HD}}$), and the distance modulus ($\mu_{\mathrm{SNe}}$) with its uncertainty ($\sigma_{\mu,\mathrm{SNe}}$). The SN Ia dataset contains more than 1500 uniformly processed events and is among the largest public SN Ia datasets currently available. The redshift coverage extends from the nearby Universe ($z \sim 0.001$) out to $z \sim 2.261$, providing a solid basis for constructing the distance-redshift relation. It provides the low-redshift distance scale from which ReFANN reconstructs $d_L(z)$ and $\mu(z)$ without assuming a background cosmological model. These reconstructed relations are then used to calibrate the low-redshift GRBs. In the subsequent GRB analysis, the Pantheon$+$ information therefore enters through the ANN-reconstructed distance scale rather than being added again as an independent copy of the same SN Ia likelihood.

The Pantheon$+$ SH0ES data set provides a covariance matrix $\mathbf{C}_{\mu}$ for the SNe Ia distance moduli, including both statistical and systematic uncertainties. The luminosity distance $d_{L,i}$ and its uncertainty $\sigma_{d_L,i}$ were obtained by error propagation:
\begin{equation}
	\begin{aligned}
		d_{L,i} &= 10^{\left(\frac{\mu_i-25}{5}\right)}\,\mathrm{Mpc},\\
		\sigma_{d_L,i} &= \sqrt{\left(\mathbf{C}_{d_L}\right)_{ii}}
		= \frac{\ln 10}{5}\, d_{L,i}\, \sqrt{\left(\mathbf{C}_{\mu}\right)_{ii}},
	\end{aligned}
	\label{eq:dl_from_mu}
\end{equation}
where the propagated covariance matrix is $\mathbf{C}_{d_L}=\mathbf{J}\,\mathbf{C}_{\mu}\,\mathbf{J}^{\mathrm{T}}$, and the Jacobian matrix is defined by $J_{ij}=\partial d_{L,i}/\partial\mu_j=(\ln 10/5)\,d_{L,i}\,\delta_{ij}$.

The GRB sample used in this paper is primarily based on observations from \textit{Fermi} GBM and \textit{Swift}-XRT, combining prompt emission and X-ray afterglow data from \citet{2015A&A...582A.115I} and \citet{2021ApJ...908..181M}. This provides the quantities needed to calibrate both the Amati relation and the Combo relation, the latter of which relies specifically on \textit{Swift}-XRT afterglow measurements. With its broad energy coverage and time resolution, \textit{Fermi} GBM can robustly provide spectral parameters. Since the Amati correlation is mainly established and tested for long GRBs ($\mathrm{T}90>2\mathrm{~s}$), we imposed the following physical and data-quality criteria: (a) Reliable redshift measurements. Each GRB must have a well-determined redshift $z_{\mathrm{GRB}}$, explicitly reported in the literature or in curated GRB databases. (b) Complete spectral information. We filtered the data provided by the observed fluence $S_{\mathrm{obs}}$ and the observed-frame spectral peak energy $E_{p, \mathrm{obs}}$ together with their $1 \sigma$ statistical uncertainty $\sigma_{S_{\mathrm{obs}}}$ and $\sigma_{E_{p, \mathrm{obs}}}$, respectively. Moreover, the time-integrated spectrum of the GRB must be well fitted by a Band function, and the fit must provide the low-energy spectral index $\alpha$ and high-energy spectral index $\beta$ together with their $1 \sigma$ statistical uncertainty $\sigma_{\alpha}$ and $\sigma_{\beta}$, respectively. (c) Strict data-quality control. We discarded events with failed spectral fits or with huge relative uncertainties, which are relative errors of order $\geq 100\%$, as mentioned above, so as to ensure the robustness of the input data. The Amati catalogue used in this paper contains 77 GRBs. Our Amati sample was constructed by cross-matching the C174 GRB catalogue with bursts having available \textit{Fermi} prompt-emission measurements and selecting the matched events for the Amati analysis. The Combo catalogue adopted here corresponds to the C174 sample of \citet{2021ApJ...908..181M}. We used their published and uniformly processed values of the rest-frame peak energy $E_{p,i}$, plateau flux $F_0$, rest-frame plateau duration $\tau$, and post-plateau decay index $\alpha_{PL}$. One event, GRB171205A, has the published rounded value $\alpha_{PL}=-1.00$, for which the factor $|1+\alpha_{PL}|$ vanishes. We therefore excluded this event, leaving 173 usable Combo GRBs. Before ANN calibration, we used the A118 Amati sample of \citet{2021JCAP...09..042K} as an independent benchmark, analysed with the same likelihood and parameter conventions as our 77-event sample. We further identified 43  GRBs common to A118 and our catalogue, which  fell to 41 after the strict-quality cut, to assess data-processing and sample-selection effects. External consistency is tested with 42 OHD of $H(z)$ over $0.07\leq z\leq1.965$. The data used in the Amati, Combo, and OHD analyses are provided as machine-readable electronic Tables~E1--E3.

To convert the observables into the rest-frame quantities required by the Amati relation, we first computed the rest-frame spectral peak energy
\begin{equation}
	E_{p, i}=E_{p, \mathrm{obs}}\left(1+z_{\mathrm{GRB}}\right),\label{sppe}
\end{equation}
where $E_{p, \mathrm{obs}}$ is the observed-frame peak energy and $z_{\mathrm{GRB}}$ is the GRB redshift. The corresponding uncertainty in logarithmic space was approximated by
\begin{equation}
	\sigma_{\log_{10} E_{p, i}}=\frac{1}{\ln 10} \frac{\sigma_{E_{p, i}}}{E_{p, \mathrm{obs}}\left(1+z_{\mathrm{GRB}}\right)},
\end{equation}
with $\sigma_{E_p}$ the $1 \sigma$ uncertainty on $E_{p, \mathrm{obs}}$. Next, to obtain the bolometric fluence $S_{\mathrm{bol}}$, we followed the procedure outlined by Schaefer (2006). Given the observed fluence $S_{\mathrm{obs}}$ in the instrumental energy band and an appropriate spectral model that fits the data, the bolometric fluence was estimated via \citep{2007ApJ...660...16S}
\begin{equation}
	S_{\mathrm{bol}}=S_{\mathrm{obs}} \frac{\int_{1 /(1+z)}^{10^4 /(1+z)} E \Phi(E) \mathrm{d} E}{\int_{E_{\min }}^{E_{\max }} E \Phi(E) \mathrm{d} E},\label{bolof}
\end{equation}
where $\left(E_{\min}, E_{\max}\right)$ denotes the detector energy range, and for the \textit{Fermi} GBM bursts used in this paper it corresponds to the nominal GBM energy band $\left(10,1000\right)\mathrm{keV}$. The function $\Phi(E)$ is the photon spectrum adopted to fit the GRB data. In this paper, we used the Band function, written as \citep{1993ApJ...413..281B}
\begin{equation}
	\Phi(E)= \begin{cases}\tilde{A} E^\alpha \exp\left[-(2+\alpha) E/E_{p, \mathrm{obs}}\right], & \mathrm{if~} E \leq\left[\frac{\alpha-\beta}{2+\alpha}\right] E_{p, \mathrm{obs}}, \\ \tilde{B} E^\beta, & \mathrm{otherwise,}\end{cases}
\end{equation}
where $\tilde{A}$ and $\tilde{B}$ are normalisation constants, $\alpha$ and $\beta$ are the low- and high-energy spectral indices.

Once the bolometric fluence $S_{\mathrm{bol}}$ had been estimated for each GRB in the rest-frame energy band of $1$--$10^{4}~\mathrm{keV}$, we computed the isotropic-equivalent radiated energy $E_{\mathrm{iso}}$ using the relation between luminosity distance and fluence \citep{2002A&A...390...81A,2006MNRAS.372..233A,2008MNRAS.391..577A}
\begin{equation}
	E_{\mathrm{iso}}=\frac{4 \pi d_L^2 S_{\mathrm{bol}}}{1+z_{\mathrm{GRB}}}.\label{amati}
\end{equation}
The determination of luminosity distances depends on the redshift of GRB. For low-redshift GRBs $(z_{\mathrm{GRB}}<z_{\mathrm{cut}}=1)$, the luminosity distance $d_L$ is obtained from the $d_L(z)$ relation reconstructed through ANN from the Pantheon$+$ SN Ia dataset, as described in Sect. 3. This choice uses the redshift range where the SN Ia sample is dense and well measured than in the $z>1$ regime. For high-redshift GRBs $(z_{\mathrm{GRB}}\geq z_{\mathrm{cut}})$, the luminosity distance $d_L$ is derived in Sect. 4 from the calibrated Amati relation using the rest-frame peak energy $E_{p,i}$ and bolometric fluence $S_{\mathrm{bol}}$.

For GRBs at low redshift, the distance modulus $\mu(z)$ was also obtained from the relation reconstructed by the ANN. We then used the low-redshift GRBs with available measurements of $E_{p,i}$, $F_0$, $\tau$, and $\alpha_{PL}$ to calibrate the Combo relation \citep{2015A&A...582A.115I,2021ApJ...908..181M}
\begin{equation}
	\begin{aligned}
		\mu_{\mathrm{GRB}}= & -97.45+\frac{5}{2}\left[\log \left(\frac{A_C}{\operatorname{erg~s}^{-1}}\right)+\gamma_C \log \left(\frac{E_{\mathrm{p}, \mathrm{i}}}{\mathrm{keV}}\right)\right. \\
		& \left.-\log \left(\frac{\tau / \mathrm{s}}{|1+\alpha_{PL}|}\right)-\log \left(\frac{F_0}{\operatorname{erg~cm}^{-2} \mathrm{~s}^{-1}}\right)-\log 4 \pi\right],\label{combo}
	\end{aligned}
\end{equation}
thereby determining its slope $\gamma_C$, normalisation $A_C$, and intrinsic scatter $\sigma_{\mathrm{ex}}$. For GRBs at high redshift ($z \geq z_{\mathrm{cut}}$), the distance modulus $\mu(z)$ was derived in Sect. 4 by applying the calibrated Combo relation. This procedure follows the same calibration logic as the Amati relation, but uses a distinct set of prompt and afterglow observables, providing an independent and complementary distance extension of the Hubble diagram.

\section{ANN model optimisation and distance reconstruction}
This section presents the methodology to reconstruct the luminosity distance $d_L(z)$ and distance modulus $\mu(z)$ with an ANN. It demonstrates the integration of ABC-rejection with a risk function to facilitate systematic hyper-parameter optimisation and model selection within the ReFANN framework. The reconstruction uncertainty combines the observational uncertainty of the SN Ia data with the variance among the accepted networks. Full mathematical details and explicit definitions of this procedure are provided in Appendix~\ref{app:sect3}.

Conventional GRB distance-calibration procedures often assume a parametric cosmology to compute $d_L(z)$ before fitting GRB luminosity correlations. While practical, this approach leaves the calibration dependent on the assumed cosmological model. Here we instead calibrated low-redshift GRBs directly against the distance scale provided by Pantheon$+$ SNe Ia, using ReFANN to reconstruct $d_L(z)$ non-parametrically and then evaluating it at the redshifts of low-$z$ GRBs. We restricted this step to $z\leq1$, where the Pantheon$+$ SH0ES sample is dense and the distance--redshift relation is empirically well constrained. Importantly, we did not treat the supernova distances as fixed inputs: reconstruction uncertainties propagate into the GRB calibration, and all cosmological and correlation parameters were inferred jointly within a single hierarchical Bayesian framework. Combined with ABC model selection and risk-function ranking for the network, this approach reduces subjective network choices and propagates calibration uncertainty into the final GRB distance estimates.

However, the ANN itself, such as the number of hidden layers $N_{\mathrm{hl}}$, number of neurones $N_{\mathrm{node}}$, activation functions and optimiser, creates a new hyper-parameter space. Without a systematic model comparison and selection procedure, these design choices become implicit priors and can compromise the objectivity of the results. To address this, we imposed two key optimisation criteria for ReFANN, especially the number of layers and the number of neurones.

First, within the ABC rejection, we treat different networks as distinct candidate models. For each, we compute three summary statistics on a validation set: the goodness of fit $\chi^2$ \citep{2021JCAP...08..027B}, the Euclidean distance between predictions and observations \citep{2017A&C....19...16J,2023ApJS..266...27Z}, and an approximate LML (\citealt{10.3389/fbuil.2017.00052,2018JCAP...04..051G,2021JCAP...08..027B}). Candidate networks are then accepted or rejected based on an observable-dependent tolerance threshold $\varepsilon$. Detailed definitions of these statistics and the acceptance criterion are provided in Appendix~\ref{app:sect3}. The tolerance $\varepsilon$ controls the balance between model discrimination and sample retention. If $\varepsilon$ is too large, most candidate networks are accepted and the hyper-parameter posteriors become weakly informative. If $\varepsilon$ is too small, too few networks are retained, leading to unstable posterior estimates and poor exploration of the hyper-parameter space. In practice, we therefore adopt an iterative procedure: starting from a high value, we gradually lower $\varepsilon$ until further reductions no longer produce significant shifts in the hyper-parameter posteriors. We identify this value as the optimal tolerance, $\varepsilon^{\star}$. This approach balances reliable inference with computational efficiency in the ABC rejection sampling. Second, following the original ReFANN, we further rank the accepted models using a risk function and Bayes-factor comparisons to identify a compact set of optimal hyper-parameters \citep{1999ITNN...10..988V,2001astro.ph.12050W}. The definitions of these quantities and the adopted interpretation scale are given in Appendix~\ref{app:sect3}.

Separate ReFANN networks were trained for $d_L(z)$ and $\mu(z)$ using the same ABC-rejection criteria, risk-function evaluation, Bayes-factor comparisons, and uncertainty propagation. Their reconstruction uncertainties include both the data contribution $\sigma_{\mu,\mathrm{data}}$ and the model contribution $\sigma_{\mu,\mathrm{model}}$.

\begin{figure}[ht!]
    \centering
    \includegraphics[width=1.0\linewidth]{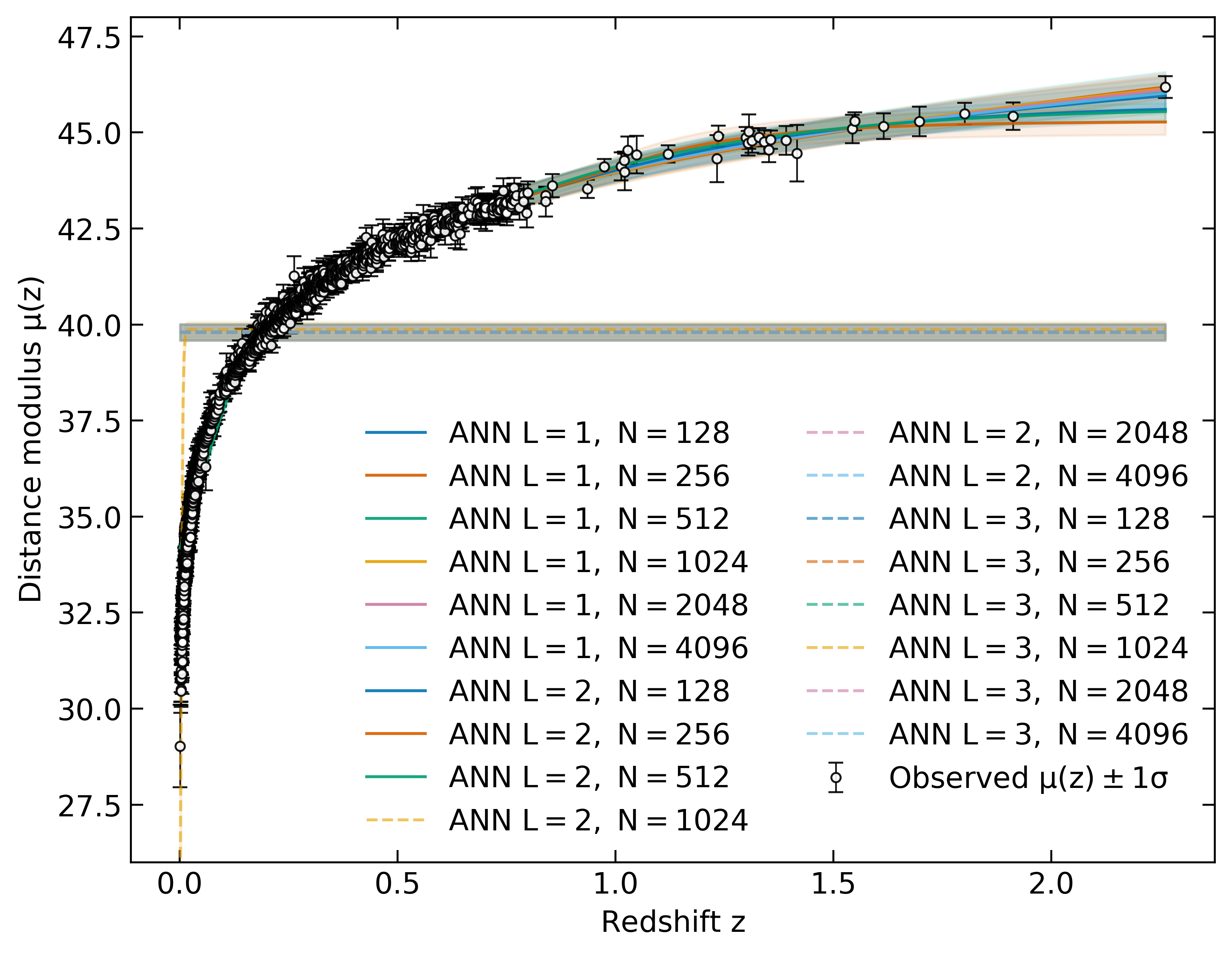}
    \caption{Reconstructed distance modulus, $\mu(z),$ obtained with ReFANN for different network hyperparameters. Each curve represents a different combination of the number of hidden layers ($N_{\rm hl}$) and the number of nodes in each layer ($N_{\rm node}$). The observed data are shown as orange points with error bars, while the shaded regions indicate the $1\sigma$ uncertainties of the ANN reconstructions. The corresponding luminosity-distance reconstruction, $d_L(z)$, is presented in Appendix Fig.~\ref{figANN-dL}.}
    \label{figANN-all}
\end{figure}

Figure~\ref{figANN-all} illustrates the ensemble of ReFANN reconstructions for the distance modulus $\mu(z)$ obtained from different hyper-parameter choices. Most candidate networks reproduce the Pantheon$+$ relation well over the calibration range $z\leq1$, whereas some become unstable when extrapolated to higher redshift. Although this extrapolated regime is not used for GRB calibration, it provides a useful training diagnostic. We therefore applied preliminary loss cuts of $\leq0.03$ for $\mu(z)$ and $\leq0.4$ for $d_L(z)$ before the ABC selection. The corresponding $d_L(z)$ reconstructions are provided in Fig.~\ref{figANN-dL}.

Figure~\ref{figABCJ-all} summarises the resulting posterior probabilities after applying the ABC-rejection procedure described above  for the candidate ReFANN hyper-parameters. For each hyper-parameter labelled $\mathrm{HL}i-\mathrm{N}j$, the probability was obtained by averaging over 10000 ABC posterior-probability evaluations, so the bars represent ensemble-averaged hyper-parameter preferences. For $d_L(z)$ shown in the top panel, the $\chi^2$- and Euclid-based posteriors are relatively flat. Most candidates cluster around $10\%$, indicating that these two criteria provide only weak discrimination among the surviving architectures. Moreover, the LML criterion yields a more peaked posterior, with $\mathrm{HL}2-\mathrm{N}1024$ reaching the highest probability, followed by $\mathrm{HL}1-\mathrm{N}1024$ and a group of models around the $\sim 10\%$ level. In contrast, hyper-parameters such as $\mathrm{HL}2-\mathrm{N}4096$ receive essentially zero posterior weight in two distance criteria. For $\mu(z)$ shown in the bottom panels, the trend is even more pronounced. Criteria such as $\chi^2$ and the Euclidean distance favour only a narrow subset of hyper-parameters. For example, under the $\chi^2$ criterion, HL2-N128 and HL2-N256 receive identical weights of $17.4 \%$, while under the Euclidean distance they each receive $13.5\%$. In contrast, the LML posterior is strongly concentrated, assigning $26\%$ of the weight to $\mathrm{HL} 2-\mathrm{N} 128$ and $\mathrm{HL} 2-\mathrm{N} 256$. In contrast, hyper-parameters such as $\mathrm{HL}2-\mathrm{N}512$ receive essentially zero posterior weight in all three distance criteria.

To more clearly compare the advantages and disadvantages of the two hyper-parameters, we further transform the posterior distribution histogram into a Bayes factor $\mathcal{B}_f$ between the two hyper-parameters displayed in the form of heat map in Fig. \ref{figBHM-all}. The top panels correspond to $d_L(z)$, while the bottom panels correspond to $\mu(z)$. Larger Bayes factors indicate stronger relative support for the model on the column $i$ against the model on the row $j$, following the definition $\mathcal{B}_{ij}$ in Eq.~(\ref{bayseq}) under equal model priors. When forming these ratios, hyper-parameters with ABC posterior probabilities consistent with zero were excluded to avoid ill-defined or numerically unstable Bayes factors. This exclusion explains why the heat maps can display fewer hyper-parameter combinations than the full posterior histograms, as illustrated by the case of $\mathrm{HL}2-\mathrm{N}4096$ for $d_L(z)$ and $\mathrm{HL}2-\mathrm{N}512$ for $\mu(z)$.

The Bayes-factor analysis shows different behaviours for the two reconstructed distances. For $d_L(z)$, the Bayes-factor contrasts under the $\chi^2$ and Euclidean criteria are `barely worth mentioning', with values remaining very close to unity, consistent with the nearly flat posteriors in Fig. \ref{figABCJ-all}. Even under the LML criterion, the separation remains moderate, indicating that several hyper-parameters are statistically comparable once they pass the initial training-quality pre-screening. For $\mu(z)$, however, the Bayes factors are substantially more informative. The LML comparison shows clear preferences, with ratios reaching the `substantial' and `strong' regime for the most favoured models. This aligns with the sharp concentration of the LML posterior on $\mathrm{HL}2-\mathrm{N}128$ and $\mathrm{HL}2-\mathrm{N}256$ observed in Fig. \ref{figABCJ-all}, respectively.

\begin{figure}[ht!]
	\centering
	\includegraphics[width=1.0\linewidth]{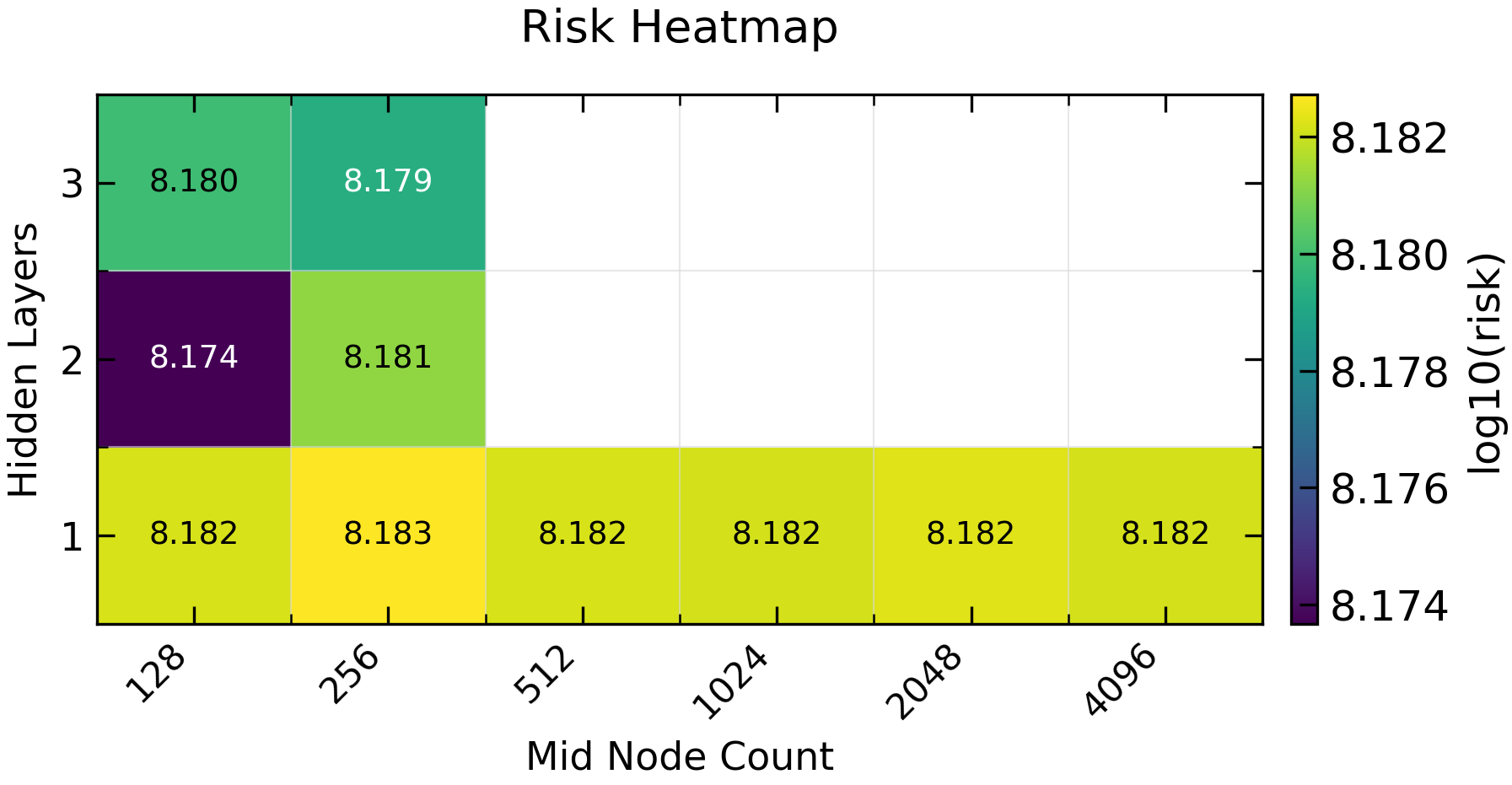}
	\includegraphics[width=1.0\linewidth]{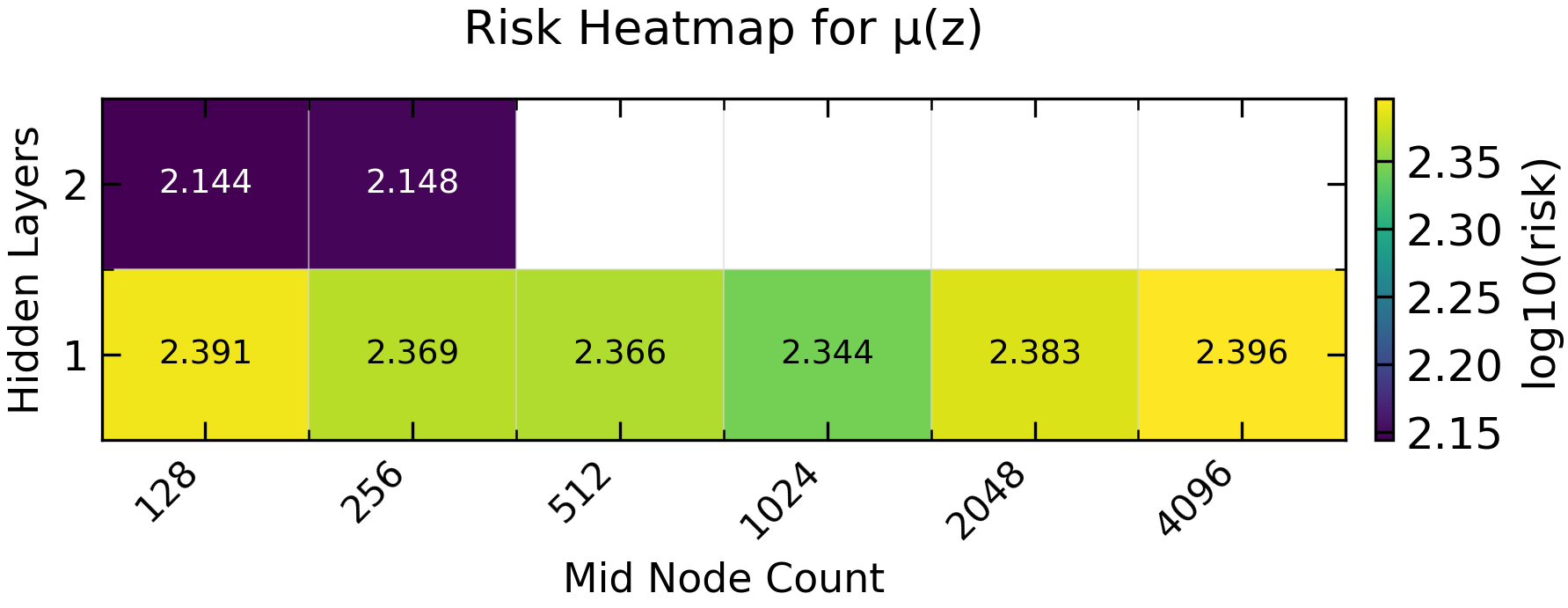}
	\caption{Risk heat maps for the luminosity distance $d_L(z)$ (top) and distance modulus $\mu(z)$ (bottom). The maps show the base-10 logarithm of the risk values for different combinations of the number of hidden layers and the number of nodes per hidden layer in the ANN architecture. The colour bar indicates the range of $\log_{10}(\mathrm{risk})$, with purple representing lower risk and green representing higher risk.}
	\label{figRHM-all}
\end{figure}

Nevertheless, for a subset of the surviving hyper-parameters, the Bayes factors remain close to $\mathcal{B}_f \approx 1$, with their ABC posterior probabilities differing only slightly. This near-degeneracy appears, for example, among several candidates for $d_L(z)$ with probabilities around the $\sim 10\%$ level, and in the nearly tied cases for $\mu(z)$ under the $\chi^2$ and Euclidean criteria. In such situations, Bayes factors and ABC posteriors alone lack sufficient discriminative power. Therefore, we incorporated the estimated risk function defined by Eq. (\ref{risk-func}) and shown in Fig. \ref{figRHM-all} as the final decision criterion. Concretely, the risk heatmap for $d_L(z)$ shows a comparatively flat landscape, with most candidates sharing a similar $\log_{10}(\mathrm{risk})$ value around 8.182. But it identifies a clear minimum at $\mathrm{HL}2-\mathrm{N}128,$ whose $\log_{10}(\mathrm{risk})$ value is 8.174. In contrast, for $\mu(z)$, the analysis identifies a clear minimum at $\mathrm{HL}2-\mathrm{N}128$, which corresponds to a $\log_{10}(\mathrm{risk})$ value of approximately 2.144, and a similarly low value of about 2.148 at $\mathrm{HL}2-\mathrm{N}256$. Several $\mathrm{HL}1$ configurations exhibit significantly higher risk, for instance, with $\log_{10}(\mathrm{risk})$ values greater than 2.3. Thus, the risk function mainly acts as a consistency check and tie-breaker when the ABC posteriors and Bayes factors do not uniquely select an hyper-parameter set.

\begin{figure}[ht!]
    \centering
    \includegraphics[width=1.0\linewidth]{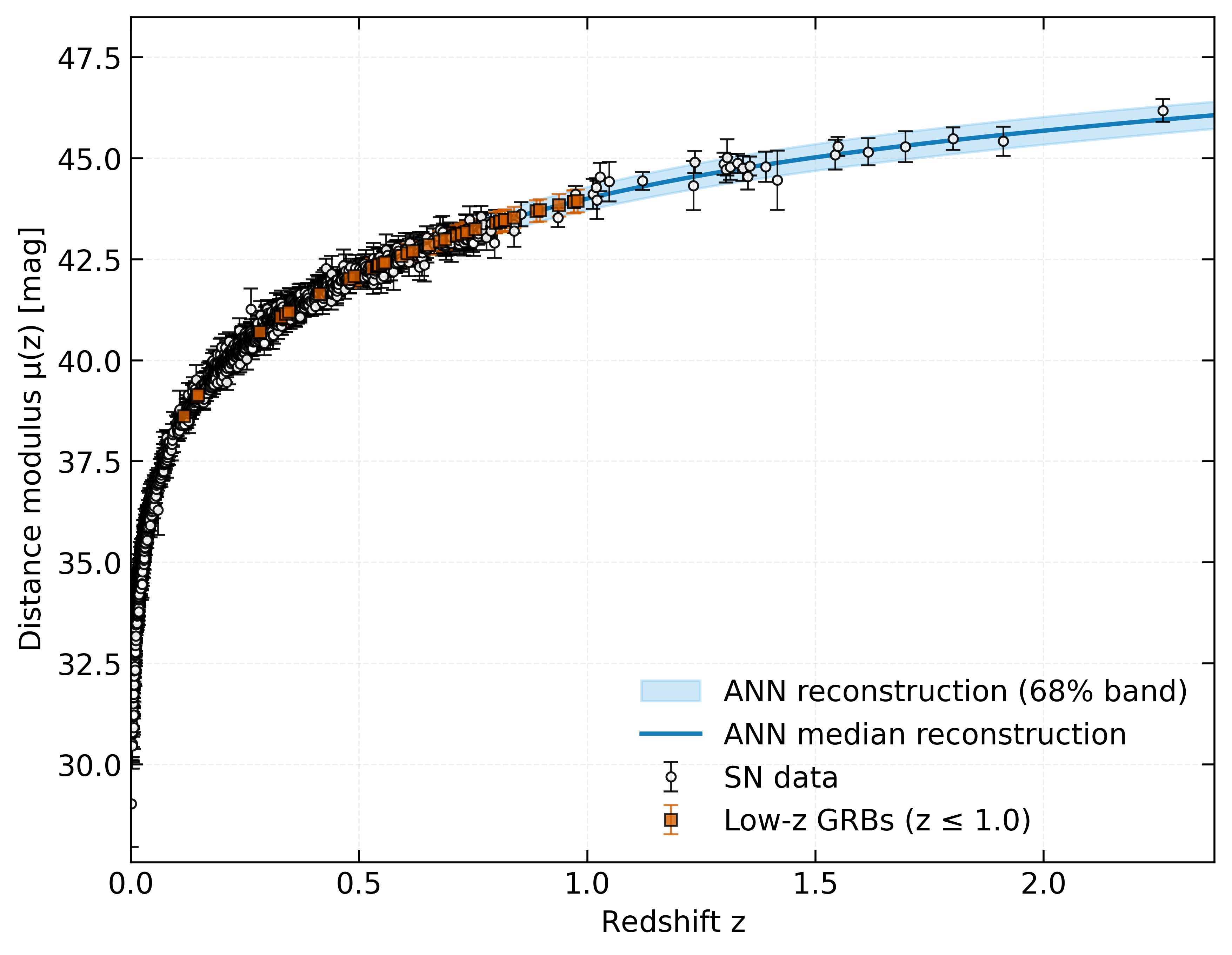}
    \caption{Comparison of the ANN-reconstructed distance modulus $\mu(z)$ for Pantheon$+$ SNe Ia with the low-redshift GRB calibration subsample. The solid line shows the reconstructed distance–redshift relation, and the shaded band represents the 68\% reconstruction uncertainty. The Pantheon$+$ SN Ia data are shown as points with observational error bars, and the low-redshift GRBs ($z\leq 1$) are overplotted to illustrate the consistency between the GRB calibration anchor and the SN Ia distance ladder. The corresponding comparison for the luminosity distance, $d_L(z)$, is shown in Fig.~\ref{figc-dL}.}
    \label{figc}
\end{figure}

Following the model selection method, we identify HL2-N128 as the optimal hyper-parameters for both $d_L(z)$ and $\mu(z)$. From the ensemble of hyper-parameters satisfying the ABC rejection and risk-function criteria, we generated the final ANN reconstruction and its covariance matrix. Figure~\ref{figc} presents the resulting $\mu(z)$ reconstruction alongside the low-redshift GRB subset $(z \leq 1)$. The corresponding $d_L(z)$ comparison is provided in Fig.~\ref{figc-dL}.

\section{GRB calibrations and cosmological constraints}
This section builds on the data and ANN reconstruction methods described above to show how we constructed GRB Hubble diagrams from the Amati and Combo relations and used them for cosmological inference. Before applying the ANN calibration, we first tested whether the adopted GRB correlations depend significantly on the assumed background cosmology. We then calibrated the low-redshift GRBs using the ReFANN reconstruction, inferred distances for the high-redshift events, and jointly constrained the GRB correlation and cosmological parameters.

An external SN Ia calibration reduces the cosmology dependence of the GRB distance scale, but does not by itself establish the model stability of the GRB correlations. We therefore performed a pre-calibration correlation--cosmology analysis following the strategy of \citet{2020MNRAS.499..391K,2021JCAP...09..042K}. This test did not use the reconstructed ANN distance. Instead, the correlation and cosmological parameters are fitted simultaneously in flat and non-flat $\Lambda$CDM and XCDM. Since the absolute distance scale is degenerate with the correlation intercept in a GRB-only analysis, we fixed $H_0=70~\mathrm{km,s^{-1},Mpc^{-1}}$ for these model stability tests. The same procedure was applied to the 77-event Amati samples, the A118 benchmark, and the Combo/C174 sample.

Among the empirical GRB luminosity correlations, the Amati relation is one of the most widely used. It connects the rest-frame spectral peak energy $E_{p,i}$ with the isotropic-equivalent radiated energy $E_{\mathrm{iso}}$ and is written as
\begin{equation}
	\log_{10}\left(\frac{E_{\mathrm{iso},i}}{\mathrm{erg}}\right)=A_A+\gamma_A \log_{10}\left(\frac{E_{p,i}}{\mathrm{keV}}\right)+\eta_A,
\end{equation}
where $A_A$ is the intercept, $\gamma_A$ is the slope, and $\eta_A$ represents the intrinsic scatter of the relation. We modelled this scatter as a Gaussian random variable, $\eta_A\sim\mathcal{N}(0,\sigma_{v_A}^2)$. For each GRB, the rest-frame peak energy ($E_{p,i}$) was computed from Eq.~(\ref{sppe}), and the bolometric fluence ($S_{\mathrm{bol}}$) is obtained after the K-correction described in Sect.~2. Given a luminosity distance ($d_L$), the isotropic-equivalent energy follows from Eq.~(\ref{amati}). We define the Amati residual as
\begin{equation}
	\Delta_i^A(d_L)
	=\log_{10}\left(\frac{E_{\mathrm{iso},i}(d_L)}{\mathrm{erg}}\right)-A_A-\gamma_A\log_{10}\left(\frac{E_{p,i}}{\mathrm{keV}}\right),
	\label{eq:amati_residual}
\end{equation}
with total variance $\left(\sigma_{\mathrm{tot},i}^{A}\right)^2=\sigma_{\log_{10}E_{\mathrm{iso},i}}^2+\gamma_A^2\sigma_{\log_{10}E_{p,i}}^2+\left(\sigma_{\mathrm{int}}^A\right)^2$. In the pre-calibration test, $E_{\mathrm{iso}}$ was evaluated using the theoretical luminosity distance $d_L^{\mathrm{th}}(z;\boldsymbol{\theta})$ for all GRBs. In the ANN-calibrated analysis, the distance was instead supplied by ReFANN for the low-redshift calibration sample and by the cosmological prediction for the high-redshift sample.

Similarly, the Combo relation is written in terms of the X-ray plateau luminosity $L_0$ as
\begin{equation}
\begin{aligned}
\log_{10}\left(\frac{L_{0,i}}{\mathrm{erg\,s^{-1}}}\right)
=&\log_{10}\left(\frac{A_C}{\mathrm{erg\,s^{-1}}}\right)+\gamma_C\log_{10}\left(\frac{E_{p,i}}{\mathrm{keV}}\right)\\
&-\log_{10}\left(\frac{\tau_i/\mathrm{s}}{|1+\alpha_{PL,i}|}\right)+\eta_C ,
\end{aligned}
\end{equation}
where $\eta_C\sim\mathcal{N}(0,\sigma_{v_C}^2)$. The corresponding residual is $\Delta_i^C(\mu)=\log_{10}\left(\frac{L_{0,i}(\mu)}{\mathrm{erg\,s^{-1}}}\right)-\log_{10}\left(\frac{A_C}{\mathrm{erg\,s^{-1}}}\right)-\gamma_C\log_{10}\left(\frac{E_{p,i}}{\mathrm{keV}}\right)+\log_{10}\left(\frac{\tau_i/\mathrm{s}}{|1+\alpha_{PL,i}|}\right)$, where $L_{0,i}$ is obtained from the plateau flux $F_0$ and the luminosity distance corresponding to $\mu$. The total variance is $\left(\sigma_{\mathrm{tot},i}^{C}\right)^2=\sigma_{\log_{10}L_{0,i}}^2+\gamma_C^2\sigma_{\log_{10}E_{p,i}}^2+\sigma^2_{\log_{10}\left(\tau_i/|1+\alpha_{PL,i}|\right)}+\left(\sigma_{\mathrm{int}}^C\right)^2$. The Combo parameters are tested in the same four background cosmologies before the ANN calibration. To model residual variance beyond the measured uncertainties and the intrinsic scatter of the GRB correlation, we introduced an additional calibration-scatter parameter, $\sigma_{\mathrm{cal}}$. This term was added in quadrature to the GRB distance uncertainties in the cosmological likelihood, absorbing potential un-modelled systematic effects from calibration and inference.

As an independent cosmological comparison, we constrained the same background models using the OHD described in Sect.~2. These data were not used in the ANN reconstruction or GRB calibration. Their role was solely to compare the GRB-only cosmological constraints with an external late-time probe. For the distance calibration, ReFANN provides non-parametric reconstructions of $d_L(z)$ and $\mu(z)$ from the Pantheon$+$ SH0ES sample. We split the GRB sample at $z_{\mathrm{cut}}=1$, where the SN Ia Hubble diagram remains comparatively well sampled. For the low-redshift GRBs, the ANN predictions at their redshifts provide the calibration distances. We repeated the analysis with $z_{\mathrm{cut}}=0.8$ and $1.2$ to test the sensitivity to this choice, with the results given in Sect.~5 and Appendix~C.

The calibration assumes that sources at the same redshift share the same luminosity distance, as expected in a homogeneous and isotropic cosmology. Based on this assumption, we treated the ANN predictions at the redshifts of low-$z$ GRBs as the expected distances for these events. In our paper, these predictions are described by a multivariate normal distribution with a full covariance matrix, capturing both the individual uncertainties and the correlations between predictions at different redshifts, $\mathbf{d}_{L,{\rm cal}}^{\rm ANN}\sim \mathcal{N}\!\left(\bar{\mathbf{d}}_{L,{\rm cal}}^{\rm ANN},\,\mathbf{C}_{L,{\rm cal}}^{\rm ANN}\right)$ and $\boldsymbol{\mu}_{\rm cal}^{\rm ANN}\sim \mathcal{N}\!\left(\bar{\boldsymbol{\mu}}_{\rm cal}^{\rm ANN},\,\mathbf{C}_{\mu,{\rm cal}}^{\rm ANN}\right)$. The ReFANN reconstruction uncertainty is propagated to the GRB calibration through joint Monte Carlo samples of the reconstructed distance at the calibration redshifts $\mathbf{z}_{\mathrm{cal}}$. This carries the covariance of the distance-redshift relation into the low-redshift GRB calibration. The GRB observational uncertainties are treated separately and propagated through the Amati and Combo likelihoods. As a consistency check, we verified that the marginal sampled distances are consistent with the Gaussian form implied by the multivariate normal reconstruction, using a $\chi^2$ goodness-of-fit test with rejection region $W=\{\chi^2>\chi_{0.95}^2\}$ \citep{1986gft..book.....D,1992nrca.book.....P,2003drea.book.....B}.

We applied a consistent Monte Carlo procedure to construct high-redshift GRB distance datasets for both correlations. For each high-redshift event $\left(z_j > z_{\mathrm{cut}}\right)$, we sampled the calibrated correlation parameters from their joint posterior and combine them with the observed GRB quantities and their uncertainties to infer the corresponding distance. For the Amati relation, this yields posterior predictive estimates of luminosity distance $d_{L, j}^{(\mathrm{high})}$ with associated uncertainties $\sigma_{d_L, j}^{(\mathrm{high})}$. For the Combo relation, we followed the same procedure to obtain distance modulus samples $\mu_j^{\mathrm{(high)}}$ and $\sigma_{\mu, j}^{(\mathrm{high})}$. These high-redshift estimates were then combined with the low-redshift values from the ReFANN reconstruction, which produced two GRB Hubble diagrams spanning $z \approx 0.03$ to $z \approx 8-9$: one from the Amati relation in the luminosity distance, and one from the Combo relation in the distance modulus,
\begin{equation}
	\left\{\left(z_i, d_{L, i}^{(\mathrm{low})}, \sigma_{d_L, i}^{(\mathrm{low})}\right)\right\}_{z_i\leq z_{\mathrm {cut}}}
	\cup
	\left\{\left(z_j, d_{L, j}^{(\mathrm{high})}, \sigma_{d_{L, j}}^{(\mathrm{high})}\right)\right\}_{z_j > z_{\mathrm{cut}}},
\end{equation}
and the other is the distance-modulus data set from the Combo correlation
\begin{equation}
	\left\{\left(z_i, \mu_i^{(\mathrm{low})}, \sigma_{\mu, i}^{(\mathrm{low})}\right)\right\}_{z_i\leq z_{\mathrm {cut}}}
	\cup
	\left\{\left(z_j, \mu_j^{(\mathrm{high})}, \sigma_{\mu, j}^{(\mathrm{high})}\right)\right\}_{z_j > z_{\mathrm{cut}}}.
\end{equation}

Following the low and high redshift sample construction described above, we constrained the cosmological parameters and the GRB-relation parameters within a hierarchical Bayesian analysis. For a given cosmological model, the theoretical luminosity distance can be written as \citep{1995ApJ...450...14G}
\begin{equation}
	d_L^{\mathrm{th}}(z ; \boldsymbol{\theta})=\frac{c(1+z)}{H_0} \int_0^z \frac{\mathrm{~d} z^{\prime}}{E\left(z^{\prime} ; \boldsymbol{\theta}\right)} ,
\end{equation}
where $\boldsymbol{\theta}$ denotes the set of cosmological parameters. In the case of a flat $\Lambda \mathrm{CDM}$ model, we have $\boldsymbol{\theta}=\left(H_0, \Omega_m\right)$, while for the $w_0w_a$CDM model we adopted $\boldsymbol{\theta}=\left(H_0, \Omega_m, w_0, w_a\right)$. In our final inference, the Pantheon$+$ information enters through the ANN covariance matrix, rather than an additional SNe Ia term. 

Given that the GRB distances are inferred through an empirical correlation, the correlation parameters must be inferred together with the cosmological parameters, and their uncertainties including intrinsic scatter must be propagated consistently into the GRB error estimation. We treated the Amati and Combo relations as two separate calibration inferences. The Amati inference is built on the luminosity distance indicator $d_L(z)$ inferred from $(E_{p,i},S_{\rm bol})$, whereas the Combo inference is naturally formulated in terms of the distance modulus $\mu(z)$ through $L_0$ and the corresponding Combo observables. Accordingly, we define two total log-likelihoods:
\begin{equation}
	\ln \mathcal{L}_{\mathrm{tot}}^{\mathrm{(Amati)}}=\ln \mathcal{L}_{\mathrm{cal}}^{(d_L)}+\ln \mathcal{L}_{\mathrm{cos}}^{(d_L,\mathrm{high})},
	\label{eq:lnLtot_amati}
\end{equation}
and
\begin{equation}
	\ln \mathcal{L}_{\mathrm{tot}}^{\mathrm{(Combo)}}=\ln \mathcal{L}_{\mathrm{cal}}^{(\mu)}+\ln \mathcal{L}_{\mathrm{cos}}^{(\mu,\mathrm{high})}.
	\label{eq:lnLtot_combo}
\end{equation}
In the first case, $\ln \mathcal{L}_{\mathrm{cal}}^{(d_L)}$ calibrates the Amati parameters $(A_A,\gamma_A,\sigma_{v_A})$ and cosmological parameters hierarchically
\begin{equation}
	\ln \mathcal{L}_{\mathrm{cal}}^{(d_L)}=-\frac{1}{2}\sum_{i\in\mathcal{D}_{\mathrm{cal}}} \frac{\left(d_{L,\mathrm{GRB}, i}^{\mathrm{obs}}-d_L^{\mathrm{th}}\left(z_i ;\boldsymbol{\theta}\right)\right)^2}{\sigma_{\mathrm{cal}, i}^2}-\frac{1}{2}\sum_{i\in\mathcal{D}_{\mathrm{cal}}}\ln\!\left(2\pi \sigma_{\mathrm{cal}, i}^2\right),\label{amatical}
\end{equation}
where $\sigma_{\mathrm{cal}, i}^2= \sigma_{\mathrm{ANN}, i}^2+\left(\sigma_{\mathrm{tot}, i}^{A}\right)^2$. The term $\ln \mathcal{L}_{\mathrm{cos}}^{(d_L,\mathrm{high})}$ compares the high-redshift GRB $d_{L,\mathrm{GRB}}^{\mathrm{obs}}$ with the theoretical prediction $d_L^{\mathrm{th}}(z;\boldsymbol{\theta})$ 
\begin{equation}
	\ln \mathcal{L}_{\mathrm{cos}}^{(d_L,\mathrm{high})}=-\frac{1}{2} \sum_{j\in\mathcal{D}_{\mathrm{cos}}} \frac{\left(d_{L,\mathrm{GRB}, j}^{\mathrm{obs}}-d_L^{\mathrm{th}}\left(z_j ; \boldsymbol{\theta}\right)\right)^2}{\sigma_{\mathrm{cos}, j}^2}-\frac{1}{2} \sum_{j\in\mathcal{D}_{\mathrm{cos}}}\ln\!\left(2\pi \sigma_{\mathrm{cos}, j}^2\right),\label{amaticos}
\end{equation}
where $\mathcal{D}_{\rm cos}=\{j\,|\,z_j>z_{\rm cut}\}$ and $\sigma_{\mathrm{cos}, j}^2= \left(\sigma_{\mathrm{tot}, j}^{A}\right)^2$. In the second case, $\ln\mathcal{L}_{\mathrm{cal}}^{(\mu)}$ plays the same role for the Combo parameters $(A_C,\gamma_C,\sigma_{v_C})$, and $\ln\mathcal{L}_{\mathrm{cos}}^{(\mu,\mathrm{high})}$ compares the high-redshift GRB $\mu_{\mathrm{GRB}}^{\mathrm{obs}}$ with $\mu^{\mathrm{th}}(z;\boldsymbol{\theta})$. The Combo likelihoods have the same form as Eqs.~(\ref{amatical}) and (\ref{amaticos}), with the luminosity distance replaced by the distance modulus.

\section{Results}

We sampled the joint posterior of cosmological parameters and the relevant correlation parameters using the affine-invariant ensemble Markov chain Monte Carlo sampler (EMCEE; \citealt{2010CAMCS...5...65G,2013PASP..125..306F}). Before applying the ANN calibration, we first examined whether the adopted GRB correlations depend on the assumed background cosmology. Across flat and non-flat $\Lambda$CDM and XCDM, the correlation parameters remain stable. The maximum joint shifts are $0.14\sigma$ for the A118 benchmark, $0.24\sigma$ for the 77-event Amati sample, and $0.59\sigma$ for Combo/C174. For A118, the recovered correlation parameters are also consistent with the published values within $1\sigma$ \citep{2021JCAP...09..042K}. We therefore find no statistically significant correlation-parameter dependence on the four cosmologies considered here.

The stability of the correlation parameters does not by itself establish cosmological consistency. We therefore compared the GRB-only results with the independent OHD measurements. In flat $\Lambda$CDM, using the same fixed value $H_0=70~\mathrm{km\,s^{-1}\,Mpc^{-1}}$, OHD gives $\Omega_m=0.292^{+0.026}_{-0.026}$. A118 shows the closest agreement with OHD, while the 77-event Amati sample is shifted towards higher $\Omega_m$ by $2.26\sigma$. Because the Amati posterior is broad and asymmetric, however, the OHD value remains within its 95\% credible interval. The discrepancy increases to $4.40\sigma$ for C174, for which the OHD value lies outside the 95\% credible interval. The GRB-only samples also do not independently establish current acceleration: in flat $\Lambda$CDM, $P(q_0<0)$ is 0.527, 0.342, and 0.116 for A118, Amati-77, and C174, respectively, compared with unity for OHD.
\begin{table}[htbp]
	\centering
	\caption{Pre-calibration validation of the GRB samples.}
	\label{tab:precal_validation}
	\begin{tabular}{lccc}
		\hline\hline
		Sample & $N$ & $\Delta_{\rm corr}^{\rm max}$ & $\Delta_{\rm OHD}$ \\
		\hline
		A118 & 118 & $0.14\sigma$ & $1.46\sigma$ \\
		Amati & 77 & $0.24\sigma$ & $2.26\sigma$ \\
		Combo/C174 & 173 & $0.59\sigma$ & $4.40\sigma$ \\
		\hline
	\end{tabular}
	\par\smallskip
	\begin{minipage}{\linewidth}
		\footnotesize
		\raggedright
		Notes: $\Delta_{\rm corr}^{\rm max}$ denotes the maximum joint shift in the correlation parameters among the four tested cosmologies. $\Delta_{\rm OHD}$ denotes the difference between the marginal constraints on $\Omega_m$ from the GRB and OHD data.
	\end{minipage}
\end{table}

As an additional check, we compared our Amati measurements with A118 using the same events. We identified 43 common GRBs, of which 41 remain after the strict-quality cut. For a fixed event set, using the two sets of prompt-emission measurements changes the cosmological parameters by less than about $0.4\sigma$ across the tested models. The more noticeable difference occurs in the inferred intrinsic scatter, at approximately $1.4$--$1.5\sigma$. The detailed matched-event analysis is presented in Appendix~\ref{app:matched}.

We next considered the cosmological constraints obtained after the ANN calibration. The marginalised constraints are shown in Figs.~\ref{figcorneram} and \ref{figcornerco} for the Amati and Combo analyses, respectively, with the top panels corresponding to the $\Lambda$CDM fits and the bottom panels to the $w_0w_a$CDM fits. The results are summarised in Table~\ref{tab:cosmo_results}. To test the dependence on the calibration threshold, we repeated the full analysis with $z_{\text{cut}}=0.8$ and $1.2$ and found no qualitative change in the inferred parameters. The corresponding results are presented in Figs.~\ref{figcorneramzc} and \ref{figcornercozc} and Table~\ref{tab:appC_zcut_cosmo}. We also tested alternative sample selections based on prompt brightness and spectral or measurement quality. The resulting shifts in the cosmological and GRB-correlation parameters remain comparable to, or smaller than, the baseline uncertainties. These tests are shown in Figs.~\ref{fig:robust_amati_combo_lcdm} and \ref{fig:robust_amati_combo_w0wa}.

\begin{table*}[htbp]
	\centering
	\caption{ANN-calibrated cosmological and correlation parameters for the Amati and Combo relations in $\Lambda$CDM and $w_0w_a$CDM.}
	\label{tab:cosmo_results}
	\begin{tabular}{lcccc}
		\hline\hline
		Parameter & Amati ($\Lambda$CDM) & Amati ($w_0w_a$CDM) & Combo ($\Lambda$CDM) & Combo ($w_0w_a$CDM) \\
		\midrule
		\multicolumn{5}{l}{Cosmological parameters} \\
		\cmidrule(r){1-1}\cmidrule(lr){2-2}\cmidrule(lr){3-3}\cmidrule(lr){4-4}\cmidrule(l){5-5}
		$H_0$ &
		$70.75_{-2.53,(-4.11)}^{+2.71,(+4.94)}$ &
		$70.58_{-3.19,(-5.17)}^{+4.51,(+9.41)}$ &
		$70.91_{-2.07,(-3.66)}^{+1.78,(+3.23)}$ &
		$71.03_{-2.68,(-4.30)}^{+3.03,(+6.03)}$ \\
		$\Omega_m$ &
		$0.553_{-0.247,(-0.410)}^{+0.274,(+0.418)}$ &
		$0.588_{-0.238,(-0.472)}^{+0.247,(+0.387)}$ &
		$0.531_{-0.164,(-0.281)}^{+0.220,(+0.405)}$ &
		$0.552_{-0.191,(-0.437)}^{+0.214,(+0.393)}$ \\
		$w_0$ &
		-- &
		$-1.287_{-0.993,(-1.59)}^{+1.04,(+2.02)}$ &
		-- &
		$-1.174_{-0.744,(-1.54)}^{+0.634,(+1.28)}$ \\
		$w_a$ &
		-- &
		$-0.585_{-2.96,(-4.18)}^{+3.16,(+5.09)}$ &
		-- &
		$-0.030_{-1.98,(-2.81)}^{+1.84,(+2.78)}$ \\
		\addlinespace
		\multicolumn{5}{l}{Correlation parameters} \\
		\cmidrule(r){1-1}\cmidrule(lr){2-2}\cmidrule(lr){3-3}\cmidrule(lr){4-4}\cmidrule(l){5-5}
		$A$ &$50.06_{-0.96\,(-1.98)}^{+0.90\,(+1.82)}$ &$50.07_{-0.95\,(-1.93)}^{+0.90\,(+1.80)}$ &$49.75_{-0.55\,(-1.12)}^{+0.56\,(+1.11)}$ &$49.79_{-0.42\,(-0.836)}^{+0.428\,(+0.855)}$ \\
		$\gamma$ &
		$1.061_{-0.362\,(-0.721)}^{+0.378\,(+0.778)}$ &
		$1.050_{-0.360\,(-0.719)}^{+0.375\,(+0.770)}$ &
		$0.759_{-0.214\,(-0.424)}^{+0.212\,(+0.431)}$ &
		$0.731_{-0.163\,(-0.327)}^{+0.160\,(+0.323)}$ \\
		\addlinespace
		
		\multicolumn{5}{l}{Intrinsic scatter parameters} \\
		\cmidrule(r){1-1}\cmidrule(lr){2-2}\cmidrule(lr){3-3}\cmidrule(lr){4-4}\cmidrule(l){5-5}
		$\sigma_{\mathrm{cal}}$ &
		$0.932_{-0.138\,(-0.256)}^{+0.161\,(+0.337)}$ &
		$0.930_{-0.138\,(-0.255)}^{+0.159\,(+0.331)}$ &
		$0.659_{-0.133\,(-0.239)}^{+0.155\,(+0.330)}$ &
		$0.729_{-0.119\,(-0.213)}^{+0.138\,(+0.292)}$ \\
		$\sigma_{\mathrm{cos}}$ &
		$0.234_{-0.163\,(-0.223)}^{+0.240\,(+0.500)}$ &
		$0.233_{-0.162\,(-0.222)}^{+0.243\,(+0.507)}$ &
		$0.642_{-0.127\,(-0.232)}^{+0.149\,(+0.312)}$ &
		$0.706_{-0.113\,(-0.204)}^{+0.127\,(+0.269)}$ \\
		\hline
	\end{tabular}
	\par\smallskip
	\begin{minipage}{\linewidth}
	\footnotesize
	\raggedright
Notes: Uncertainties are reported as $X_{-a(-c)}^{+b(+d)}$, where $a$ and $b$ denote the lower and upper uncertainties at the 68\% credible level, respectively, and $c$ and $d$ denote those at the 95\% credible level. $H_0$ is given in units of $\mathrm{km\,s^{-1}\,Mpc^{-1}}$. The parameter $A$ denotes the intercept of the corresponding correlation. $\sigma_{\mathrm{cal}}$ and $\sigma_{\mathrm{cos}}$ denote the intrinsic scatter terms used in the calibration and cosmological stages of the hierarchical inference, respectively.
	\end{minipage}
\end{table*}

The ANN-calibrated Amati and Combo analyses yield overlapping cosmological constraints within the current uncertainties. In flat $\Lambda$CDM, both relations give broad matter-density constraints. When extended to $w_0w_a$CDM, the posteriors broaden substantially because of the degeneracy between $w_0$ and $w_a$, and both remain compatible with the $\Lambda$CDM limit $(w_0,w_a)=(-1,0)$ within the 95\% credible regions. However, the overlap of the ANN-calibrated posteriors should not be interpreted as evidence that the two original GRB samples have the same cosmological reliability. As shown by the pre-calibration external-consistency test, the C174 sample shows a significant discrepancy with OHD. We therefore used the 77-event Amati sample for the main cosmological interpretation, while the Combo result is retained primarily as a calibration cross-check.

\begin{figure}[ht!]
	\centering
	\includegraphics[width=1.0\linewidth]{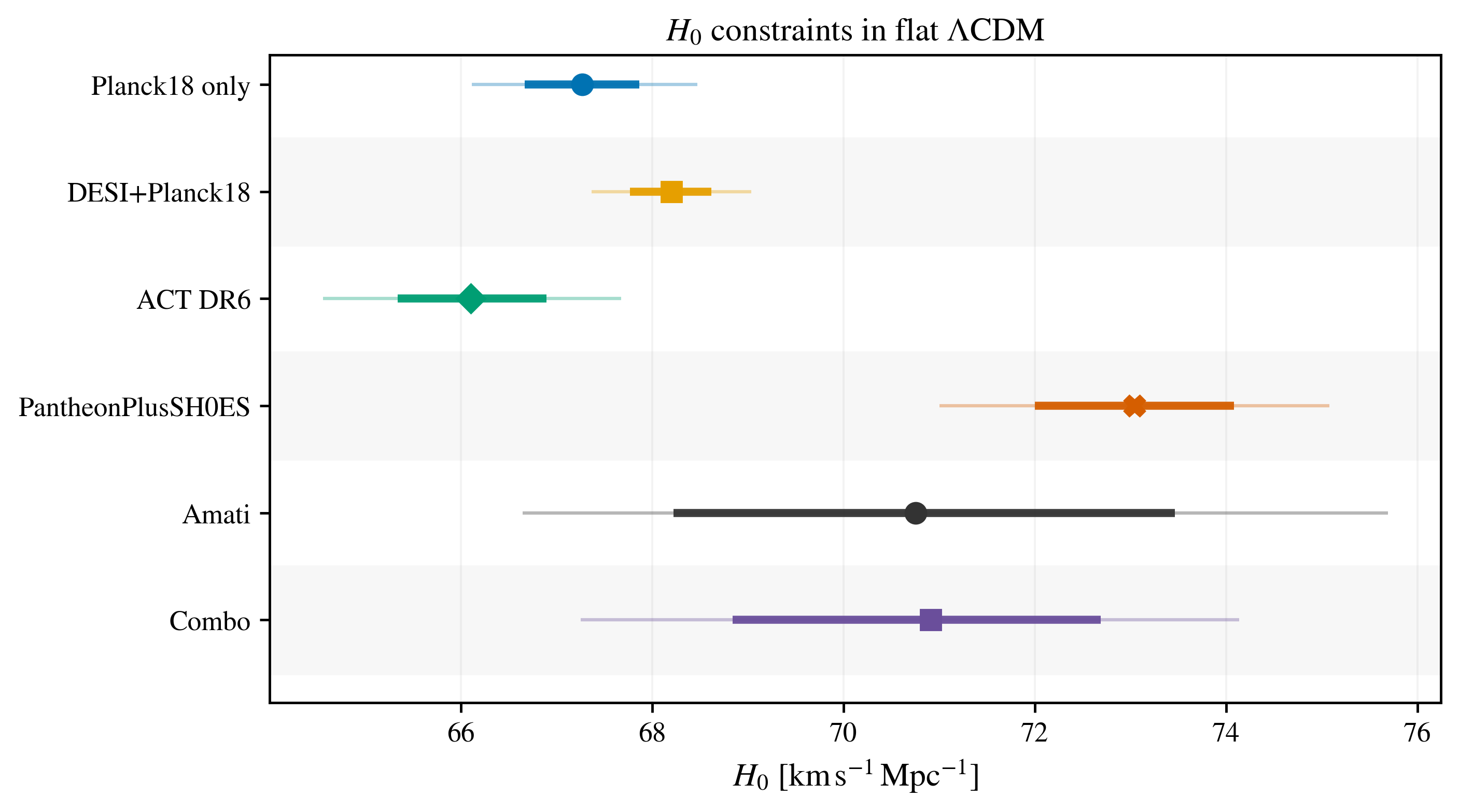}
	\includegraphics[width=1.0\linewidth]{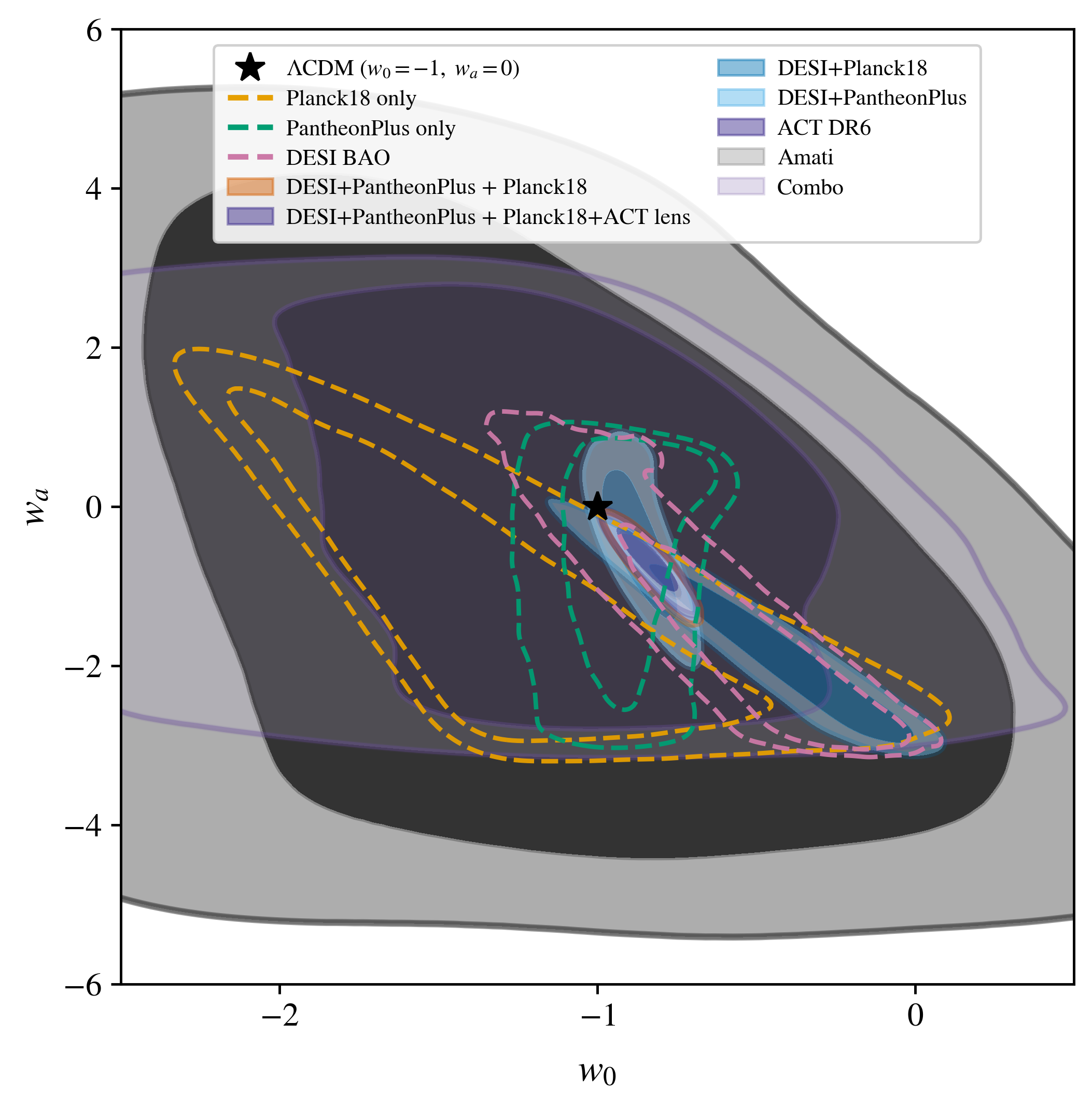}
	\caption{Summary and comparison of our constraints on $H_0$ and $(w_0,w_a)$. Top: Comparison of flat-$\Lambda$CDM constraints on $H_0$ from Planck18 \citep{2020A&A...641A...6P}, DESI+Planck18 \citep{2025JCAP...02..021A}, ACT DR6 \citep{2025JCAP...11..062L,2025JCAP...11..063C,2024ApJ...962..113M}, PantheonPlus+SH0ES \citep{2022ApJ...938..110B}, and our Amati and Combo analyses. Thick and thin bars indicate the 68\% and 95\% credible intervals, respectively. Bottom: Corresponding constraints in the $(w_0,w_a)$ plane, with the black star marking the $\Lambda$CDM point, $(w_0,w_a)=(-1,0)$. The Amati and Combo contours are shown as lightly shaded regions for comparison.}
	\label{figH0w0wa}
\end{figure}

Figure~\ref{figH0w0wa} compares the calibrated GRB constraints with representative results from recent cosmological probes. In the top panel, the calibrated GRB constraints lie between the lower CMB-based values of $H_0$ and the higher local-distance-ladder value, but have substantially larger uncertainties. Because the GRB distance scale is inherited from Pantheon$+$ SH0ES through the ANN calibration, these $H_0$ constraints are conditional on the adopted SN Ia absolute scale and should not be interpreted as independent local measurements. Their main role here is to test the extension of the calibrated Hubble diagram to higher redshift. We further tested this dependence by repeating the Amati analysis with alternative SN Ia distance zero points. The change is absorbed mainly by the Amati intercept, while the remaining correlation and cosmological parameters shift by less than $0.12\sigma$. Thus, within the tested range, the inferred shape of the expansion history is insensitive to the absolute zero point, although the overall distance scale remains calibration dependent. The detailed results are presented in Appendix~\ref{app:zero_point}.

The bottom panel of Fig.~\ref{figH0w0wa} presents the constraints in the $(w_0,w_a)$ plane. Individual datasets show extended degeneracy directions, while combinations of multiple probes substantially reduce the allowed parameter region. The GRB-only posteriors obtained from the Amati and Combo relations are much broader than those from combined CMB, baryon acoustic oscillations (BAO), and SN Ia analyses and remain compatible with the $\Lambda$CDM point $(w_0,w_a)=(-1,0)$. The present GRB samples therefore provide only weak constraints on dynamical dark energy, with the posterior widths largely reflecting parameter degeneracies and the limited precision of the high-redshift distances. Given the breadth of the GRB contours, we display them as lightly shaded background regions for clarity. Although these constraints are not competitive with current multi-probe combinations, the GRBs extend the calibrated Hubble diagram beyond the main SN Ia redshift range and provide complementary information at high redshift.

Using the GRB Hubble diagrams obtained from the Amati and Combo relations with the low-$z$ ANN calibration, we examined the distance moduli and residuals presented in Figs.~\ref{figresidualam} and \ref{figresidualco}. In each figure, the first two panels show $\mu_{\rm GRB}(z)$ together with the best-fitting $\Lambda$CDM and $w_0w_a$CDM predictions, while the lower panels show the corresponding residuals, $\Delta\mu\equiv\mu_{\rm GRB}-\mu_{\rm th}(z;\boldsymbol{\theta})$. The low-redshift GRBs follow the calibrated distance relation, whereas the high-redshift events extend the Hubble diagram to $z\sim9$. In $w_0w_a$CDM, the distance--redshift relation becomes less constrained at high redshift because of the additional $(w_0,w_a)$ degeneracy, and the present GRB samples do not provide tight constraints on departures from $\Lambda$CDM.

For both the Amati and Combo relations, the residuals are distributed around zero and show no statistically significant trend with redshift under either cosmological model. We therefore find no clear evidence for a redshift-dependent bias in the calibrated GRB distances within the current uncertainties. The largest residuals occur at high redshift, where the measurement uncertainties and intrinsic scatter are also larger. Subtler redshift evolution, selection effects, or differences between the two GRB correlations cannot be excluded with the present samples and will require larger and more homogeneous high-redshift datasets \citep{2021JCAP...09..042K}.

\begin{figure}[ht!]
	\centering
	\includegraphics[width=1.0\linewidth]{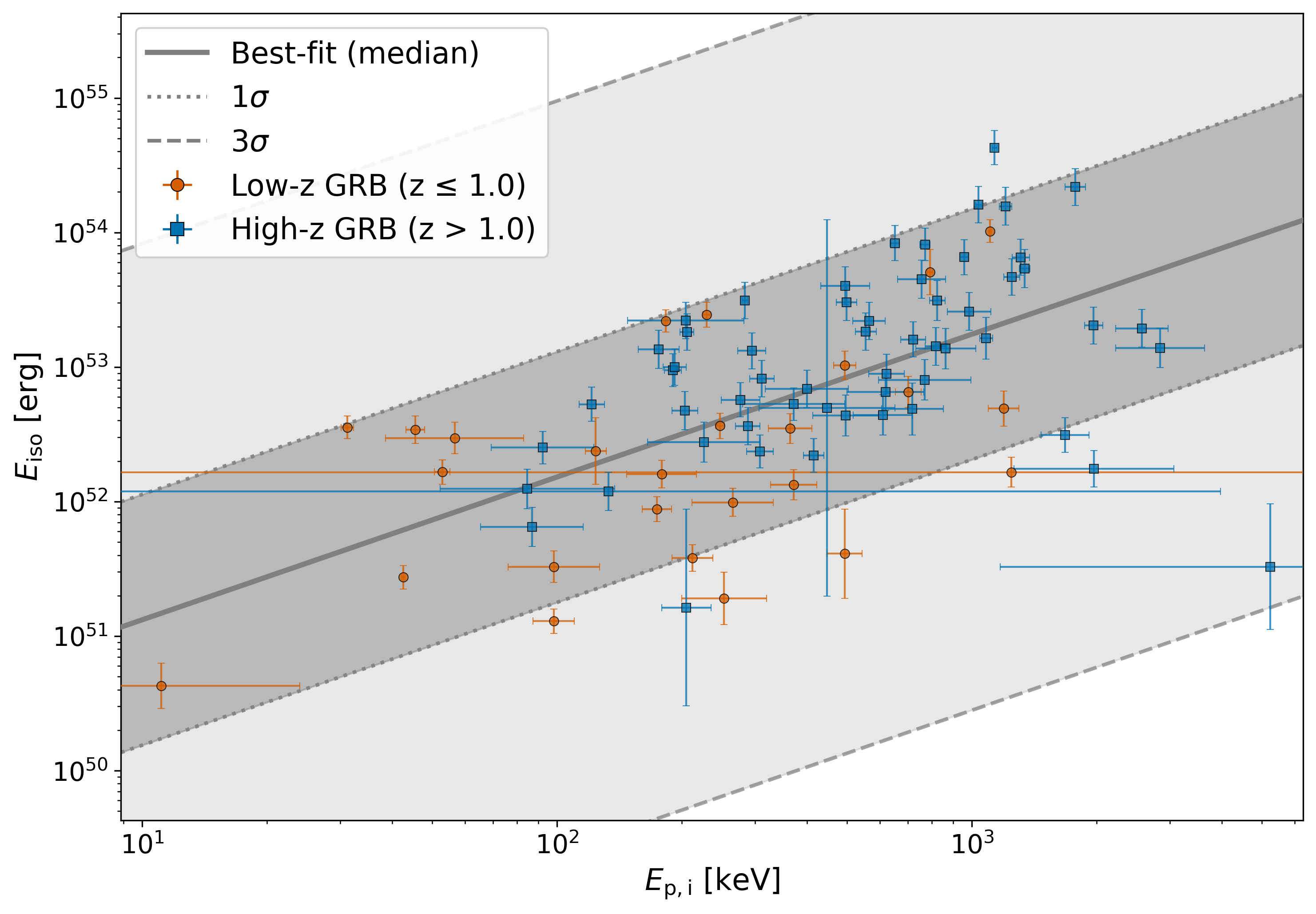}
	\includegraphics[width=1.0\linewidth]{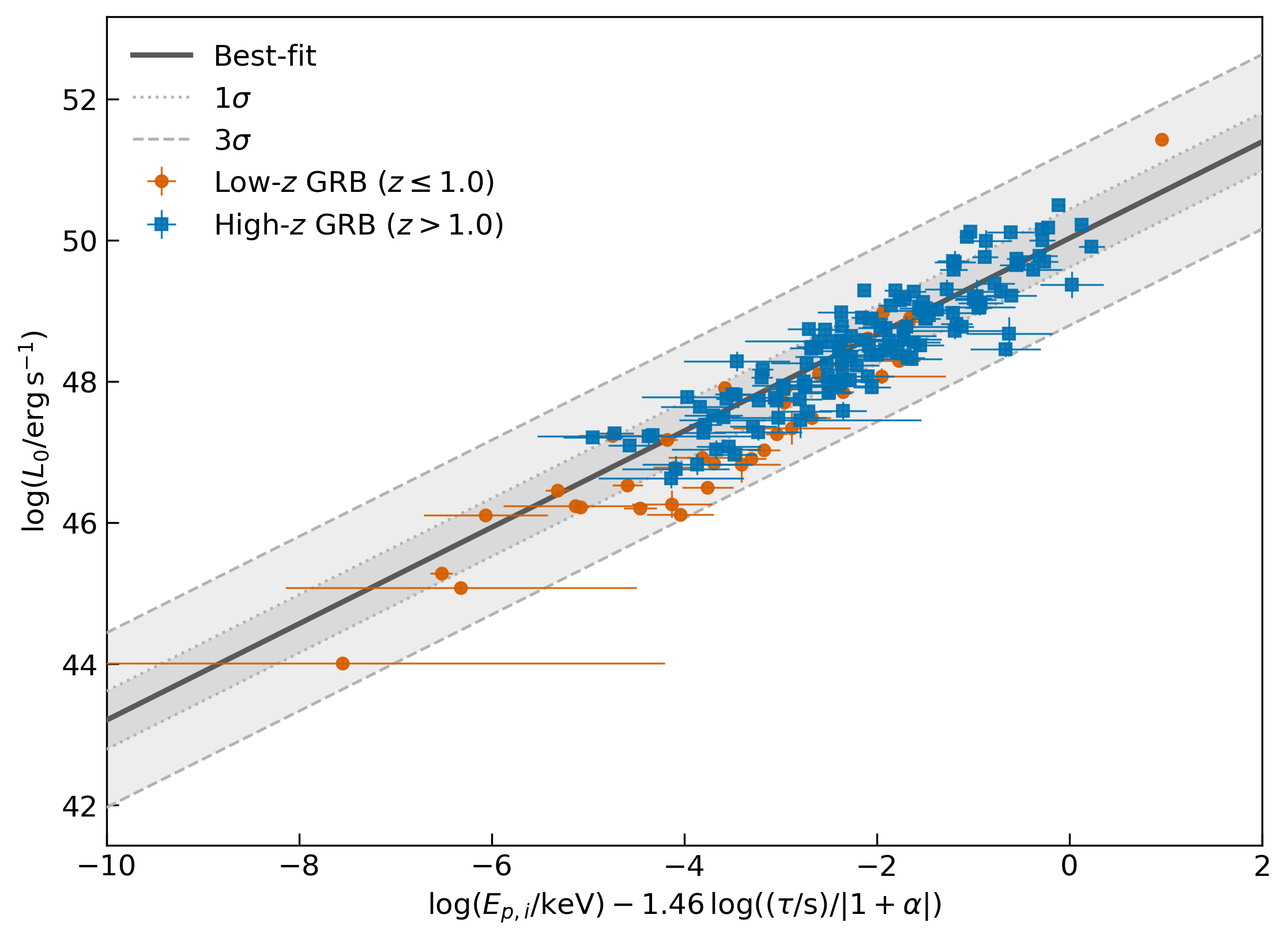}
	\caption{Empirical relations for standardising GRBs as distance indicators. Top: Amati relation between the isotropic-equivalent energy, $E_{\mathrm{iso}}$, and the intrinsic peak energy, $E_{p,i}$. Bottom: Combo relation in the parameter space of $\log L_0$ versus $\log(E_{p,i}/\mathrm{keV})-1.46\log[(\tau/\mathrm{s})/|1+\alpha|]$. Low- and high-redshift GRBs are shown in orange and blue, respectively.  The solid lines show the best-fitting relations, and the grey bands indicate the corresponding $1\sigma$ and $3\sigma$ intrinsic-scatter regions.}
	\label{figtworelation}
\end{figure}

Finally, Fig.~\ref{figtworelation} compares the Amati and Combo correlations, which rely on different observables but are calibrated against the same low-redshift distance scale provided by the ANN reconstruction. The Amati relation uses prompt-emission quantities, whereas the Combo relation additionally includes X-ray plateau information. The figure shows the best-fitting correlations and their intrinsic scatter, with the high-redshift events included for comparison rather than as independent calibrators. The two relations can respond differently to sample selection, intrinsic scatter, and observational systematics because of their different observables and sample properties \citep{2021JCAP...09..042K}.

\section{Discussion}

Our methodology consists of two main steps. First, we used ReFANN to reconstruct $d_L(z)$ and $\mu(z)$ from Pantheon$+$ SH0ES data, providing the low-redshift distance scale used to calibrate the Amati and Combo relations. To reduce dependence on subjective network choices, we combined ABC rejection based on $\chi^2$, Euclidean distance, and approximated LML with a risk-function evaluation, identifying HL2-N128 as the preferred architecture for both reconstructions. The final reconstructions include observational uncertainties and model variance estimated from the accepted network ensemble. Second, the reconstructed distances are used to calibrate the GRB correlations below $z_{\mathrm{cut}}=1$, after which the high-redshift GRBs are incorporated into the cosmological inference. The correlation and cosmological parameters are sampled jointly so that calibration uncertainty, intrinsic scatter, and cosmological uncertainty are propagated through the inference.

An important distinction emerging from our analysis is that removing the assumed cosmological model from the calibration step does not by itself establish that a GRB sample is suitable for cosmological inference. We therefore tested the uncalibrated correlations directly across flat and non-flat $\Lambda$CDM and XCDM before applying the ANN calibration. The correlation parameters show no statistically significant dependence on the tested cosmologies, with maximum joint shifts of $0.14\sigma$, $0.24\sigma$, and $0.59\sigma$ for A118, the 77-event Amati sample, and Combo/C174, respectively. The A118 analysis also reproduces the published correlation parameters within $1\sigma$ \citep{2021JCAP...09..042K}. These results support the model stability of the correlations themselves, but the external comparison with OHD shows that model stability and cosmological consistency are not equivalent. A118 agrees most closely with the OHD matter-density constraint, while the 77-event Amati sample shows a moderate shift towards higher $\Omega_m$, although the OHD value remains within its broad 95\% credible interval. In contrast, C174 differs from OHD by $4.40\sigma$, with the OHD value lying outside its 95\% credible interval. We therefore base the cosmological interpretation primarily on the Amati sample and retain Combo as an exploratory calibration cross-check. Similar sample-dependent behaviour has been reported in previous GRB analyses, where some datasets yield cosmological constraints that are not fully consistent with the currently accelerating expansion \citep{2021JCAP...09..042K,2024JCAP...10..093C,2025JCAP...09..081C}.

After ANN calibration, the Amati and Combo analyses yield overlapping cosmological posteriors within the current uncertainties. Under flat $\Lambda$CDM, both give broad constraints on $\Omega_m$, while the extension to $w_0w_a$CDM produces substantially wider posteriors because of the additional degeneracy between $w_0$ and $w_a$. In both cases, the $\Lambda$CDM limit $(w_0,w_a)=(-1,0)$ remains within the 95\% credible region, and the present GRB samples therefore provide no significant evidence for an evolving dark-energy equation of state. The high-redshift GRBs extend the calibrated Hubble diagram beyond the main SN Ia redshift range, but their present precision is insufficient for competitive constraints on dynamical dark energy. Their role is consequently complementary rather than competitive, providing high-redshift distance information in a regime where SNe Ia become sparse.

The interpretation of $H_0$ requires particular care. Because the ANN reconstruction is anchored to the Pantheon$+$ SH0ES distance scale, the inferred GRB values of $H_0$ are conditional on that absolute calibration and are not independent local measurements. This explains why $H_0$ remains relatively stable when the cosmological model is extended from $\Lambda$CDM to $w_0w_a$CDM. Our controlled zero-point test further shows that changing the adopted SN Ia absolute scale primarily shifts the Amati intercept, while the remaining correlation and cosmological parameters vary by less than $0.12\sigma$. The shape information inferred from the calibrated Hubble diagram is therefore insensitive to the tested zero-point changes, although its absolute scale remains calibration dependent. This distinction is important when comparing the GRB-derived $H_0$ posteriors with CMB, BAO, or local-distance-ladder measurements.

The Amati and Combo relations also respond differently to observational uncertainties because they involve different GRB observables. The Amati relation uses prompt-emission quantities, whereas Combo additionally incorporates X-ray plateau information through $L_0$, $\tau$, and $\alpha$. The calibrated Combo relation exhibits a smaller scatter in some stages of the inference, but this should not be interpreted as evidence of greater cosmological reliability. Its additional observables introduce different measurement uncertainties, selection requirements, and possible afterglow systematics. The pre-calibration OHD comparison demonstrates that a relatively stable or narrow empirical correlation does not necessarily translate into cosmological consistency. The overlap between the ANN-calibrated Amati and Combo posteriors is therefore best regarded as a comparison of the common calibration procedure rather than as an independent validation of both GRB samples. More pronounced differences among GRB correlations can emerge with larger and more homogeneous samples and with explicit modelling of selection effects \citep{2021JCAP...09..042K}.

The present results are consistent with $\Lambda$CDM and do not provide significant evidence for departures from it within the $w_0w_a$CDM framework. This is consistent with recent studies showing that apparent preferences for dynamical dark energy can depend sensitively on the adopted data combination and modelling assumptions \citep{2025PhRvD.111b3512C,2024A&A...690A..40L}. Model-independent GRB analyses have likewise found no strong evidence against a cosmological constant when calibration uncertainties are propagated \citep{2020A&A...641A.174L}, while analyses calibrated directly with Dark Energy Spectroscopic Instrument (DESI) survey, OHD, or BAO data can favour mildly evolving dark-energy solutions for some data combinations \citep{2024JCAP...12..055A,2025JHEAp..4600348A}. The range of conclusions in the current literature therefore appears to reflect, at least in part, differences in calibration strategy, sample construction, and likelihood treatment. Our results do not support interpreting the present GRB data alone as robust evidence for an evolving $w(z)$.

Several limitations should nevertheless be kept in mind. The ANN reconstruction is a non-parametric empirical estimate constrained by the redshift coverage and precision of the SN Ia sample, and more explicitly Bayesian neural-network approaches could provide an alternative treatment of predictive uncertainty. We also assumed a redshift-independent intrinsic scatter. Allowing $\sigma_v$ to vary with redshift would provide a direct test for possible evolution of the GRB population or of the empirical correlations. In addition, detector thresholds, spectral-selection requirements, redshift completeness, and follow-up selection can truncate the observed distributions in the $E_{p,i}$--$E_{\mathrm{iso}}$ and Combo planes. A complete treatment would require joint modelling of the underlying GRB population and the relevant detector and follow-up selection functions. Such modelling is beyond the scope of the present analysis, and the remaining selection effects should therefore be regarded as a systematic limitation rather than as effects excluded by the absence of a visible redshift trend in the residuals. Future larger and more homogeneous GRB samples, together with improved population and selection modelling, will be essential for determining whether the high-redshift reach of GRBs can be converted into substantially stronger cosmological constraints.

\section{Conclusions}
We combined the Pantheon$+$ sample of SNe Ia, an ANN optimised through ABC rejection and a dedicated risk function, and two empirical GRB correlations, namely the Amati and Combo relations. From our analysis, a number of significant results emerge:

(i) Our methodology calibrates GRB correlations without assuming a cosmological model and is based on an ANN distance reconstruction correlations without assuming a specific cosmological model. The resulting distance ladder extends cosmological measurements up to redshift $z \sim 9$. 

(ii) Before applying the ANN calibration, we tested the model dependence of the GRB correlations across flat and non-flat $\Lambda$CDM and XCDM. The correlation parameters show no statistically significant dependence on the tested cosmologies. The A118 benchmark gives the closest agreement with the independent OHD matter-density constraint, while the 77-event Amati sample shows a moderate shift towards higher $\Omega_m$, although the OHD value remains within its 95\% credible interval. In contrast, Combo/C174 shows a substantially larger discrepancy with OHD. We therefore based the main cosmological interpretation on the Amati sample and retain Combo primarily as a calibration cross-check.

(iii) After ANN calibration, the Amati and Combo analyses yield overlapping cosmological posteriors within the current uncertainties. In $w_0w_a$CDM, the constraints remain broad and are compatible with the $\Lambda$CDM limit $(w_0,w_a)=(-1,0)$ within the 95\% credible regions. The present GRB data therefore provide no significant evidence for an evolving dark-energy equation of state.

(iv) The inferred $H_0$ values are conditional on the absolute distance scale inherited from Pantheon$+$ SH0ES and should not be interpreted as independent GRB measurements of the local Hubble constant. A controlled zero-point test shows that changing the adopted SN Ia absolute scale mainly shifts the Amati intercept, while the remaining correlation and cosmological parameters vary by less than $0.12\sigma$. The inferred shape of the high-redshift expansion history is therefore stable against the tested zero-point changes, although its absolute scale remains calibration dependent. The comparatively high $\Omega_m$ values inferred from the pre-calibration GRB-only fits should not be regarded as independent evidence for a higher cosmic matter density, given the sample-dependent discrepancies revealed by the comparison with OHD and the remaining observational and selection-related systematics.

(v) The present analysis is mainly limited by the small size of the available high-redshift GRB sample and by residual selection effects, particularly Malmquist bias. With a larger number of well-observed high-$z$ GRBs from ongoing and future missions such as the Space-based multi-band astronomical Variable Objects Monitor (SVOM) \citep{2011CRPhy..12..298P,2016arXiv161006892W,2022IJMPD..3130008A}, Einstein Probe (EP) \citep{2022hxga.book...86Y,2025NatAs...9..564L}, and the Transient High-Energy Sky and Early Universe Surveyor (THESEUS) \citep{2018AdSpR..62..191A,2018AdSpR..62..662S,2021arXiv210208702A,2021ExA....52..183A}, the statistical uncertainties on the cosmological parameters are expected to decrease. Larger and more homogeneous samples will help determine whether the relatively high $\Omega_m$ estimates obtained from some present GRB-only fits persist after accounting for sample selection and other observational systematics. Further progress will also require improved modelling of selection effects, together with cross-checks against other high-redshift probes, including BAO and future gravitational-wave standard sirens.

\section*{Data availability}
Tables E1--E3 are available in electronic form at the Centre de Données astronomiques de Strasbourg (CDS) via anonymous FTP to cdsarc.u-strasbg.fr (130.79.128.5) or via \url{http://cdsweb.u-strasbg.fr/cgi-bin/qcat?J/A+A/}.

\begin{acknowledgements}
Wei Hong is supported by China Scholarship Council (File No. 202506040057). Luca Izzo acknowledges financial support from the INAF Data Grant Program 'YES' (PI: Izzo) {\it Multi-wavelength and multi-messenger analysis of relativistic supernovae}. Tong-Jie Zhang is supported by National SKA Program of China, No. 2022SKA0110202, the China Manned Space Program with grant No. CMS-CSST-2025-A01 and China Scholarship Council (File No. 202606040159). 

\end{acknowledgements}

\bibliographystyle{aa}
\bibliography{GRBCC}

\begin{appendix}
\nolinenumbers
\section{Methodological appendix to the ANN reconstruction}
\label{app:sect3}

This appendix provides the technical details supporting the ANN-based reconstruction methodology described in Sect.~3. While the main text focuses on the distance-modulus ($\mu$) reconstruction for conciseness, we present here the corresponding luminosity-distance ($d_L$) results, together with the ANN-selection diagnostics and explicit definitions of the uncertainty decomposition, ABC distance measures, and model-ranking criteria used in the ReFANN pipeline.
\begin{equation}
	\sigma_{d_L, \mathrm{ANN}}^2(z) = \sigma_{d_L, \mathrm{data}}^2(z)+\sigma_{d_L, \mathrm{model}}^2(z),
\end{equation}
where the error of the luminosity distance $\sigma_{d_L, \mathrm{data}}(z)$ can be obtained from Eq.~(\ref{eq:dl_from_mu}), and $\sigma_{d_L, \mathrm{model}}(z)$ is the standard deviation between networks,
\begin{equation}
	\sigma_{d_L, \mathrm{model}}(z)=\sqrt{\frac{1}{N-1} \sum_i\left[d_{L,i}^{(k)}(z)-\bar{d}_L(z)\right]^2},
\end{equation}
where $N$ is the number of the SN Ia data and $\bar{d}_L(z)=\frac{1}{N} \sum_i d_{L,i}^{(k)}(z)$, while $d_{L,i}^{(k)}$ represents the predicted values of ANN.

Within the ABC framework, different network architectures are treated as distinct candidate models. For each network, we evaluated three summary statistics on a validation set. The first is the goodness of fit $\chi^2$ \citep{2021JCAP...08..027B},
\begin{equation}
	D_{\chi^2}^{(k)}=\sum_i \frac{\left[d_{L,i}^{\mathrm{obs}}-d_{L,i}^{(k)}\right]^2}{\sigma_{d_{L,i}}^2},
\end{equation}
where $d_{L,i}^{\mathrm{obs}}$ represents the observed SN Ia data. The second is the Euclidean distance between predictions and observations \citep{2017A&C....19...16J,2023ApJS..266...27Z},
\begin{equation}
	D_{\mathrm{Euclid}}^{(k)}=\frac{\sqrt{\sum_i\left[d_{L,i}^{\mathrm{obs}}-d_{L,i}^{(k)}\right]^2}}{N},
\end{equation}
and the third is an approximate LML (\citealt{10.3389/fbuil.2017.00052,2018JCAP...04..051G,2021JCAP...08..027B}),
\begin{equation}
	D_{\mathrm{LML}}^{(k)}=\frac{1}{2}\left[\left(\mathbf{r}^{(k)}\right)^T \mathbf{C}^{-1} \mathbf{r}^{(k)}+\ln |\mathbf{C}|+N \ln (2 \pi)\right].
\end{equation}
For a Gaussian likelihood, the residual vector for the $k$-th network is defined as $\mathbf{r}^{(k)} = \boldsymbol{d_L}^{\mathrm{SNe}} - \boldsymbol{d_L}^{(k)}$, with the covariance matrix given by $\mathbf{C} = \mathbf{C}_{\mathrm{SNe}} + \sigma_{\mathrm{int}}^2 \mathbf{I}$. Here, $\mathbf{C}_{\mathrm{SNe}}$ denotes the full covariance matrix from Pantheon$+$, $\sigma_{\mathrm{int}}$ represents the extra variability of the ANN, and $\mathbf{I}$ is the identity matrix. The ABC-rejection algorithm uses these statistics to select architectures that are consistent with the Pantheon$+$ data. For each observable, an appropriate tolerance $\varepsilon$ is adopted to determine the accepted hyper-parameters $(N_{\mathrm{hl}},N_{\mathrm{node}})$.

Following the original ReFANN approach, we further ranked the accepted models using the risk function \citep{1999ITNN...10..988V,2001astro.ph.12050W}
\begin{equation}
	\begin{aligned}
		\mathrm{risk} & =\sum_{i}\operatorname{Bias}_i^2+\sum_{i} \operatorname{Variance}_i \\
		& =\sum_{i}\left[d_{L,i}^{\mathrm{obs}}-d_{L,i}^{(k)}\right]^2+\sum_{i=1} \sigma_{d_{L,i}}^2.\label{risk-func}
	\end{aligned}
\end{equation}
The ABC acceptance behaviour, risk scores, and Bayes-factor comparisons were then combined to identify the preferred ReFANN hyper-parameters \citep{24ce203a-855a-3aa9-952f-976d23b28943,10.1093/oso/9780198503682.001.0001}. The Bayes factor is defined as
\begin{equation}
	\mathcal{B}_f=\frac{P\left(\boldsymbol{d_L} \mid M_1\right)}{P\left(\boldsymbol{d_L} \mid M_2\right)}
	=\frac{P\left(M_1 \mid \boldsymbol{d_L}\right)}{P\left(M_2 \mid \boldsymbol{d_L}\right)}
	\frac{P\left(M_2\right)}{P\left(M_1\right)},\label{bayseq}
\end{equation}
where $M_1$ and $M_2$ represent alternative models, and $\boldsymbol{d_L}$ denotes the observed data. A value of $\mathcal{B}_f>1$ favours $M_1$ over $M_2$. The adopted interpretation of the Bayes factor is summarised in Table~\ref{table:baysjeffreys}.
\begin{table}[htbp]
	\centering
	\caption{Correspondence between the Bayes factor $\mathcal{B}_f$ and the strength of evidence.}
	\label{table:baysjeffreys}
	\begin{tabular}{c c c}
		\hline\hline
		$\mathcal{B}_f$ & $\log_{10} \mathcal{B}_f$ & Strength of evidence \\
		\hline
		$<10^0$ & $<0$ & Negative (supports $M_2$) \\
		$10^0$ to $10^{1/2}(\approx 3.16)$ & 0 to 1/2 & Barely worth mentioning \\
		$10^{1/2}(\approx 3.16)$ to $10^1$ & 1/2 to 1 & Substantial \\
		$10^1$ to $10^{3/2}(\approx 31.62)$ & 1 to 3/2 & Strong \\
		$10^{3/2}(\approx 31.62)$ to $10^2$ & 3/2 to 2 & Very strong \\
		$>10^2$ & $>2$ & Decisive \\
		\hline
	\end{tabular}
\end{table}

\begin{figure}[ht!]
	\centering
	\includegraphics[width=1.0\linewidth]{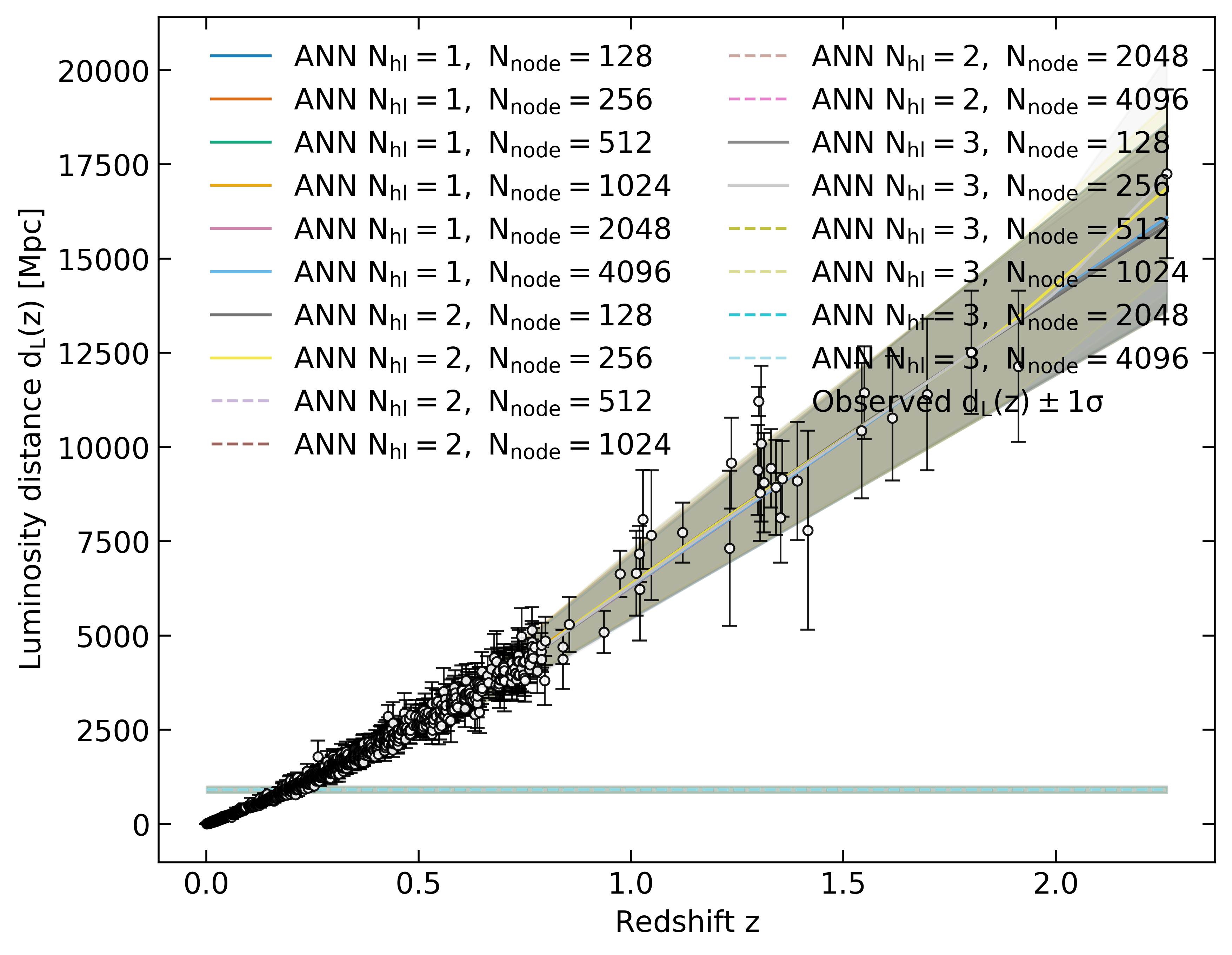}
	\caption{Same as Fig.~\ref{figANN-all}, but for the luminosity distance $d_L(z)$.}
	\label{figANN-dL}
\end{figure}

\begin{figure}[ht!]
	\centering
	\includegraphics[width=1.0\linewidth]{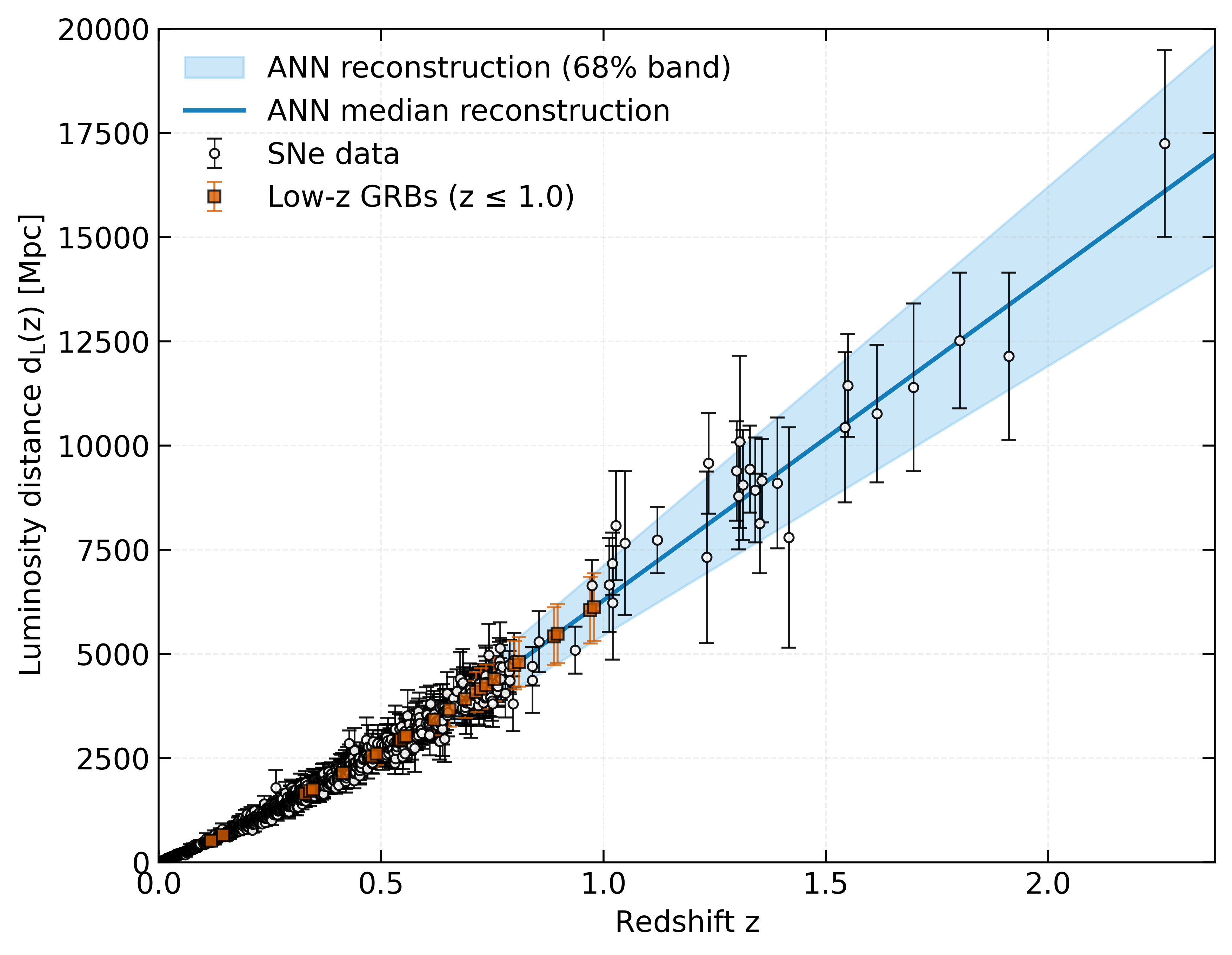}
	\caption{Same as Fig.~\ref{figc}, but for the luminosity distance $d_L(z)$.}
	\label{figc-dL}
\end{figure}

\begin{figure*}[ht!]
	\centering
	\includegraphics[width=0.87\linewidth]{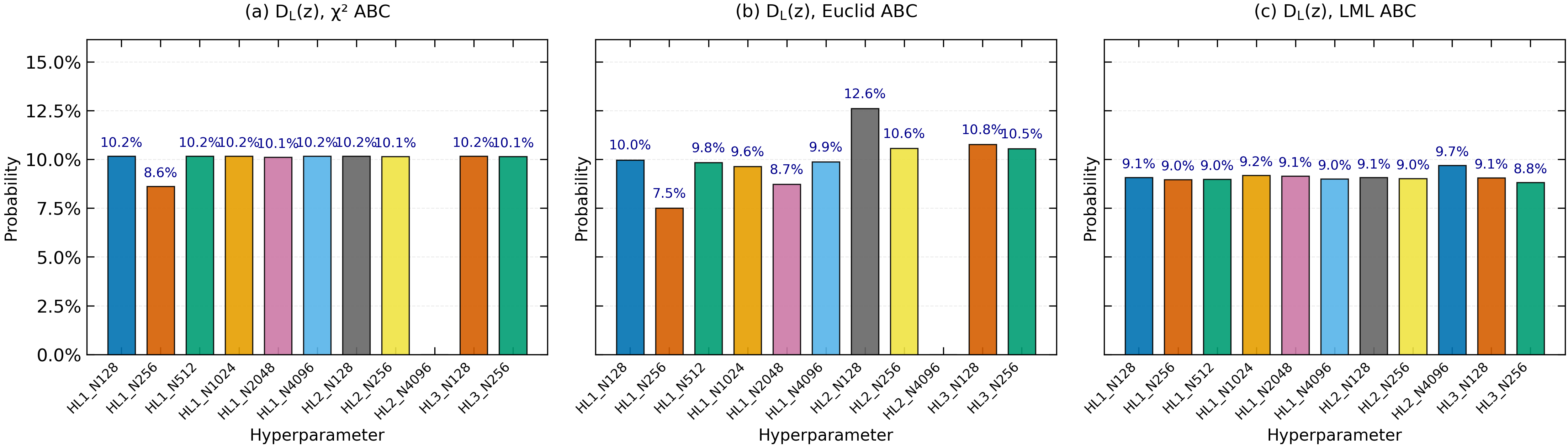}
	\includegraphics[width=0.87\linewidth]{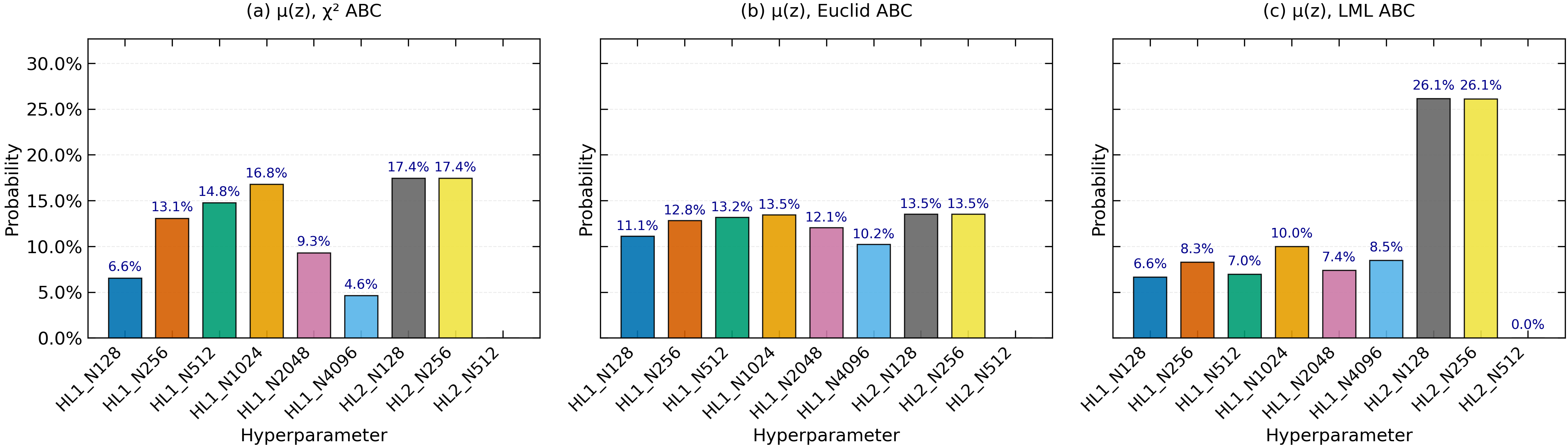}
	\caption{Posterior probabilities obtained using ABC rejection for the luminosity distance $d_L(z)$ (upper panel) and distance modulus $\mu(z)$ (lower panel), evaluated with $\chi^2$ (left column), Euclidean distance (middle column), and LML (right column). The ANN hyperparameters are denoted by $\mathrm{HL}i$-$\mathrm{N}j$, where $i$ and $j$ represent the number of hidden layers and the number of nodes in each layer, respectively.}
	\label{figABCJ-all}
\end{figure*}

\begin{figure*}[ht!]
	\centering
	\includegraphics[width=0.87\linewidth]{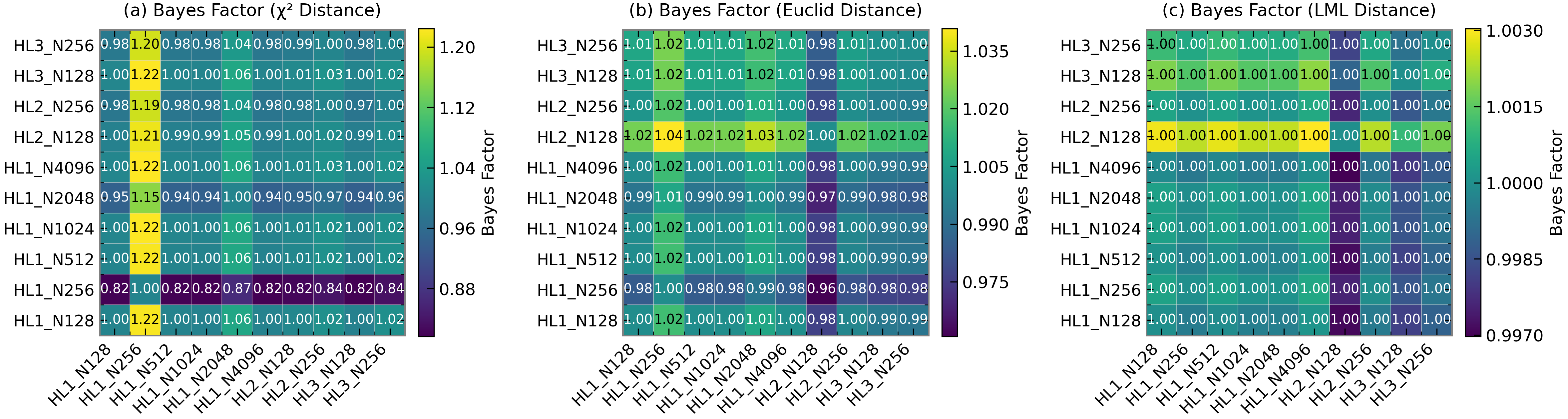}
	\includegraphics[width=0.87\linewidth]{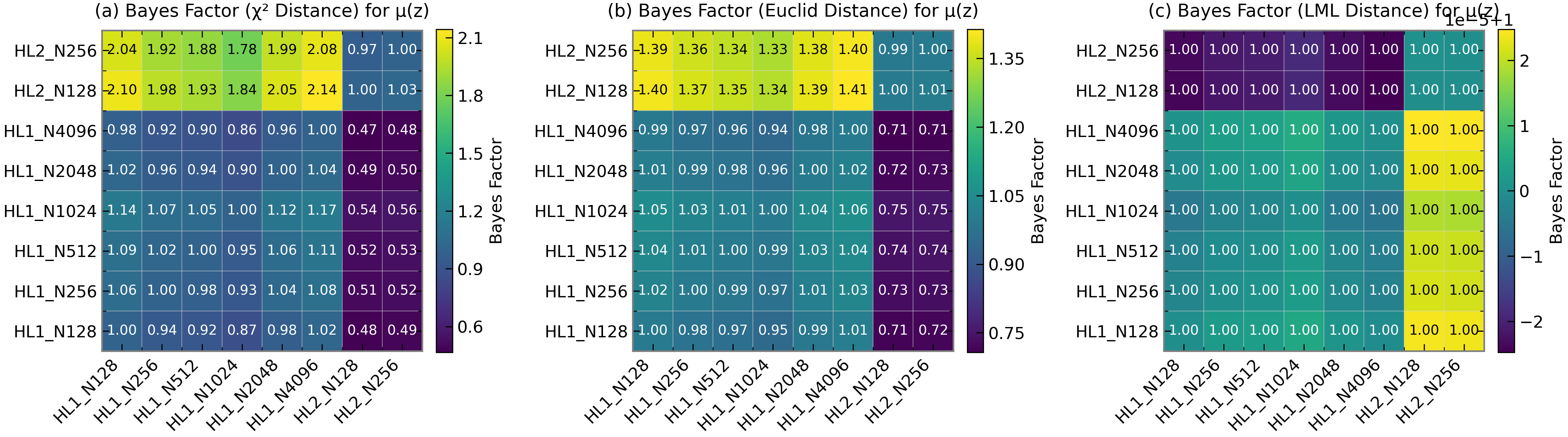}
	\caption{Bayes-factor comparison of ANN architectures for the luminosity distance $d_L(z)$ (upper panel) and distance modulus $\mu(z)$ (lower panel), using $\chi^2$ (left column), Euclidean distance (middle column), and LML (right column). The ANN architectures are denoted by $\mathrm{HL}i$-$\mathrm{N}j$.}
	\label{figBHM-all}
\end{figure*}

\section{Posterior distributions and distance modulus residuals}
\label{app:sect4}

To maintain clarity in the main text, the posterior corner plots and the GRB distance modulus residual figures are presented in this appendix.
\begin{figure}[ht!]
	\centering
	\includegraphics[width=1.0\linewidth]{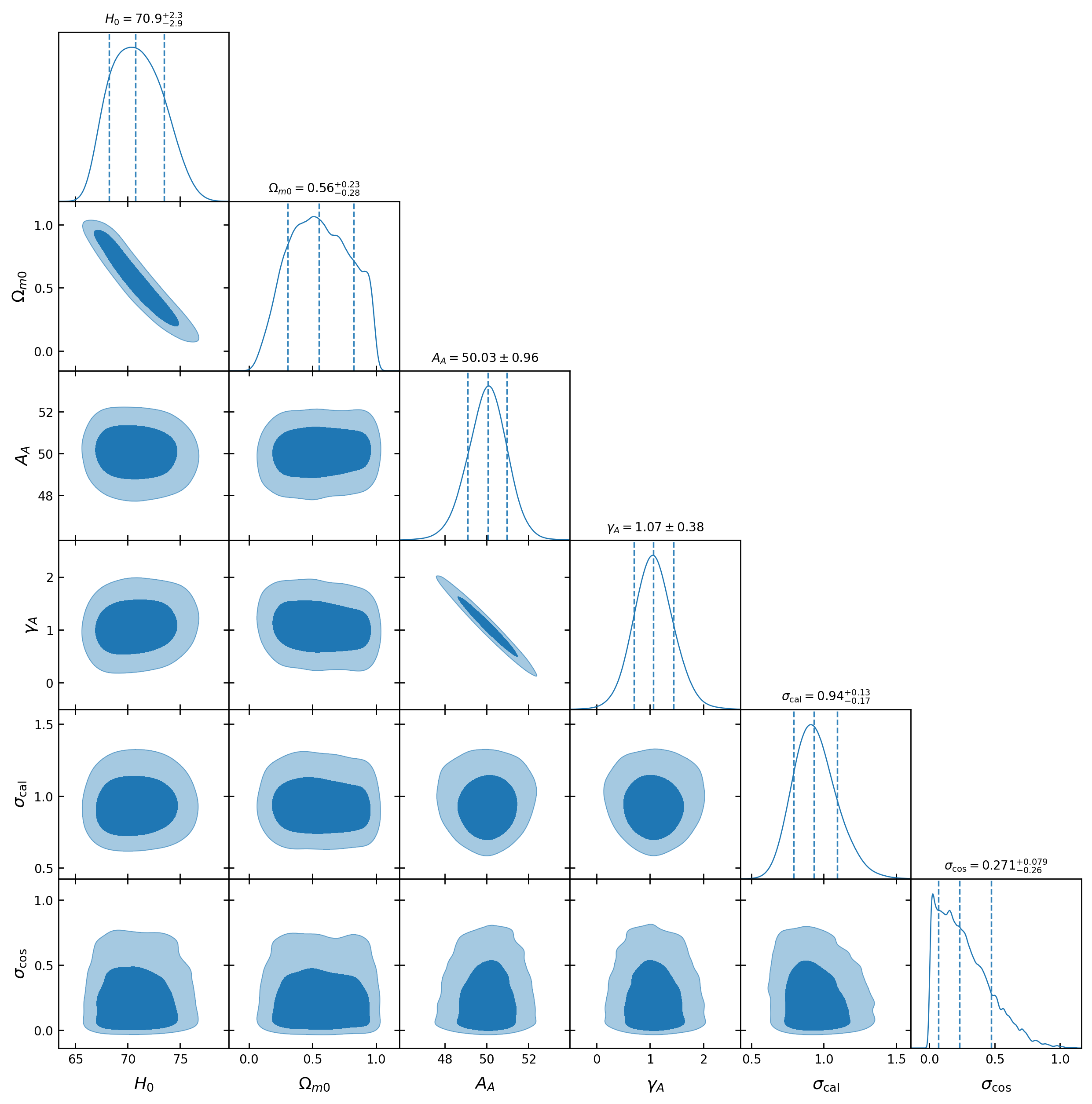}
	\includegraphics[width=1.0\linewidth]{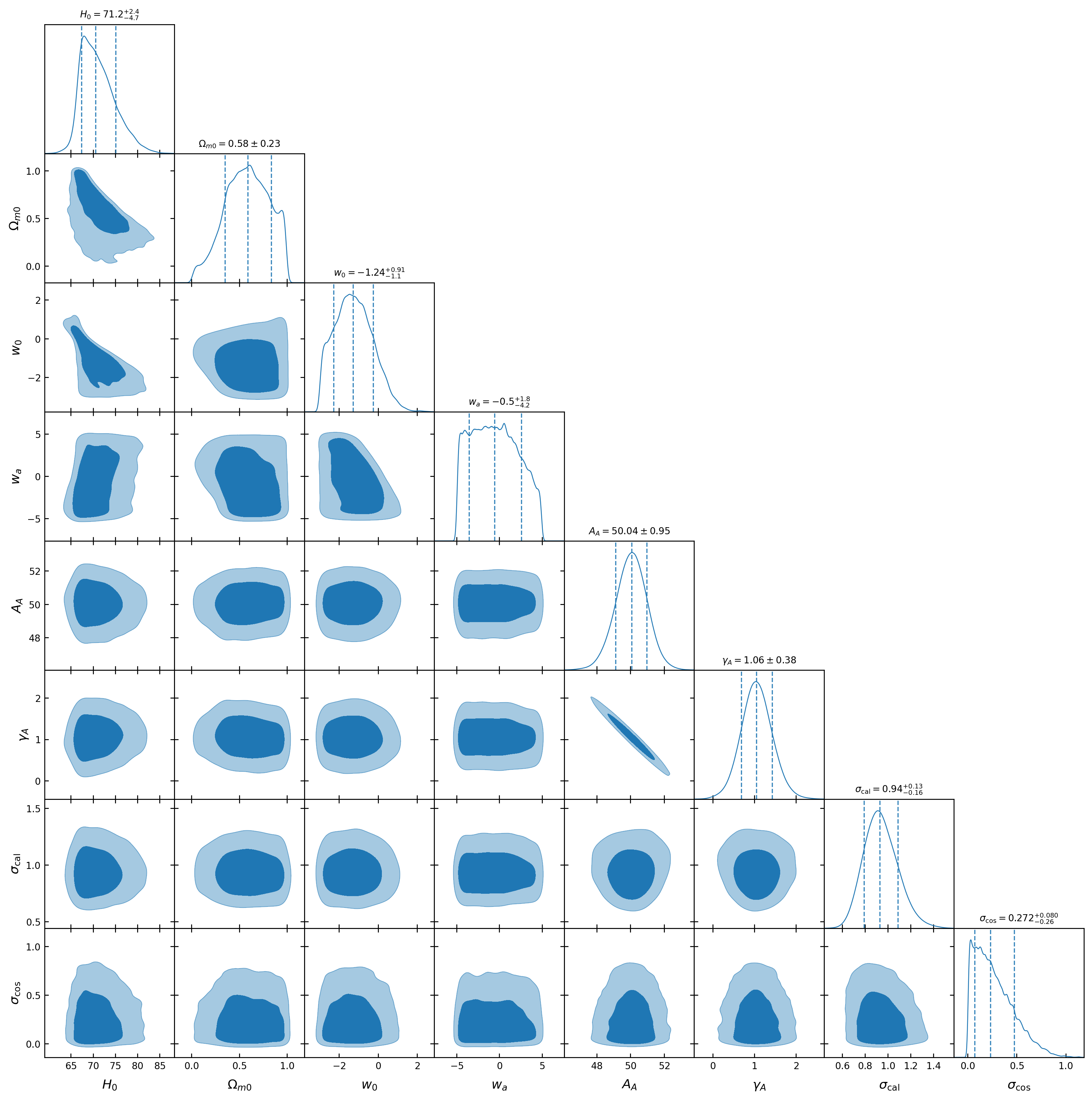}
	\caption{Joint and marginalised posterior distributions for the Amati analysis. The contours indicate the $68\%$, $95\%$, and $99\%$ credible regions. Top: Results for $\Lambda$CDM. Bottom: Results for $w_0w_a$CDM. The sampled parameters comprise the cosmological parameters, the Amati correlation parameters $(A_A,\gamma_A)$, and the intrinsic-scatter parameters $(\sigma_{\mathrm{cal}},\sigma_{\mathrm{cos}})$.}
	\label{figcorneram}
\end{figure}

\begin{figure}[ht!]
	\centering
	\includegraphics[width=1.0\linewidth]{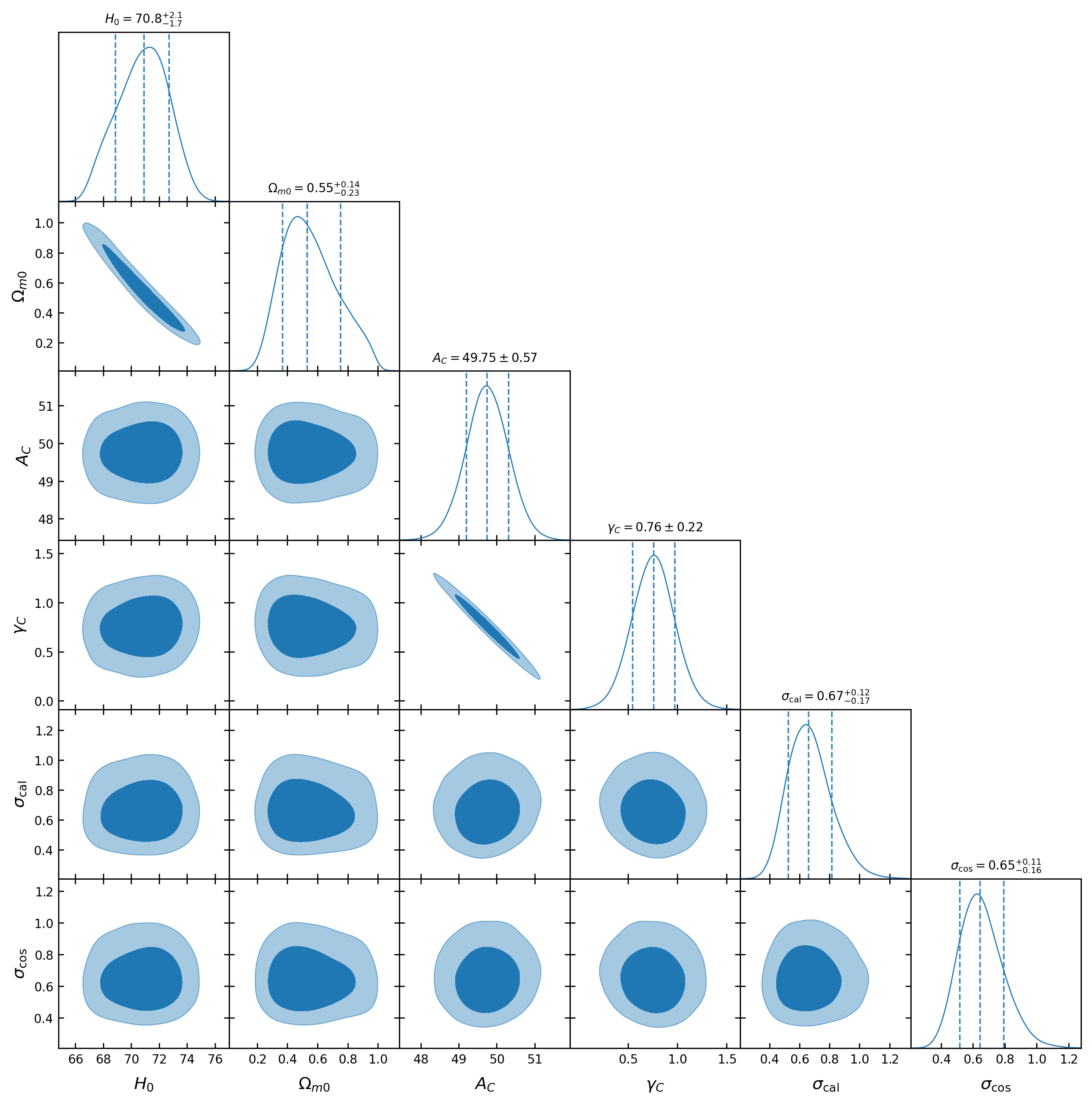}
	\includegraphics[width=1.0\linewidth]{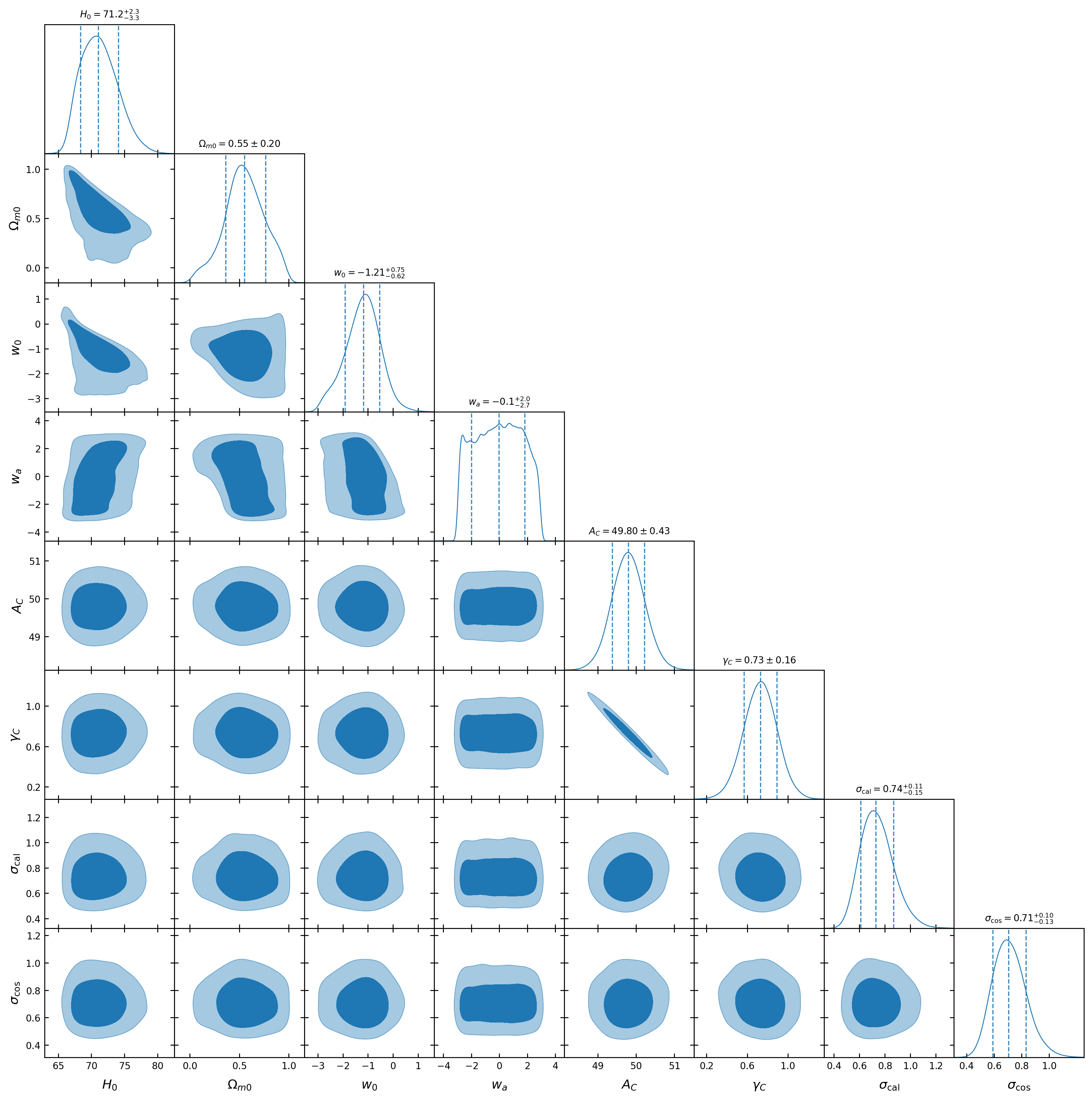}
	\caption{Same as Fig.~\ref{figcorneram}, but for the Combo analysis, with correlation parameters $(A_C,\gamma_C)$.}
	\label{figcornerco}
\end{figure}

\begin{figure}[ht!]
	\centering
	\includegraphics[width=0.9\linewidth]{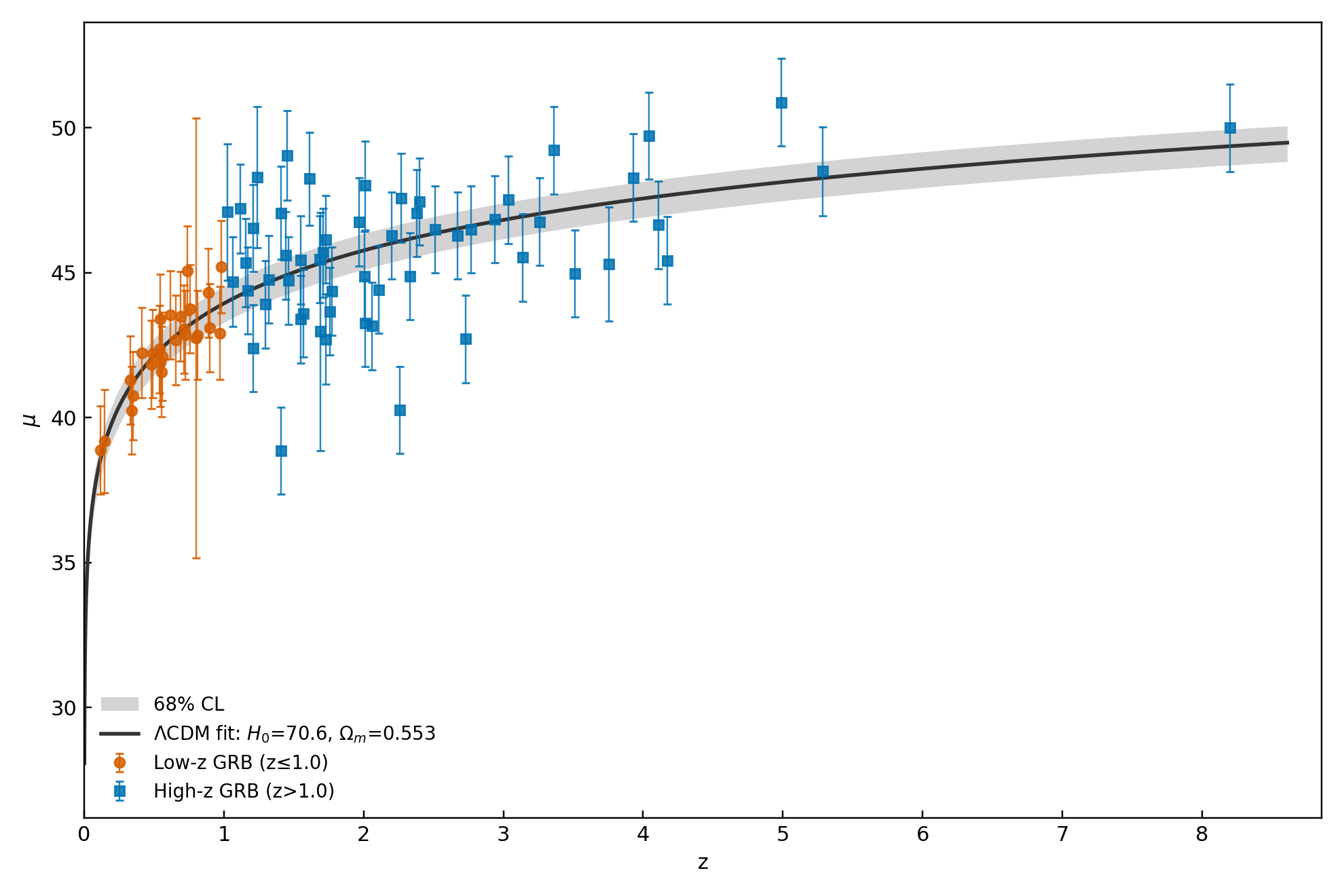}
	\includegraphics[width=0.9\linewidth]{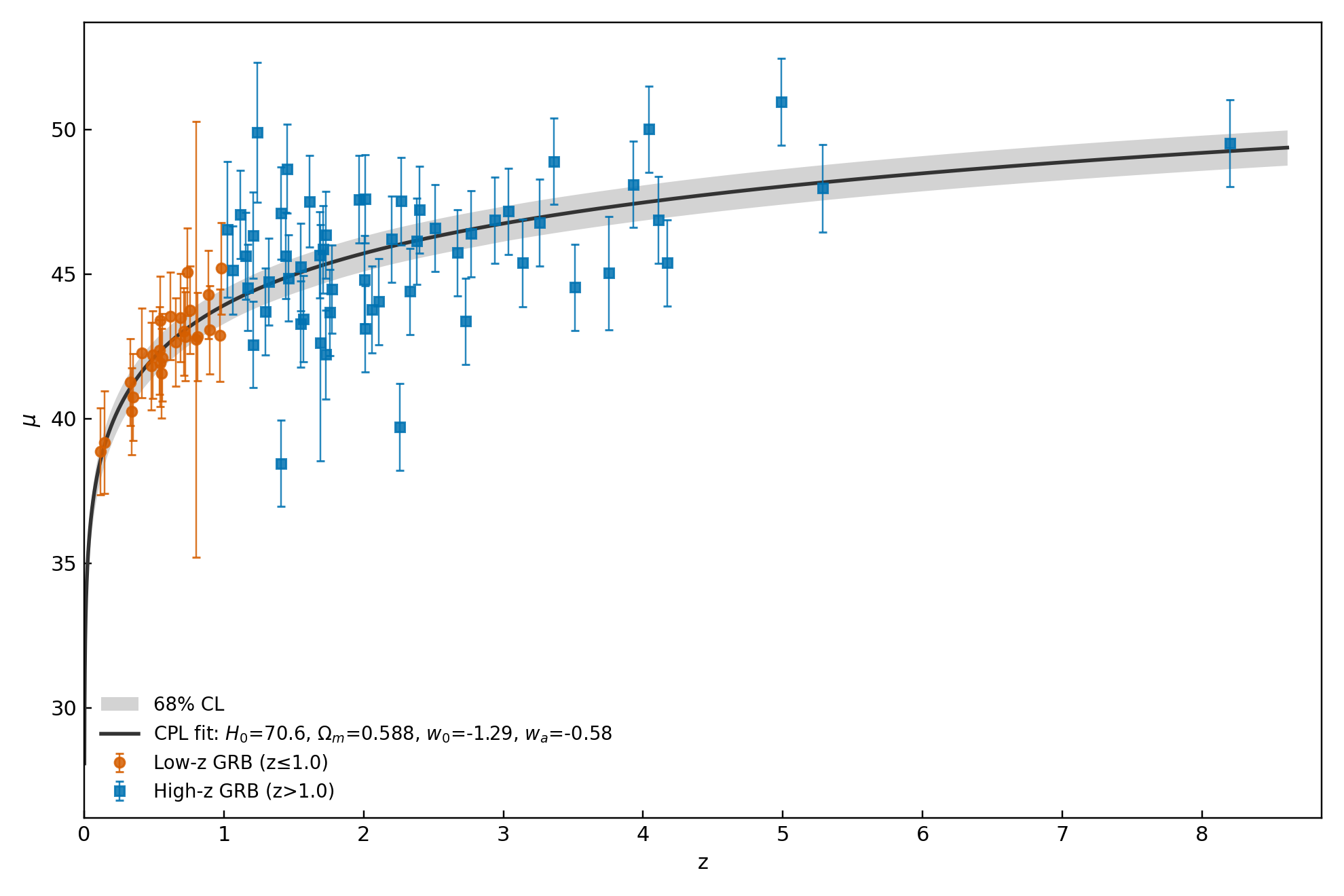}
	\includegraphics[width=0.9\linewidth]{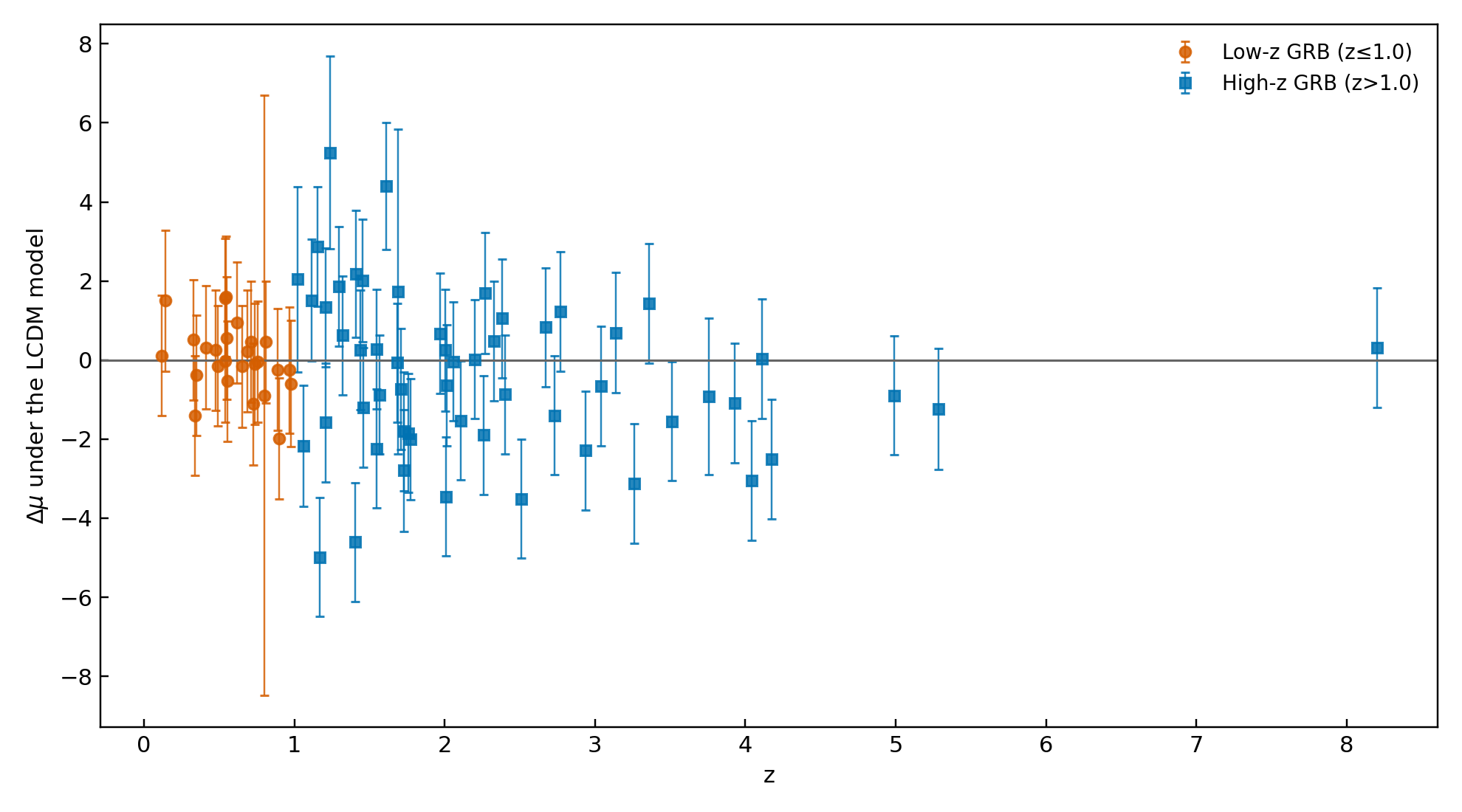}
	\includegraphics[width=0.9\linewidth]{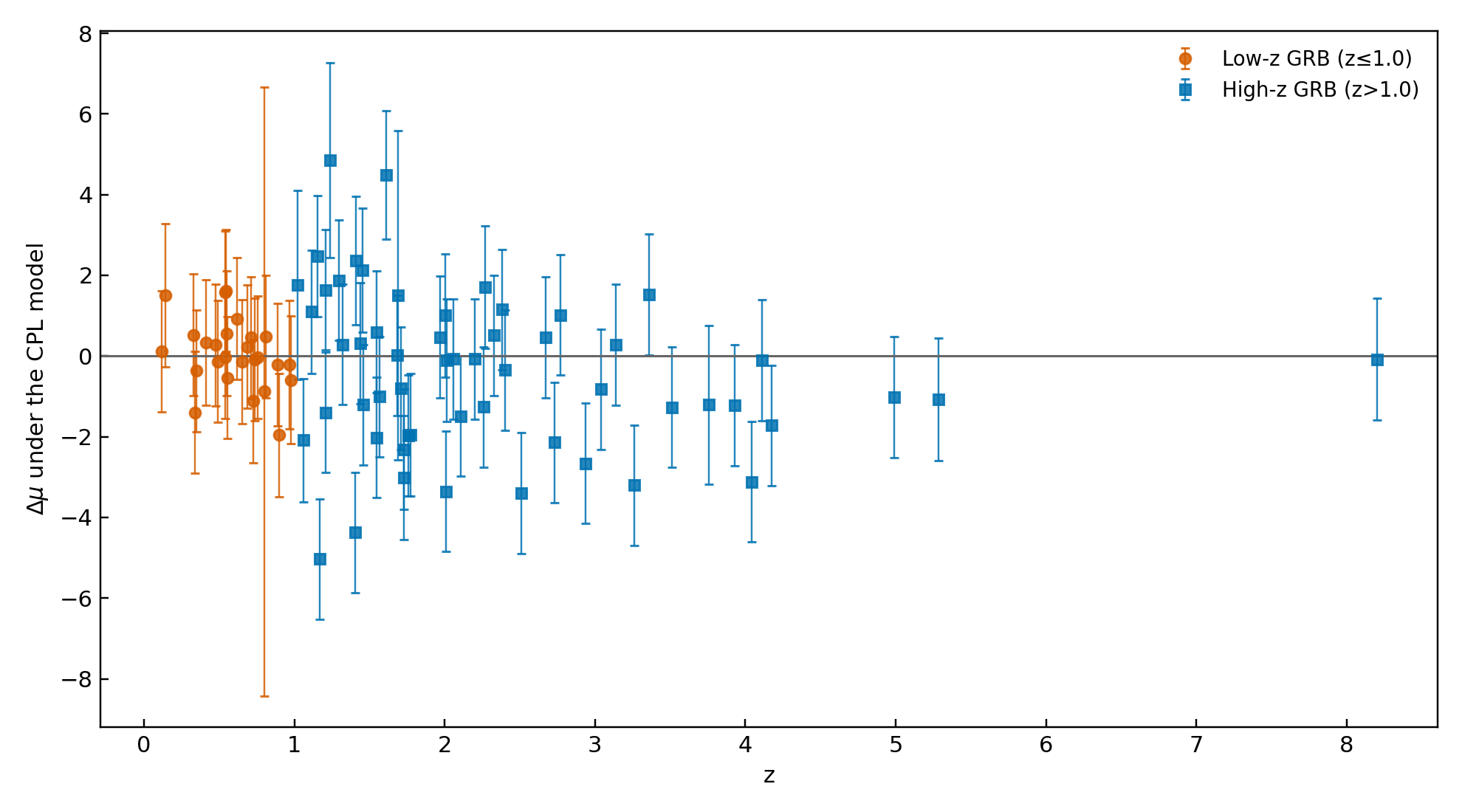}
	\caption{GRB distance moduli and residuals for the Amati relation. From top to bottom, the panels show the GRB distance moduli together with the best-fitting $\Lambda$CDM and $w_0w_a$CDM models, followed by the corresponding residuals, $\Delta\mu\equiv\mu_{\mathrm{GRB}}-\mu_{\mathrm{th}}(z;\boldsymbol{\theta})$. The residual panels provide a visual assessment of possible redshift-dependent trends.}
	\label{figresidualam}
\end{figure}

\begin{figure}[ht!]
	\centering
	\includegraphics[width=0.85\linewidth]{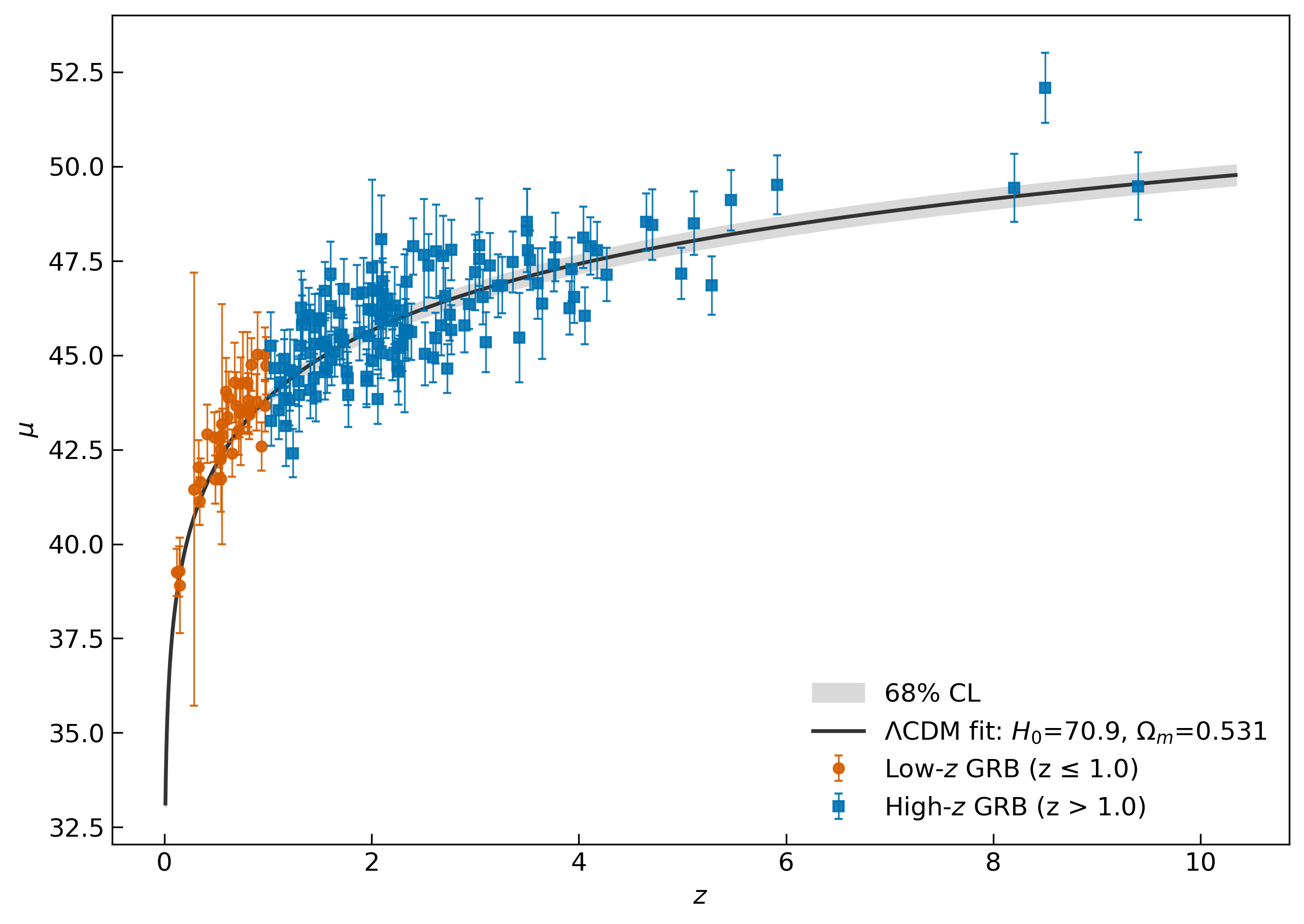}
	\includegraphics[width=0.85\linewidth]{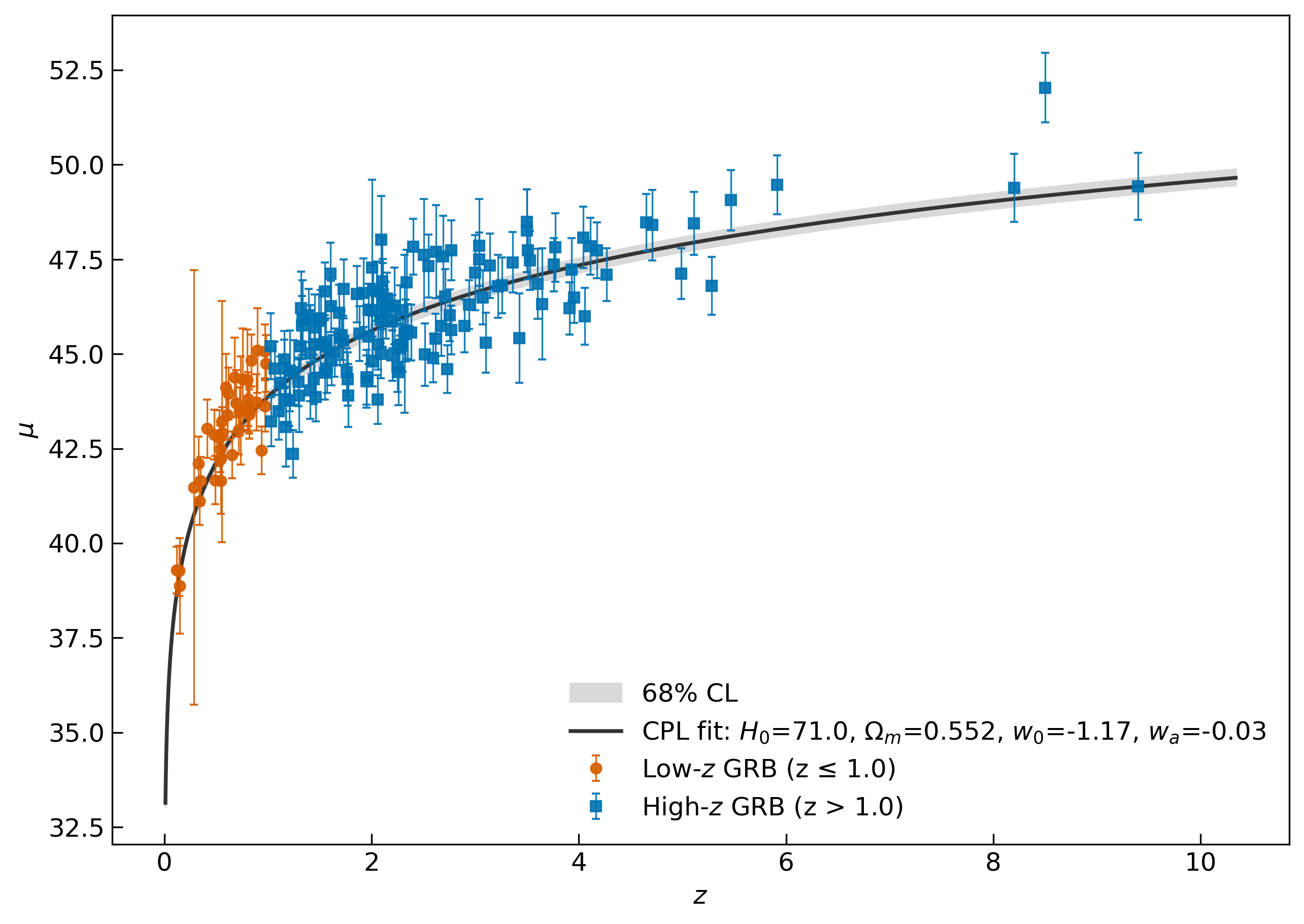}
	\includegraphics[width=0.85\linewidth]{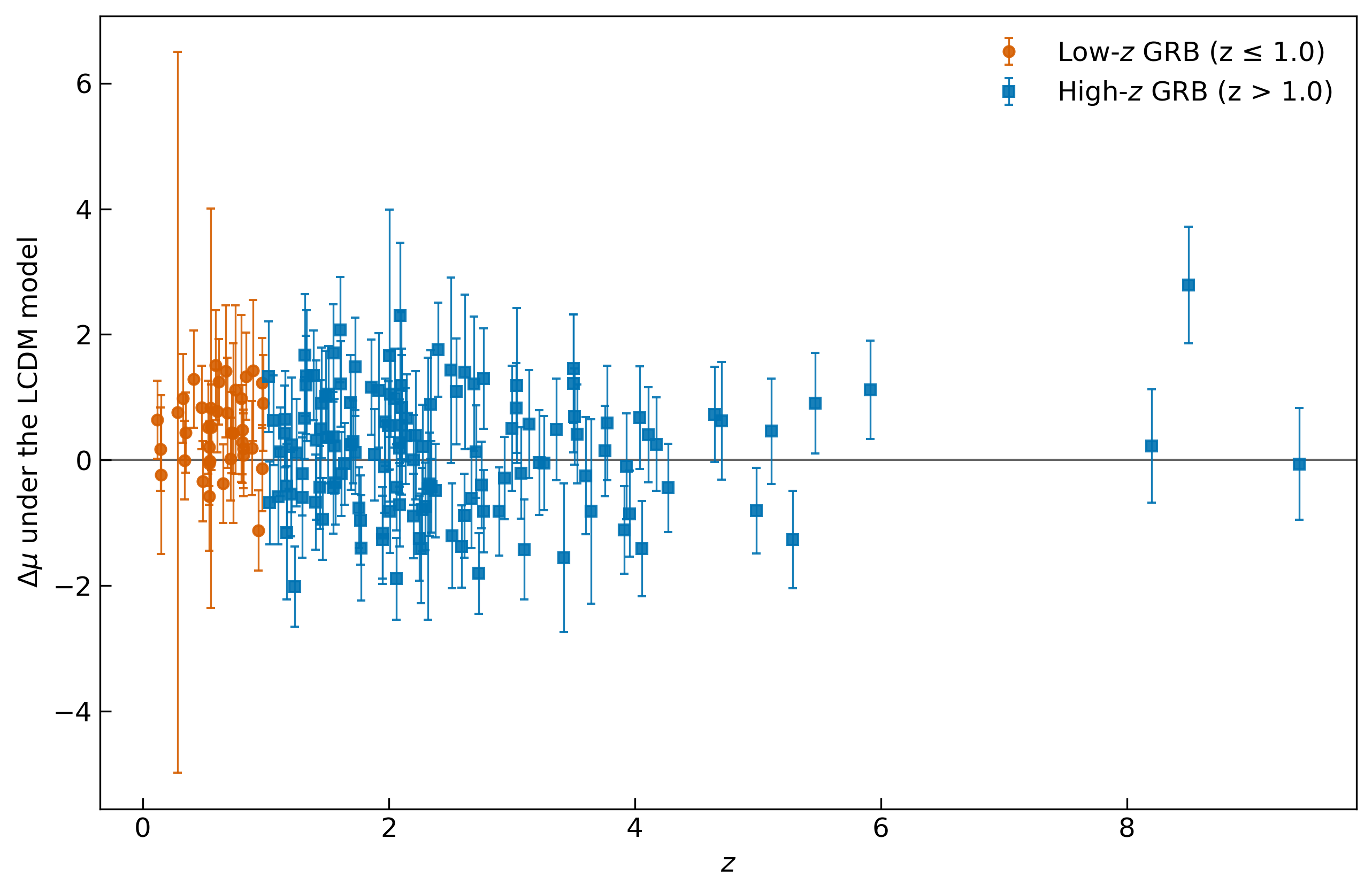}
	\includegraphics[width=0.85\linewidth]{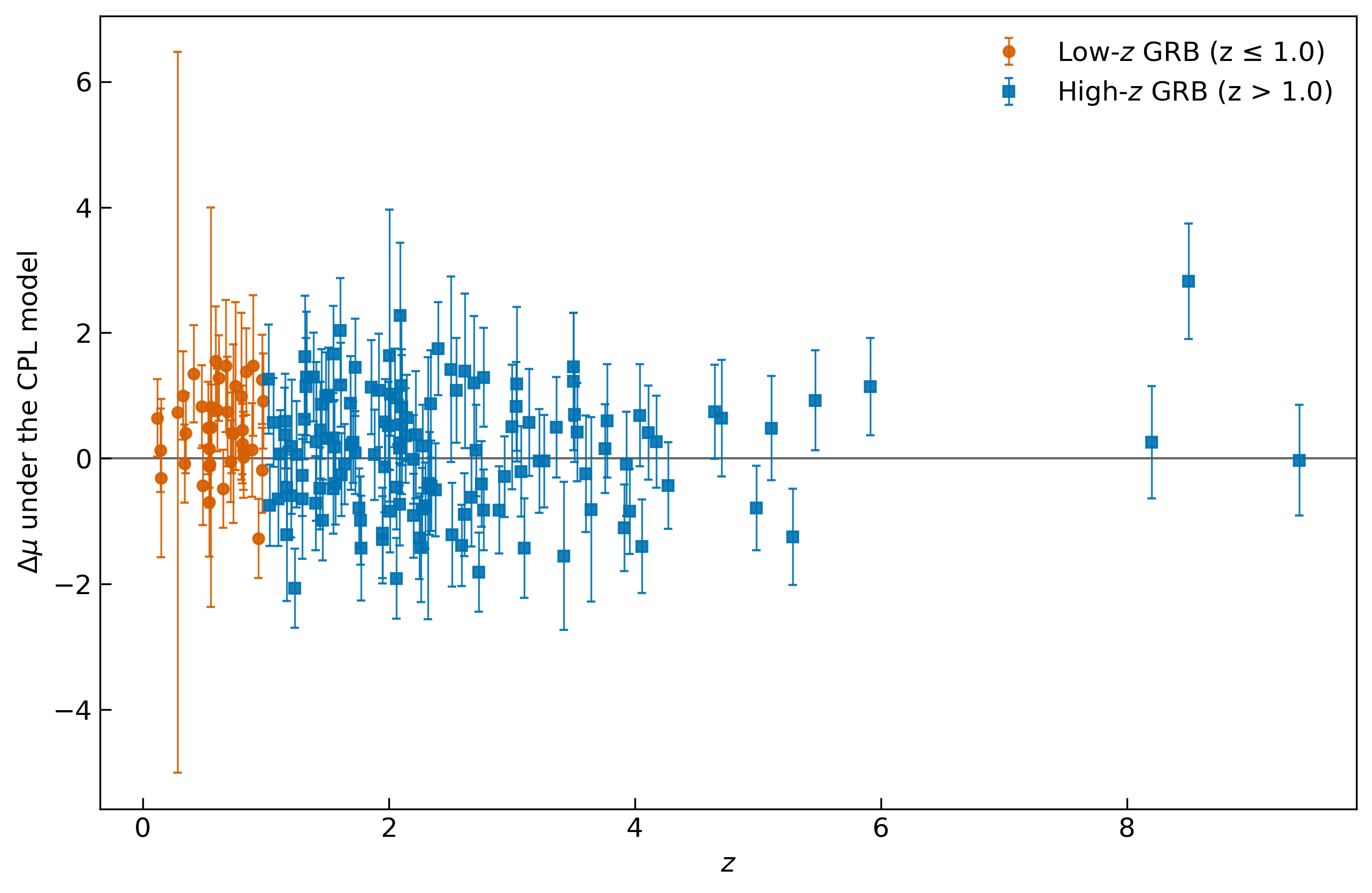}
	\caption{Same as Fig.~\ref{figresidualam}, but for the Combo relation.}
	\label{figresidualco}
\end{figure}

\section{Robustness tests}
\label{app:robustness}

\subsection{Robustness to the choice of $z_{\rm cut}$}
\label{app:robust_zcut}

\begin{figure}[ht!]
	\centering
	\includegraphics[width=1.0\linewidth]{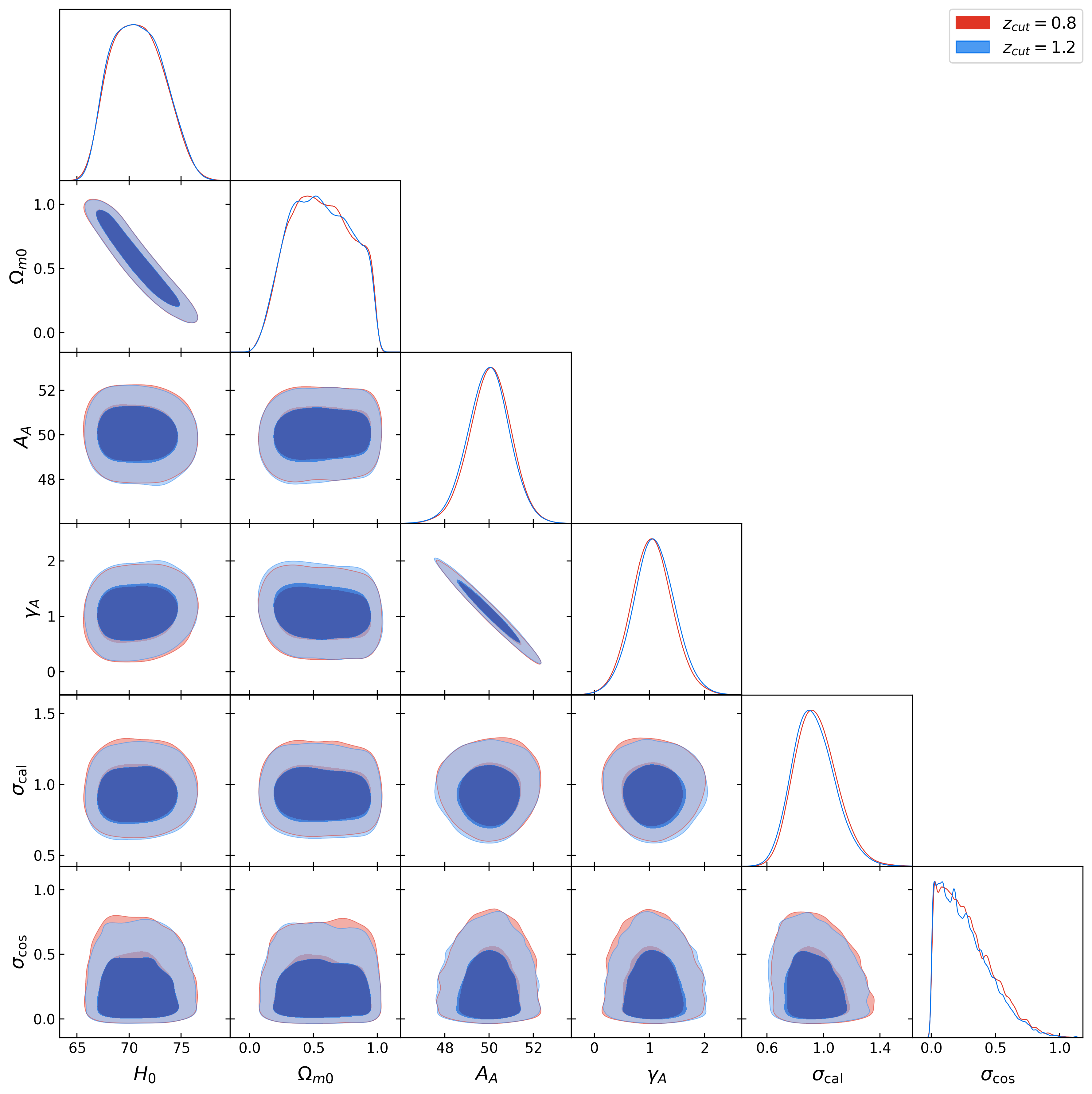}
	\includegraphics[width=1.0\linewidth]{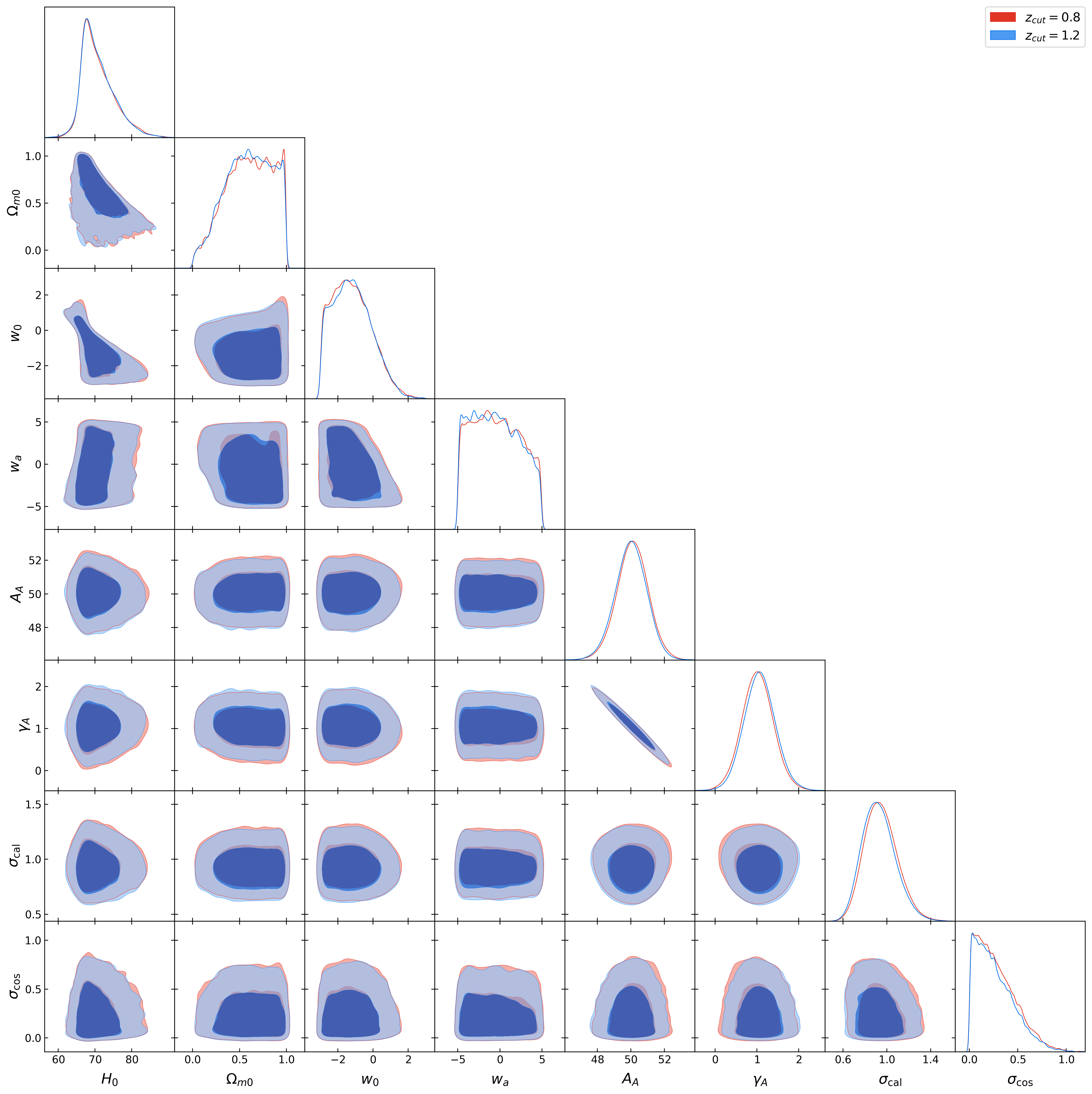}
	\caption{Posterior distributions for the Amati relation obtained with different choices of $z_{\rm cut}$. The contours indicate the $68\%$, $95\%$, and $99\%$ credible regions for $\Lambda$CDM (top) and $w_0w_a$CDM (bottom).}
	\label{figcorneramzc}
\end{figure}

\begin{figure}[ht!]
	\centering
	\includegraphics[width=1.0\linewidth]{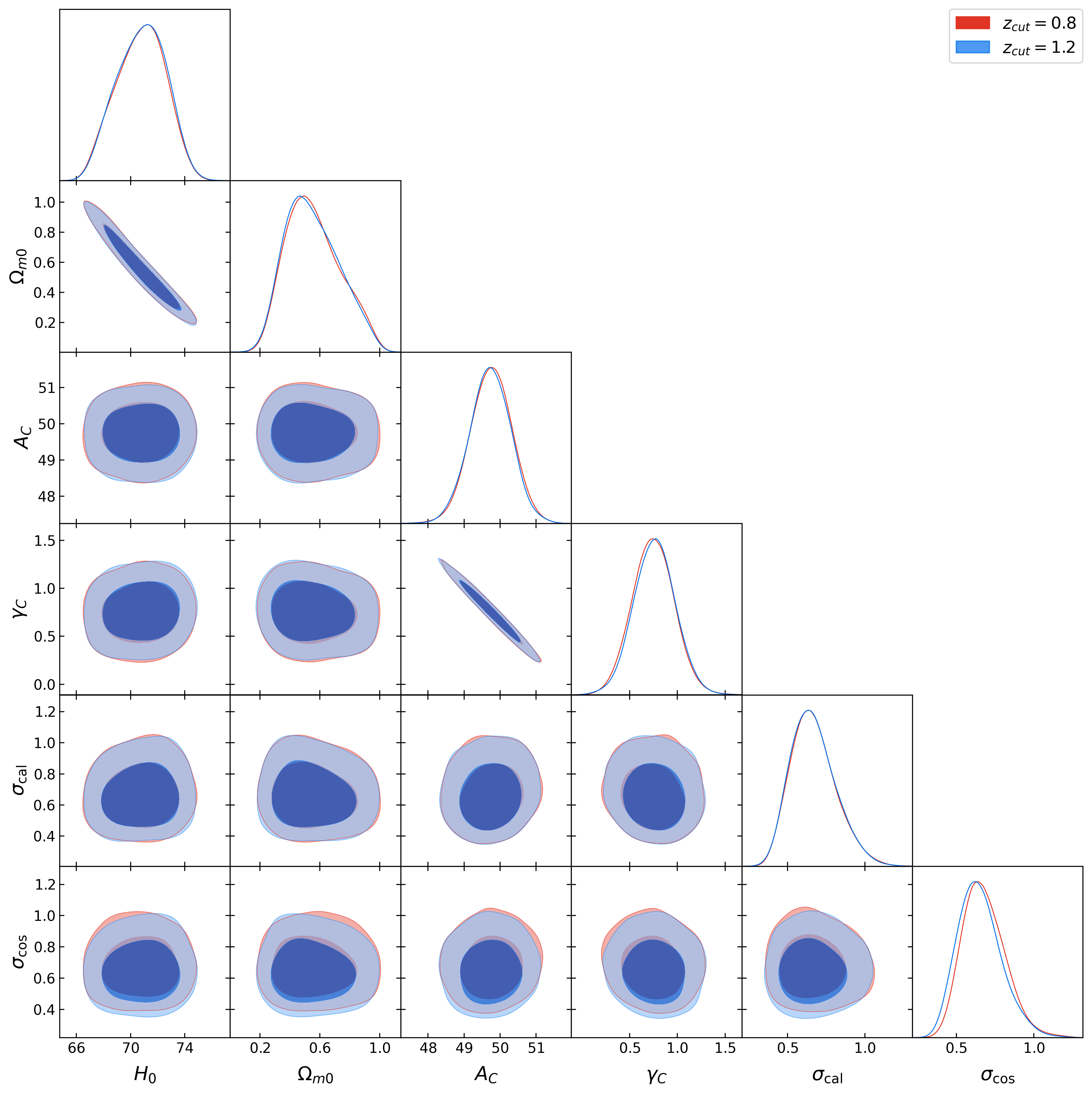}
	\includegraphics[width=1.0\linewidth]{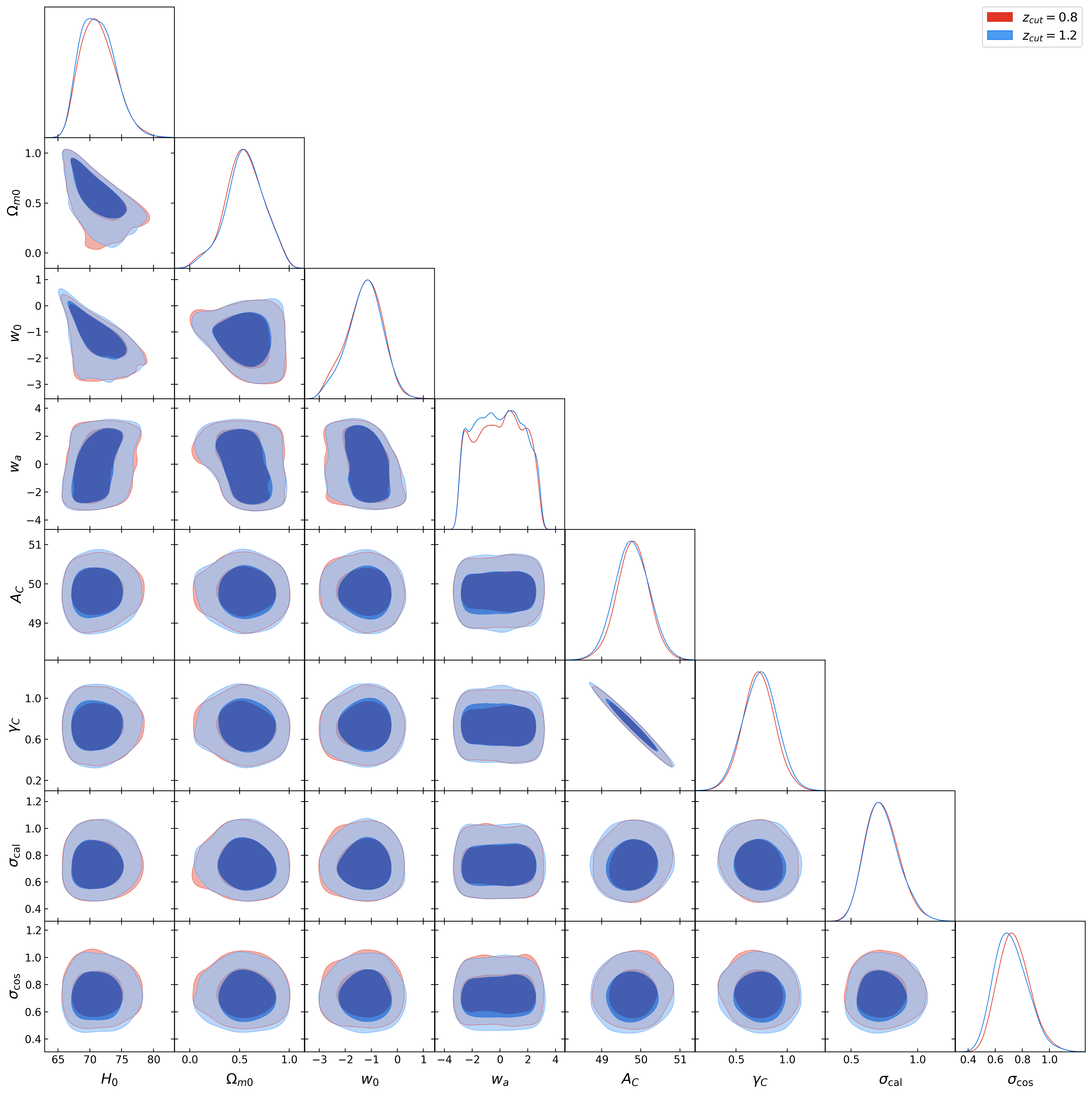}
	\caption{Same as Fig.~\ref{figcorneramzc}, but for the Combo relation.}
	\label{figcornercozc}
\end{figure}

We repeated the full hierarchical inference for $z_{\rm cut}=0.8$, $1.0$, and $1.2$. The resulting posterior constraints are summarised in Table~\ref{tab:appC_zcut_cosmo}, while Figs.~\ref{figcorneramzc} and \ref{figcornercozc} show the corresponding posterior comparisons for the Amati and Combo analyses.

\begin{table*}[htbp]
	\centering
	\caption{Robustness of the cosmological constraints to the choice of $z_{\rm cut}$.}
	\label{tab:appC_zcut_cosmo}
	\begin{tabular}{lcccc}
		\hline\hline
		Parameter & Amati ($\Lambda$CDM) & Amati ($w_0w_a$CDM) & Combo ($\Lambda$CDM) & Combo ($w_0w_a$CDM) \\
		\midrule
		\multicolumn{5}{l}{Cosmological parameters} \\
		\cmidrule(r){1-1}\cmidrule(lr){2-2}\cmidrule(lr){3-3}\cmidrule(lr){4-4}\cmidrule(l){5-5}
		$H_0$ ($z_{\rm cut}=0.8$) & $70.74_{-2.53\,(-4.09)}^{+2.69\,(+4.87)}${\scriptsize\,[0.006]} & $70.05_{-3.18\,(-5.66)}^{+5.41\,(+12.00)}${\scriptsize\,[0.139]} & $70.87_{-2.07\,(-3.61)}^{+1.78\,(+3.19)}${\scriptsize\,[0.021]} & $71.12_{-2.56\,(-4.25)}^{+3.01\,(+6.06)}${\scriptsize\,[0.031]} \\
		$H_0$ ($z_{\rm cut}=1.0$) & $70.75_{-2.53\,(-4.11)}^{+2.71\,(+4.94)}${\scriptsize\,[ref]} & $70.58_{-3.19\,(-5.17)}^{+4.51\,(+9.41)}${\scriptsize\,[ref]} & $70.91_{-2.07\,(-3.66)}^{+1.78\,(+3.23)}${\scriptsize\,[ref]} & $71.03_{-2.68\,(-4.30)}^{+3.03\,(+6.03)}${\scriptsize\,[ref]} \\
		$H_0$ ($z_{\rm cut}=1.2$) & $70.74_{-2.54\,(-4.03)}^{+2.71\,(+4.84)}${\scriptsize\,[0.004]} & $70.03_{-3.18\,(-5.88)}^{+5.36\,(+11.40)}${\scriptsize\,[0.143]} & $70.91_{-2.08\,(-3.67)}^{+1.80\,(+3.18)}${\scriptsize\,[0.000]} & $71.08_{-2.60\,(-4.24)}^{+2.90\,(+5.97)}${\scriptsize\,[0.016]} \\
		\addlinespace
		$\Omega_m$ ($z_{\rm cut}=0.8$) & $0.557_{-0.246\,(-0.405)}^{+0.272\,(+0.414)}${\scriptsize\,[0.016]} & $0.613_{-0.277\,(-0.502)}^{+0.268\,(+0.372)}${\scriptsize\,[0.102]} & $0.536_{-0.164\,(-0.275)}^{+0.223\,(+0.403)}${\scriptsize\,[0.027]} & $0.549_{-0.186\,(-0.436)}^{+0.207\,(+0.388)}${\scriptsize\,[0.014]} \\
		$\Omega_m$ ($z_{\rm cut}=1.0$) & $0.553_{-0.247\,(-0.410)}^{+0.274\,(+0.418)}${\scriptsize\,[ref]} & $0.588_{-0.238\,(-0.472)}^{+0.247\,(+0.387)}${\scriptsize\,[ref]} & $0.531_{-0.164\,(-0.281)}^{+0.220\,(+0.405)}${\scriptsize\,[ref]} & $0.552_{-0.191\,(-0.437)}^{+0.214\,(+0.393)}${\scriptsize\,[ref]} \\
		$\Omega_m$ ($z_{\rm cut}=1.2$) & $0.552_{-0.244\,(-0.403)}^{+0.274\,(+0.416)}${\scriptsize\,[0.003]} & $0.605_{-0.271\,(-0.496)}^{+0.265\,(+0.377)}${\scriptsize\,[0.071]} & $0.532_{-0.164\,(-0.279)}^{+0.215\,(+0.403)}${\scriptsize\,[0.004]} & $0.558_{-0.181\,(-0.424)}^{+0.200\,(+0.372)}${\scriptsize\,[0.027]} \\
		\addlinespace
		$w_0$ ($z_{\rm cut}=0.8$) & -- & $-1.271_{-1.090\,(-1.62)}^{+1.240\,(+2.44)}${\scriptsize\,[0.016]} & -- & $-1.225_{-0.773\,(-1.50)}^{+0.652\,(+1.20)}${\scriptsize\,[0.074]} \\
		$w_0$ ($z_{\rm cut}=1.0$) & -- & $-1.287_{-0.993\,(-1.59)}^{+1.04\,(+2.02)}${\scriptsize\,[ref]} & -- & $-1.174_{-0.744\,(-1.54)}^{+0.634\,(+1.28)}${\scriptsize\,[ref]} \\
		$w_0$ ($z_{\rm cut}=1.2$) & -- & $-1.228_{-1.120\,(-1.66)}^{+1.200\,(+2.35)}${\scriptsize\,[0.057]} & -- & $-1.205_{-0.719\,(-1.52)}^{+0.627\,(+1.24)}${\scriptsize\,[0.044]} \\
		\addlinespace
		$w_a$ ($z_{\rm cut}=0.8$) & -- & $-0.419_{-3.04\,(-4.33)}^{+3.31\,(+5.07)}${\scriptsize\,[0.054]} & -- & $-0.005_{-2.01\,(-2.84)}^{+1.85\,(+2.75)}${\scriptsize\,[0.013]} \\
		$w_a$ ($z_{\rm cut}=1.0$) & -- & $-0.585_{-2.96\,(-4.18)}^{+3.16\,(+5.09)}${\scriptsize\,[ref]} & -- & $-0.030_{-1.98\,(-2.81)}^{+1.84\,(+2.78)}${\scriptsize\,[ref]} \\
		$w_a$ ($z_{\rm cut}=1.2$) & -- & $-0.579_{-2.99\,(-4.20)}^{+3.30\,(+5.14)}${\scriptsize\,[0.002]} & -- & $-0.088_{-1.90\,(-2.76)}^{+1.88\,(+2.89)}${\scriptsize\,[0.031]} \\
		\addlinespace
		\multicolumn{5}{l}{Correlation parameters} \\
		\cmidrule(r){1-1}\cmidrule(lr){2-2}\cmidrule(lr){3-3}\cmidrule(lr){4-4}\cmidrule(l){5-5}
		$A$ ($z_{\rm cut}=0.8$) & $50.07_{-0.94\,(-1.93)}^{+0.90\,(+1.80)}${\scriptsize\,[0.014]} & $50.11_{-0.92\,(-1.90)}^{+0.91\,(+1.84)}${\scriptsize\,[0.050]} & $49.77_{-0.57\,(-1.13)}^{+0.57\,(+1.09)}${\scriptsize\,[0.035]} & $49.80_{-0.40\,(-0.84)}^{+0.41\,(+0.81)}${\scriptsize\,[0.015]} \\
		$A$ ($z_{\rm cut}=1.0$) & $50.06_{-0.96\,(-1.98)}^{+0.90\,(+1.82)}${\scriptsize\,[ref]} & $50.07_{-0.95\,(-1.93)}^{+0.90\,(+1.80)}${\scriptsize\,[ref]} & $49.75_{-0.55\,(-1.12)}^{+0.56\,(+1.11)}${\scriptsize\,[ref]} & $49.79_{-0.42\,(-0.836)}^{+0.428\,(+0.855)}${\scriptsize\,[ref]} \\
		$A$ ($z_{\rm cut}=1.2$) & $49.99_{-0.95\,(-1.94)}^{+0.91\,(+1.84)}${\scriptsize\,[0.070]} & $50.06_{-0.95\,(-1.94)}^{+0.90\,(+1.82)}${\scriptsize\,[0.051]} & $49.76_{-0.55\,(-1.12)}^{+0.56\,(+1.09)}${\scriptsize\,[0.024]} & $49.77_{-0.40\,(-0.82)}^{+0.41\,(+0.83)}${\scriptsize\,[0.056]} \\
		\addlinespace
		$\gamma$ ($z_{\rm cut}=0.8$) & $1.047_{-0.362\,(-0.734)}^{+0.380\,(+0.784)}${\scriptsize\,[0.038]} & $1.017_{-0.355\,(-0.722)}^{+0.380\,(+0.786)}${\scriptsize\,[0.089]} & $0.747_{-0.206\,(-0.417)}^{+0.210\,(+0.433)}${\scriptsize\,[0.056]} & $0.723_{-0.152\,(-0.305)}^{+0.153\,(+0.312)}${\scriptsize\,[0.052]} \\
		$\gamma$ ($z_{\rm cut}=1.0$) & $1.061_{-0.362\,(-0.721)}^{+0.378\,(+0.778)}${\scriptsize\,[ref]} & $1.050_{-0.360\,(-0.719)}^{+0.375\,(+0.770)}${\scriptsize\,[ref]} & $0.759_{-0.214\,(-0.424)}^{+0.212\,(+0.431)}${\scriptsize\,[ref]} & $0.731_{-0.163\,(-0.327)}^{+0.160\,(+0.323)}${\scriptsize\,[ref]} \\
		$\gamma$ ($z_{\rm cut}=1.2$) & $1.085_{-0.354\,(-0.713)}^{+0.388\,(+0.792)}${\scriptsize\,[0.067]} & $1.033_{-0.360\,(-0.714)}^{+0.378\,(+0.773)}${\scriptsize\,[0.050]} & $0.750_{-0.213\,(-0.431)}^{+0.209\,(+0.426)}${\scriptsize\,[0.035]} & $0.724_{-0.153\,(-0.307)}^{+0.154\,(+0.316)}${\scriptsize\,[0.048]} \\
		\addlinespace
		\multicolumn{5}{l}{Intrinsic scatter parameters} \\
		\cmidrule(r){1-1}\cmidrule(lr){2-2}\cmidrule(lr){3-3}\cmidrule(lr){4-4}\cmidrule(l){5-5}
		$\sigma_{\mathrm{cal}}$ ($z_{\rm cut}=0.8$) & $0.940_{-0.137\,(-0.257)}^{+0.160\,(+0.334)}${\scriptsize\,[0.057]} & $0.940_{-0.136\,(-0.252)}^{+0.160\,(+0.334)}${\scriptsize\,[0.069]} & $0.655_{-0.136\,(-0.244)}^{+0.156\,(+0.333)}${\scriptsize\,[0.027]} & $0.726_{-0.120\,(-0.215)}^{+0.135\,(+0.283)}${\scriptsize\,[0.022]} \\
		$\sigma_{\mathrm{cal}}$ ($z_{\rm cut}=1.0$) & $0.932_{-0.138\,(-0.256)}^{+0.161\,(+0.337)}${\scriptsize\,[ref]} & $0.930_{-0.138\,(-0.255)}^{+0.159\,(+0.331)}${\scriptsize\,[ref]} & $0.659_{-0.133\,(-0.239)}^{+0.155\,(+0.330)}${\scriptsize\,[ref]} & $0.729_{-0.119\,(-0.213)}^{+0.138\,(+0.292)}${\scriptsize\,[ref]} \\
		$\sigma_{\mathrm{cal}}$ ($z_{\rm cut}=1.2$) & $0.924_{-0.137\,(-0.252)}^{+0.162\,(+0.338)}${\scriptsize\,[0.054]} & $0.921_{-0.137\,(-0.250)}^{+0.162\,(+0.337)}${\scriptsize\,[0.073]} & $0.656_{-0.134\,(-0.240)}^{+0.156\,(+0.333)}${\scriptsize\,[0.021]} & $0.733_{-0.119\,(-0.214)}^{+0.140\,(+0.296)}${\scriptsize\,[0.032]} \\
		\addlinespace
		$\sigma_{\mathrm{cos}}$ ($z_{\rm cut}=0.8$) & $0.245_{-0.160\,(-0.221)}^{+0.247\,(+0.514)}${\scriptsize\,[0.056]} & $0.244_{-0.159\,(-0.219)}^{+0.250\,(+0.525)}${\scriptsize\,[0.051]} & $0.668_{-0.123\,(-0.222)}^{+0.148\,(+0.312)}${\scriptsize\,[0.186]} & $0.733_{-0.103\,(-0.187)}^{+0.119\,(+0.251)}${\scriptsize\,[0.232]} \\
		$\sigma_{\mathrm{cos}}$ ($z_{\rm cut}=1.0$) & $0.234_{-0.163\,(-0.223)}^{+0.240\,(+0.500)}${\scriptsize\,[ref]} & $0.233_{-0.162\,(-0.222)}^{+0.243\,(+0.507)}${\scriptsize\,[ref]} & $0.642_{-0.127\,(-0.232)}^{+0.149\,(+0.312)}${\scriptsize\,[ref]} & $0.706_{-0.113\,(-0.204)}^{+0.127\,(+0.269)}${\scriptsize\,[ref]} \\
		$\sigma_{\mathrm{cos}}$ ($z_{\rm cut}=1.2$) & $0.230_{-0.163\,(-0.223)}^{+0.239\,(+0.496)}${\scriptsize\,[0.008]} & $0.231_{-0.162\,(-0.221)}^{+0.241\,(+0.505)}${\scriptsize\,[0.009]} & $0.641_{-0.128\,(-0.229)}^{+0.148\,(+0.325)}${\scriptsize\,[0.009]} & $0.711_{-0.112\,(-0.200)}^{+0.128\,(+0.273)}${\scriptsize\,[0.033]} \\
		\hline
	\end{tabular}
	\par\smallskip
	\begin{minipage}{\linewidth}
		\footnotesize
		\raggedright
		Notes: Uncertainties are reported as $X_{-a(-c)}^{+b(+d)}$, where $a$ and $b$ denote the lower and upper uncertainties at the 68\% credible level, respectively, and $c$ and $d$ denote those at the 95\% credible level. Square brackets give $\Delta(p)/\sigma \equiv |p(z_{\rm cut})-p(z_{\rm cut}=1)|/\sigma_p(z_{\rm cut}=1)$, where $\sigma_p$ is the mean of the upper and lower 68\% uncertainties of the baseline result. The entries for $z_{\rm cut}=1$ correspond to the results in Table~\ref{tab:cosmo_results}. $H_0$ is given in units of $\mathrm{km\,s^{-1}\,Mpc^{-1}}$.
	\end{minipage}
\end{table*}

All parameter shifts remain below approximately $0.24\sigma$ relative to the baseline analysis. The inferred constraints are therefore insensitive to the tested variation of $z_{\rm cut}$.

\subsection{Robustness to the GRB sample definition}
\label{app:robust_sample}

We further examined the sensitivity of the ANN-calibrated inference to alternative GRB sample definitions. For each relation, we considered the baseline sample, prompt-brightness cuts, measurement-quality cuts, and combined brightness-plus-quality cuts, and repeated the same inference pipeline. These tests assess the dependence of the results on sample composition; they are not intended to model or correct the underlying GRB selection function.

\begin{figure}
	\centering
	\includegraphics[width=0.5\textwidth]{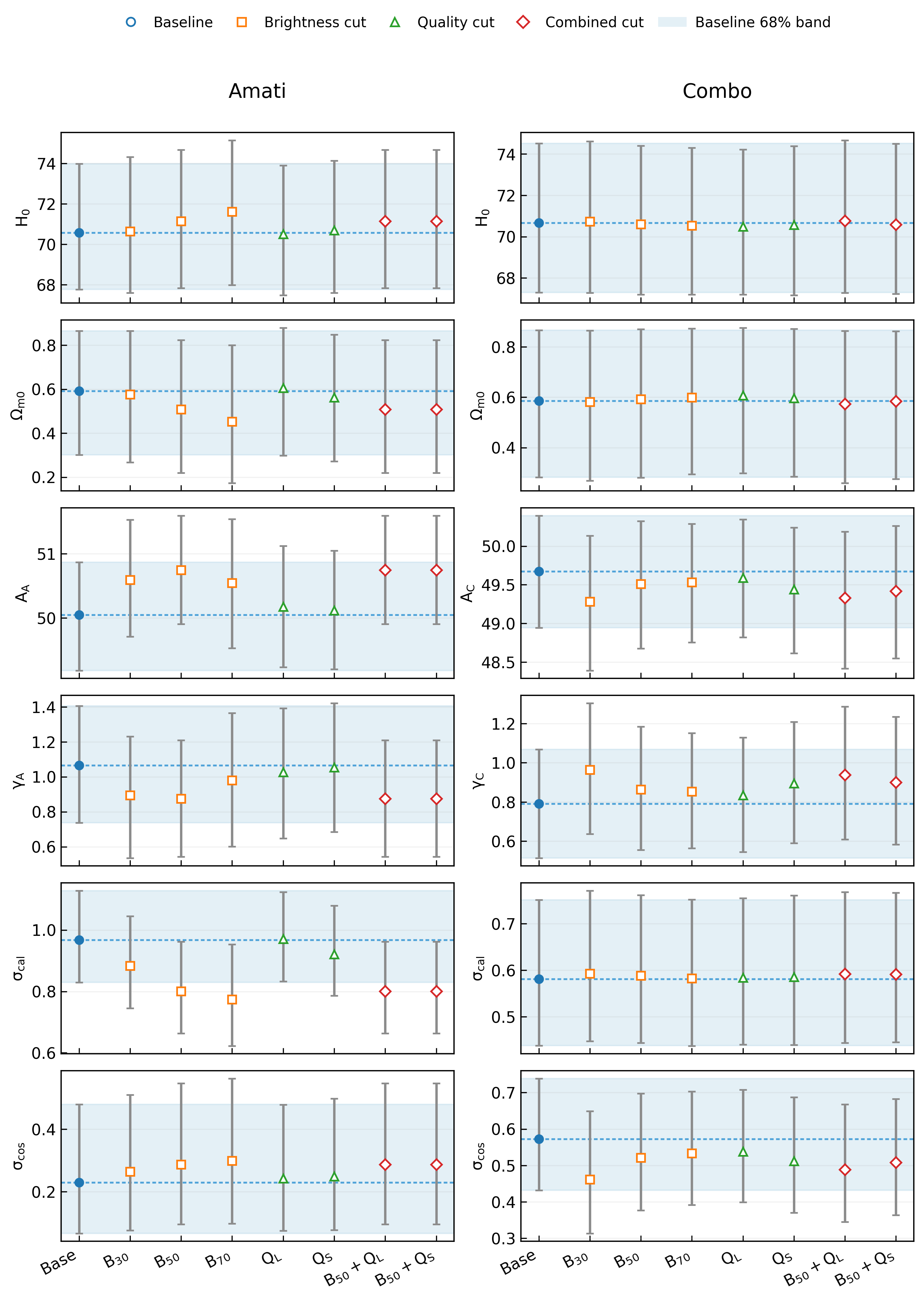}
	\caption{Robustness of the Amati (left) and Combo (right) analyses in flat $\Lambda$CDM for alternative sample definitions. Points mark the posterior medians, vertical bars indicate the corresponding credible intervals, and shaded regions show the baseline 68\% credible intervals.}
	\label{fig:robust_amati_combo_lcdm}
\end{figure}

\begin{figure}
	\centering
	\includegraphics[width=0.5\textwidth]{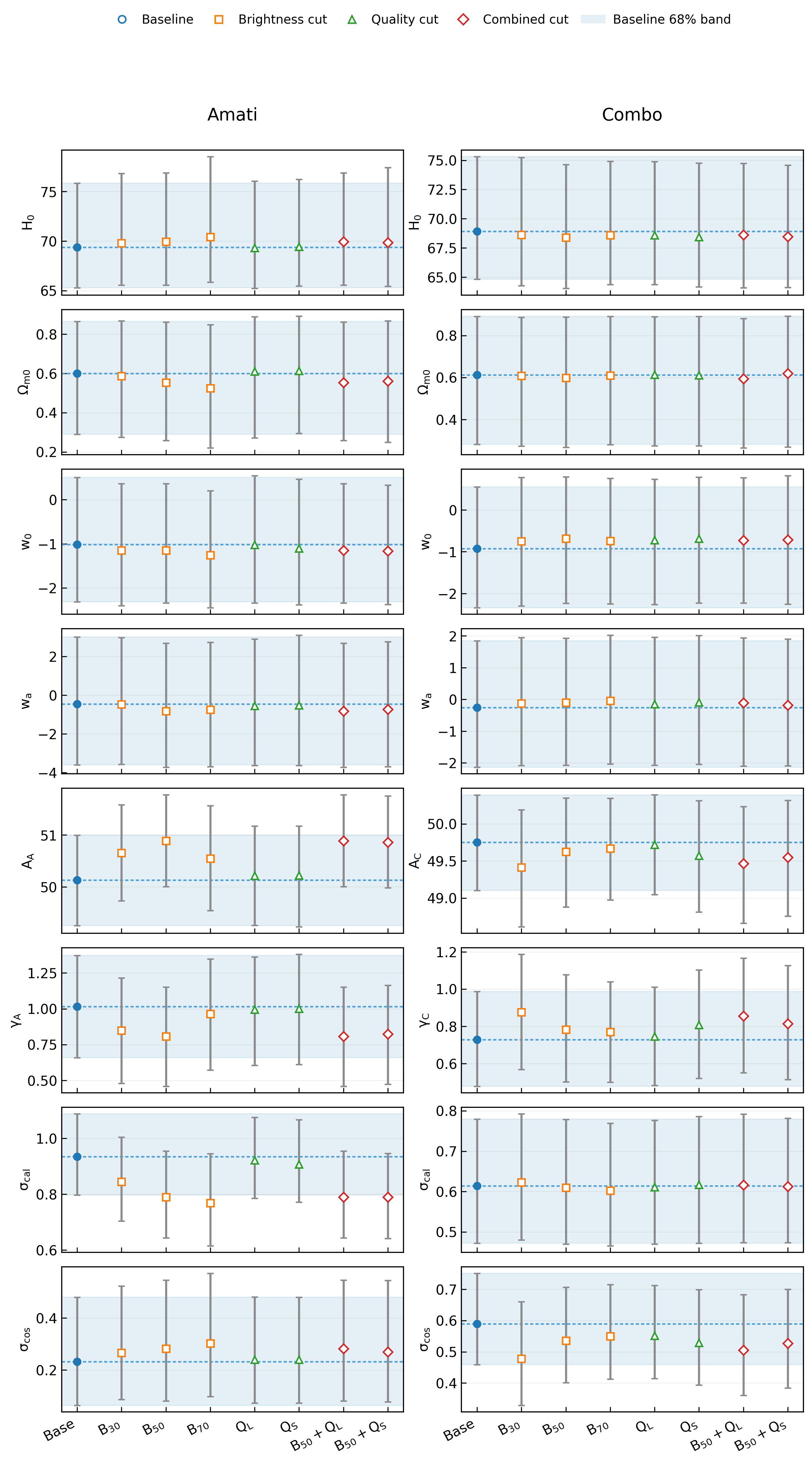}
	\caption{Same as Fig.~\ref{fig:robust_amati_combo_lcdm}, but for flat $w_0w_a$CDM.}
	\label{fig:robust_amati_combo_w0wa}
\end{figure}

Figures~\ref{fig:robust_amati_combo_lcdm} and \ref{fig:robust_amati_combo_w0wa} summarise the resulting posterior medians and credible intervals. The inferred cosmological and correlation parameters remain within the uncertainty scale of the baseline analysis across the tested subsamples. The $w_0w_a$CDM results are likewise stable, with no evidence that the broad dark-energy posteriors are driven by a particular sample definition.

\section{Sensitivity to the SN Ia distance zero point}
\label{app:zero_point}

Because the ANN reconstruction inherits the Pantheon$+$ SH0ES absolute distance scale, the $H_0$ inferred from the calibrated GRB Hubble diagram is conditional on this calibration. We therefore performed a controlled zero-point sensitivity test using the 77-event Amati sample with $z_{\rm cut}=1$. The Pantheon$+$ SH0ES value $H_0=73.04~\mathrm{km\,s^{-1}\,Mpc^{-1}}$ is adopted as the baseline, while $H_0=69.96$ and $69.03~\mathrm{km\,s^{-1}\,Mpc^{-1}}$ are used as alternative zero points motivated by the Chicago--Carnegie Hubble Program (CCHP) determinations discussed by \citet{2025ApJ...985..203F}. This test probes the response to a common distance rescaling and is not an independent $H_0$ determination or a full CCHP recalibration. For these auxiliary runs, the ANN uncertainties are represented by the diagonal reconstruction errors.

Changing the adopted absolute scale rescales the luminosity distance and therefore shifts the distance modulus and the Amati intercept. Since $E_{\rm iso}\propto d_L^2$, the expected transformations are
\begin{equation}
	\begin{aligned}
		\frac{d_L'}{d_L} &= \frac{H_{0,\rm base}}{H_{0,\rm alt}}, \\
		\Delta\mu &= 5\log_{10}\left(\frac{H_{0,\rm base}}{H_{0,\rm alt}}\right), \\
		\Delta A_A &= 2\log_{10}\left(\frac{H_{0,\rm base}}{H_{0,\rm alt}}\right)
		= \frac{2}{5}\Delta\mu .
	\end{aligned}
\end{equation}
A pure zero-point change therefore shifts mainly the normalisation of the Amati relation, while leaving its slope and redshift dependence approximately unchanged.
\begin{table}[htbp]
	\centering
	\small
	\setlength{\tabcolsep}{3pt}
	\caption{Zero-point sensitivity of the 77-event Amati sample.}
	\label{tab:zero_point}
	\begin{tabular}{@{}lcccc@{}}
		\hline\hline
		Calibration & $H_0$ & $d_L'/d_L$ & $\Delta\mu$ & $\Delta A_A^{\rm exp}$ \\
		& $(\mathrm{km\,s^{-1}\,Mpc^{-1}})$ & & (mag) & \\
		\hline
		Pantheon$+$ SH0ES & 73.04 & 1.0000 & 0 & 0 \\
		CCHP alt. 1 & 69.96 & 1.0440 & 0.0936 & 0.0374 \\
		CCHP alt. 2 & 69.03 & 1.0581 & 0.1226 & 0.0490 \\
		\hline
	\end{tabular}
	\par\smallskip
	\begin{minipage}{\linewidth}
		\footnotesize
		\raggedright
		Notes: The last three columns give the expected changes relative to the Pantheon$+$ SH0ES baseline.
	\end{minipage}
\end{table}

The recovered Amati intercept shifts in the expected direction and by an amount consistent with the analytic rescaling in Table~\ref{tab:zero_point}. The slope, intrinsic scatter, and cosmological shape parameters remain nearly unchanged. The largest displacement of any non-baseline parameter is $0.070\sigma$ in flat $\Lambda$CDM and $0.110\sigma$ in $w_0w_a$CDM, with all shifts below $0.12\sigma$.

Thus, the tested zero-point changes are absorbed primarily by the normalisation of the GRB correlation. The shape of the calibrated high-redshift Hubble diagram is correspondingly stable, whereas its absolute scale, and hence any inferred $H_0$, remains conditional on the external SNe Ia calibration. The broad $H_0$ and $\Omega_m$ posteriors from these auxiliary runs should not be interpreted as new measurements associated with the alternative zero points.
	
\section{Pre-calibration validation and external consistency}
\label{app:precal}

Calibrating the GRB correlations with an externally reconstructed distance scale avoids assuming a cosmological model during calibration, but it does not by itself establish that the correlation parameters are insensitive to the cosmological background. We therefore performed a pre-calibration analysis in which the uncalibrated GRB observables and cosmological parameters were fitted simultaneously. Following the multi-model strategy of \citet{2020MNRAS.499..391K} and \citet{2021JCAP...09..042K}, we considered flat and non-flat $\Lambda$CDM and flat and non-flat XCDM. No ANN-reconstructed distances are used. Since the absolute scale is degenerate with the correlation intercept in the GRB-only analysis, we fixed $H_0=70~\mathrm{km\,s^{-1}\,Mpc^{-1}}$.

Tables~\ref{tab:app_precal_amati} and \ref{tab:app_precal_combo} give the fitted correlation parameters for A118, our 77-event Amati sample, and Combo/C174 under the four cosmological backgrounds. The quoted values are posterior medians with 68\% credible intervals.
\begin{table}[htbp]
	\centering
	\scriptsize
	\caption{Pre-calibration correlation parameters for A118 and the 77-event Amati sample.}
	\label{tab:app_precal_amati}
	\begin{tabular}{@{}lccc@{}}
		\hline\hline
		Model & $A_A$ & $\gamma_A$ & $\sigma_{\rm int}^A$ \\
		\hline
		\multicolumn{4}{c}{A118} \\
		\hline
		Flat $\Lambda$CDM & $50.092^{+0.237}_{-0.245}$ & $1.110^{+0.084}_{-0.081}$ & $0.406^{+0.030}_{-0.027}$ \\
		Non-flat $\Lambda$CDM & $50.096^{+0.244}_{-0.239}$ & $1.114^{+0.084}_{-0.088}$ & $0.407^{+0.030}_{-0.027}$ \\
		Flat XCDM & $50.097^{+0.263}_{-0.254}$ & $1.107^{+0.084}_{-0.085}$ & $0.407^{+0.031}_{-0.027}$ \\
		Non-flat XCDM & $50.094^{+0.253}_{-0.251}$ & $1.106^{+0.087}_{-0.085}$ & $0.407^{+0.031}_{-0.027}$ \\
		\hline
		\multicolumn{4}{c}{Amati, 77 events} \\
		\hline
		Flat $\Lambda$CDM & $50.164^{+0.368}_{-0.373}$ & $1.019^{+0.146}_{-0.143}$ & $0.580^{+0.053}_{-0.046}$ \\
		Non-flat $\Lambda$CDM & $50.249^{+0.378}_{-0.379}$ & $0.997^{+0.146}_{-0.147}$ & $0.576^{+0.053}_{-0.045}$ \\
		Flat XCDM & $50.172^{+0.376}_{-0.376}$ & $1.014^{+0.147}_{-0.146}$ & $0.579^{+0.052}_{-0.045}$ \\
		Non-flat XCDM & $50.290^{+0.396}_{-0.405}$ & $0.976^{+0.146}_{-0.146}$ & $0.575^{+0.053}_{-0.044}$ \\
		\hline
	\end{tabular}
	\par\smallskip
	\begin{minipage}{\linewidth}
		\footnotesize
		\raggedright
		Notes: $H_0$ is fixed at $70~\mathrm{km\,s^{-1}\,Mpc^{-1}}$.
	\end{minipage}
\end{table}

\begin{table}[htbp]
	\centering
	\scriptsize
	\caption{Same as Table~\ref{tab:app_precal_amati}, but for Combo/C174.}
	\label{tab:app_precal_combo}
	\begin{tabular}{@{}lccc@{}}
		\hline\hline
		Model & $A_C$ & $\gamma_C$ & $\sigma_{\rm int}^C$ \\
		\hline
		Flat $\Lambda$CDM & $49.691^{+0.200}_{-0.201}$ & $0.733^{+0.076}_{-0.078}$ & $0.355^{+0.025}_{-0.022}$ \\
		Non-flat $\Lambda$CDM & $49.740^{+0.203}_{-0.202}$ & $0.719^{+0.078}_{-0.078}$ & $0.354^{+0.025}_{-0.023}$ \\
		Flat XCDM & $49.683^{+0.200}_{-0.200}$ & $0.729^{+0.074}_{-0.077}$ & $0.354^{+0.025}_{-0.023}$ \\
		Non-flat XCDM & $49.704^{+0.218}_{-0.213}$ & $0.704^{+0.077}_{-0.081}$ & $0.353^{+0.025}_{-0.022}$ \\
		\hline
	\end{tabular}
\end{table}

The correlation parameters overlap closely across the four cosmological backgrounds. Their maximum joint displacements are $0.14\sigma$, $0.24\sigma$, and $0.59\sigma$ for A118, Amati, and Combo/C174, respectively. The A118 parameters are also consistent with the published results within $1\sigma$ \citep{2021JCAP...09..042K}, providing a direct check of our implementation. We therefore find no statistically significant cosmology dependence of the fitted correlation parameters for the three samples considered here.

\begin{table}[htbp]
	\centering
	\scriptsize
	\caption{Comparison of GRB-only and OHD constraints in flat $\Lambda$CDM.}
	\label{tab:app_precal_summary}
	\begin{tabular}{@{}lccccc@{}}
		\hline\hline
		Sample & $N$ & $\Omega_m$ & $P(q_0<0)$ & $\Delta_{\rm corr}$ & $\Delta_{\rm CC}$ \\
		\hline
		A118 & 118 & $0.649^{+0.229}_{-0.262}$ & 0.527 & $0.14\sigma$ & $1.46\sigma$ \\
		Amati & 77 & $0.763^{+0.164}_{-0.249}$ & 0.342 & $0.24\sigma$ & $2.26\sigma$ \\
		Combo/C174 & 173 & $0.861^{+0.094}_{-0.159}$ & 0.116 & $0.59\sigma$ & $4.40\sigma$ \\
		OHD & 42 & $0.292^{+0.026}_{-0.026}$ & 1.000 & -- & -- \\
		\hline
	\end{tabular}
	\par\smallskip
	\begin{minipage}{\linewidth}
		\footnotesize
		\raggedright
		Notes: $H_0$ is fixed at $70~\mathrm{km\,s^{-1}\,Mpc^{-1}}$. $\Delta_{\rm corr}$ denotes the maximum joint shift in the correlation parameters among the four tested cosmologies. $\Delta_{\rm CC}$ denotes the difference between the marginal constraints on $\Omega_m$ from the GRB and OHD data.
	\end{minipage}
\end{table}

The stability of a GRB correlation does not necessarily imply cosmological consistency. We therefore compared the GRB-only constraints with the independent OHD cosmic-chronometer compilation. With the same fixed $H_0$, OHD gives $\Omega_m=0.292^{+0.026}_{-0.026}$ in flat $\Lambda$CDM and strongly favours current acceleration. A118 differs from the OHD matter-density constraint by $1.46\sigma$. The 77-event Amati sample is shifted towards higher $\Omega_m$ by $2.26\sigma$, although the OHD value remains within its broad 95\% credible interval. In contrast, Combo/C174 differs by $4.40\sigma$, with the OHD value lying outside its 95\% credible interval.

The acceleration probabilities lead to the same qualitative conclusion. In flat $\Lambda$CDM, we obtain $P(q_0<0)=0.527$, $0.342$, and $0.116$ for A118, Amati, and Combo/C174, respectively, whereas the corresponding probability for OHD is effectively unity. Thus, despite the stability of their correlation parameters, the present GRB-only samples do not independently establish current cosmic acceleration.

The fixed value of $H_0$ is adopted only to place the uncalibrated GRB and OHD analyses on the same footing. We therefore repeated the OHD analysis with $H_0$ free. Table~\ref{tab:app_cc42_freeh0} shows that OHD continues to favour current acceleration in all four cosmological models.
\begin{table}[htbp]
	\centering
	\small
	\caption{Robustness of the OHD constraints with $H_0$ treated as a free parameter.}
	\label{tab:app_cc42_freeh0}
	\begin{tabular}{@{}lcc@{}}
		\hline\hline
		Model & $H_0~(\mathrm{km\,s^{-1}\,Mpc^{-1}})$ & $P(q_0<0)$ \\
		\hline
		Flat $\Lambda$CDM & $67.52^{+2.53}_{-2.61}$ & 1.000 \\
		Non-flat $\Lambda$CDM & $67.27^{+3.56}_{-3.39}$ & 0.992 \\
		Flat XCDM & $69.05^{+6.88}_{-4.90}$ & 0.972 \\
		Non-flat XCDM & $68.00^{+5.71}_{-4.57}$ & 0.964 \\
		\hline
	\end{tabular}
	\par\smallskip
	\begin{minipage}{\linewidth}
		\footnotesize
		\raggedright
		Notes: Quoted uncertainties correspond to 68\% credible intervals.
	\end{minipage}
\end{table}

These tests distinguish correlation stability from cosmological constraining power. The Amati correlation remains stable across the tested backgrounds, while its broad matter-density posterior remains compatible with OHD at the 95\% level. Combo/C174 also shows stable correlation parameters but a substantially larger external discrepancy. This distinction motivates our use of the 77-event Amati sample for the main cosmological interpretation, with Combo/C174 retained primarily as an exploratory calibration cross-check.

\section{Matched-event comparison with A118}
\label{app:matched}

Our baseline Amati catalogue and A118 are constructed from different parent samples and use different prompt-emission measurements. To separate the effect of sample composition from that of the adopted measurements, we performed an event-by-event comparison of the two catalogues. The matching is based jointly on observing date and redshift rather than on burst name alone. This procedure identifies 43 confirmed common GRBs. One same-date object is rejected because the two catalogues assign incompatible redshifts (our GRB080916406 has $z=0.689$, whereas A118 lists GRB080916C at $z=4.35$). After applying the quality criterion used for our baseline sample, 41 common GRBs remain.

We then repeated the pre-calibration joint correlation--cosmology analysis using identical event lists but alternating between the prompt-emission quantities adopted in our catalogue and those reported for A118. This construction keeps the GRB population fixed and isolates the effect of the different measurements and data processing. Across the four cosmological models considered in Appendix~\ref{app:precal}, the inferred cosmological parameters change by less than approximately $0.4\sigma$. The same behaviour is found for both the 43-event matched sample and the stricter 41-event subset. The difference between the complete Amati and A118 cosmological posteriors therefore appears to arise mainly from the non-overlapping event populations rather than from the measurements of the common events alone.

The intrinsic scatter is more sensitive to the adopted prompt-emission measurements. For the matched samples, our measurements give $\sigma_{\rm int}\simeq0.57$--$0.59$ across the four cosmological models, whereas the A118 measurements give $\sigma_{\rm int}\simeq0.45$--$0.46$. The corresponding displacement is approximately $1.4$--$1.5\sigma$. For comparison, the full A118 sample gives $\sigma_{\rm int}\simeq0.406$ in flat $\Lambda$CDM, while our 77-event baseline sample gives $\sigma_{\rm int}\simeq0.580$. The fact that the scatter already decreases when the event list is held fixed suggests that differences in spectral processing and adopted prompt-emission quantities contribute to the tighter A118 correlation. The additional reduction obtained for the complete A118 catalogue further indicates that sample composition also plays a role.

These two effects should therefore be distinguished. Replacing the prompt-emission measurements for the same GRBs has only a small effect on the broad GRB-only cosmological constraints, but a more visible effect on the intrinsic scatter. Changing the event population can alter both the scatter and the cosmological posterior because the redshift distribution, measurement precision, and leverage of individual high-redshift bursts also change. The matched-event analysis thus suggests that the smaller A118 scatter is associated with both its measurement treatment and sample composition, whereas the larger difference between the complete cosmological posteriors is driven mainly by the non-overlapping events.

This comparison does not establish that either catalogue uniquely represents the underlying GRB population, nor does it replace an explicit treatment of detector and sample-selection effects. Its purpose is instead to identify which aspects of the inference are most sensitive to the catalogue construction. Together with the multi-model tests in Appendix~\ref{app:precal}, it shows that the Amati correlation parameters are stable against the tested cosmological backgrounds, while the quantitative scatter and the broad GRB-only cosmological constraints can still depend on the adopted measurements and event population.
	
\end{appendix}

\end{document}